\documentclass[a4paper,12pt]{article}
\usepackage[utf8]{inputenc}
\usepackage{jheppub}
\usepackage{tikz}
\usepackage[miktex]{gnuplottex}
\usetikzlibrary{shapes,arrows,positioning,automata,backgrounds,calc,er,patterns}
\usepackage{tikz-feynman}
\tikzfeynmanset{compat=1.0.0}
\usepackage{soul}
\usepackage{slashed} 
\usepackage{enumerate}
\usepackage{epsfig} 
\usepackage{float} 
\usepackage{subcaption} 
\usepackage{amsmath} 
\usepackage{wasysym} 
\usepackage{mathrsfs} 
\usepackage{amsfonts} 
\usepackage{amsbsy} 
\usepackage{amscd}    
\usepackage{multirow} 
\usepackage{tikz}
\usepackage{slashed}
\usepackage[miktex]{gnuplottex}
\usetikzlibrary{arrows,positioning,shapes.geometric} 
\usepackage{color}
\usepackage{verbatim}
\usepackage{xcolor}
\usepackage{aligned-overset}

\definecolor{person01}{rgb}{0.0, 0.64, 0.0} 
\definecolor{person02}{rgb}{2.2, 0.20, 0.60} 
\definecolor{juhi}{rgb}{1.99, 0.21, 1.33} 
\definecolor{jayita}{rgb}{2.55, 0.69, 0.0} 
\definecolor{person05}{rgb}{0.0, 1.39, 1.39} 
\definecolor{person06}{rgb}{1.58, 0.0, 0.0} 
\definecolor{julia1}{rgb}{1.0, 0.49, 0.0} 

\title{Search for Dark Matter in 2HDMS at LHC and future Lepton Colliders} 

\author[a,b]{Juhi Dutta,}   
\author[c]{Jayita Lahiri,}
\author[d]{Cheng Li,}
\author[c,e]{Gudrid Moortgat-Pick,}
\author[c]{Sheikh Farah Tabira}
\author[c]{and Julia Anabell Ziegler}
\affiliation[a]{Homer L. Dodge Department of Physics and Astronomy,
University of Oklahoma, Norman, OK 73019, USA}
\affiliation[b]{The Institute of Mathematical Sciences, 
4th Cross St, CIT Campus, Tharamani, Chennai, Tamil Nadu, India, 600113}
\affiliation[c]{II. Institut f{\"u}r Theoretische Physik,  Universit{\"a}t Hamburg, Luruper Chaussee 149, 22761 Hamburg, Germany}
\affiliation[d]{School of Science, Shenzhen Campus of Sun Yat-Sen University, Gongchang Road 66, 518107 Shenzhen, China}
\affiliation[e]{Deutsches Elektronen-Synchrotron DESY, 
Notkestr. 85, 22607 Hamburg, Germany}

\emailAdd{juhidutta@imsc.res.in}
\emailAdd{jayita.lahiri@desy.de}
\emailAdd{gudrid.moortgat-pick@desy.de}
\emailAdd{lich389@mail.sysu.edu.cn}
\emailAdd{sheikh.farah.tabira@desy.de}
\emailAdd{julia.ziegler@desy.de}
 
\abstract{We investigate the phenomenological prospects of the Two Higgs Doublet and Complex Singlet Scalar Extension (2HDMS) in the context of dark matter (DM) and Higgs phenomenology. The 2HDMS provides an enlarged Higgs sector along with a DM candidate. In this work, we perform an exhaustive scan to find representative benchmarks which are consistent with all theoretical and experimental constraints. We choose benchmarks with light, intermediate and massive DM masses and in some cases, also accommodate the 95 GeV excess in $b\bar{b}$ and $\gamma\gamma$ channels observed at the Large Electron-Positron Collider (LEP) and Large Hadron Collider (LHC). We investigate {and compare} different DM signal topologies { for different DM mass regions and focus on the relevant signatures at all current and future colliders such as }the LHC and at proposed future lepton colliders including electron-positron and muon colliders. Using a cut and count analysis, we show that while the High Luminosity LHC (HL-LHC) may hint towards new physics, future lepton colliders prove to be efficient discovery probes for {dark matter searches at colliders.}} 

\preprint{DESY-25-061}
\begin{document} 

\maketitle

\newpage
\section{Introduction}

The Standard Model (SM) of particle physics has established itself on strong grounds over time. At the same time, the existence of dark matter (DM) and indirect evidence of it via gravitational interaction has been found in galaxy rotation curves, the Bullet cluster data~\cite{Barrena:2002dp} as well as the cosmic microwave background~\cite{Planck:2018vyg}. Yet, the SM does not provide a suitable DM candidate. The most recent discovery of a 125 GeV Higgs boson, on the one hand, solidified the viability of the SM further, on the other hand, it opened up the possibility of an extended scalar sector. Therefore the Higgs boson and its properties have become a crucial sector to probe New Physics. Furthermore, searches for extra scalars are ongoing at the collider experiments. The extended scalar sector where extra doublets and singlets are appended to the SM Higgs sector, can also provide an answer to the DM problem. In this context, we have considered a complex singlet extension of the two Higgs doublet model (2HDMS)~\cite{Baum:2018zhf}. Such a scenario can provide a scalar DM candidate if suitable symmetries are imposed on the scalar potential.

In an earlier work~\cite{Dutta:2023cig}, we studied the DM phenomenology of this model and identified the allowed parameter space from DM constraints such as the observed relic density, upper bounds on direct and indirect detection cross-sections from various experiments (LUX-ZEPLIN (LZ) and Fermi-LAT (Fermi)). Direct and indirect detection experiments act as a strong constraint on DM models. Simultaneously, DM production at the collider has also been probed at the LHC. In our past work~\cite{Dutta:2023cig}, we presented the predictions for a benchmark scenario, allowed by all constraints at future colliders.

In the current work, we investigate DM search at future colliders in the context of 2HDMS, as a reference model.  We perform an exhaustive study, of the model parameter space, identifying regions that are allowed by existing data and at the same time are interesting from the point of view of DM phenomenology. We further mention that a new excess has been observed both at the former Large Electron-Positron Collider (LEP)~\cite{LEPWorkingGroupforHiggsbosonsearches:2003ing} in the $b\bar{b}$ mode as well as in the  $\gamma\gamma$ final state at the Large Hadron Collider (LHC) experiments CMS and ATLAS~\cite{CMS:2023yay,ATLAS:2023jzc}. {Such an excess can be explained in several BSM models with extended scalar sectors~\cite{Fox:2017uwr, Haisch:2017gql,  Benbrik:2022azi, Benbrik:2022dja, Azevedo:2023zkg, Belyaev:2023xnv,Biekotter:2019kde,Aguilar-Saavedra:2023vpd, Banik:2023ecr,Heinemeyer:2021msz,Biekotter:2021ovi, Biekotter:2023jld, Biekotter:2023oen,Cao:2016uwt, Cao:2019ofo, Choi:2019yrv, Biekotter:2021qbc, Li:2022etb,PhysRevD.100.035023, Biekotter:2017xmf, Liu:2018ryo, Liu:2018xsw, Biekotter:2019gtq, Richard:2017kot, Aguilar-Saavedra:2020wrj, Biekotter:2020cjs, Coloretti:2023wng, Bhattacharya:2023lmu, Ashanujjaman:2023etj, Escribano:2023hxj,Borah:2023hqw}. 
In the context of 2HDM extensions, it} was shown in~\cite{Heinemeyer:2021msz} and \cite{Biek_tter_2024}, such an excess can be accommodated within the 2HDMS scenario. In the current study, we incorporated this excess in some of our chosen benchmarks. Furthermore, we focus on choosing benchmarks for our collider study, that can be probed at future colliders. In this regard, we intend to perform a comparative analysis, in order to establish a complementarity between various proposed future colliders such as High Luminosity LHC (HL-LHC), International Linear Collider (ILC), Future Circular Collider in electron-positron mode (FCC-ee), Compact Linear Collider (CLIC) and muon collider, in terms of detection prospects and identify regions of parameter space that can be probed at certain colliders. Furthermore, we also point out the relevant channels that can probe certain scenarios best. In this analysis, we go beyond the signal cross-section and analyze the SM background at various colliders and their effects. 
We  explore the potential at HL-LHC via the  production channels: gluon fusion (GGF), vector boson fusion (VBF) and $b\bar{b}$ associated Higgs production (BBH) channels in mono-jet, di-jet and two $b$-jet along with missing transverse energy final states. While, HL-LHC does not show a promising discovery channel using cut-and-count analyses, it is important to note that BBH channels may also be competitive channels for heavy {Higgs bosons} searches. For the current scenario, we highlight the efficacy of future lepton colliders over HL-LHC to search for DM signatures in 2HDMS.

Our goal in this work is to motivate the DM search at the future colliders. We mainly focus on cases where the DM is produced via on-shell decay of non-standard scalars. Phenomenological studies have been done {for the 125 GeV SM Higgs boson decaying into DM candidates}~\cite{Curtin:2013fra,Liu:2016zki,Liu:2017lpo}. DM from the on-shell decay of non-standard heavy scalars have been studied in the context of N2HDM~\cite{Dey:2019lyr}, in which the analyses were done in the context of HL-LHC.
Our model 2HDMS gives rise to richer phenomenology compared to the aforementioned works. 
{{Our model 2HDMS, on the one hand, focusses on DM from the decay of non-standard scalars, on the other hand, it involves extra scalar degrees of freedom compared to the aforementioned works, giving rise to richer phenomenology both in the DM sector as well as collider sector.}}
In the present work we concentrate on the mono-photon, mono-$Z$ and missing energy searches, at $e^+e^-$ and muon colliders for low and intermediate DM benchmarks, and compare with the prospect at HL-LHC. For heavier DM scenarios, alternate channels such as $b\bar{b}$ and $t\bar{t}$ associated production channels are the dominant production channels which  give rise to two b jets along with missing energy final states which have previously not been considered in the context of heavy DM searches at high energy lepton colliders. We show via a detailed collider study the most sensitive channels for discovery of 2HDMS at both electron-positron and muon colliders.
We believe, we will be able to present here a guideline for future experiments through our detailed study.

The plan of our paper is as follows. In Sect.~\ref{sec:model}, we describe our model. Next, all existing constraints on the model are discussed in Sect.~\ref{sec:const}. In Sect.~\ref{sec:bp}, we explore the interesting regions of our parameter space and motivate the selection of suitable benchmarks. In Sect.~\ref{sec:dm}, we discuss the DM phenomenology and its interplay with detection prospect at colliders. In Sect.~\ref{sec:collider}, we present our collider analysis for various future colliders. Finally, we summarize our results in Sect.~\ref{sec:summary} .

\section{The model}
\label{sec:model}
In this work we consider a type II Two Higgs Doublet model with an additional complex scalar singlet (2HDMS). Compared to the SM, which contains one scalar $SU(2)$ doublet, this model contains two scalar doublets and a scalar singlet. 
This type of model was investigated for example in \cite{Baum:2018zhf, Heinemeyer:2021msz, Dutta:2022qeq, Dutta:2023cig}. The imaginary part of the scalar singlet will provide the DM candidate in our work.

The scalar potential can be written as the sum of the Two Higgs Doublet potential $V_\text{2HDM}$ and the singlet potential $V_S$. Considering a $U(1)$, $Z_2$ and $Z'_2$ symmetry, the most general expression can be written as:
\begin{subequations}
\begin{align}\label{eq:2HDMS_potential}
    V &= V_\text{2HDM} + V_{S} \\
    \label{eq:2HDM_potential}V_\text{2HDM} &= m_{11}^2 \Phi_1^{\dagger} \Phi_1 + m_{22}^2 \Phi_2^{\dagger} \Phi_2 - [m_{12}^2 \Phi_1^{\dagger} \Phi_2 + h.c. ] + \frac{\lambda_1}{2} (\Phi_1^{\dagger} \Phi_1)^2 \nonumber\\ 
    & \quad + \frac{\lambda_2}{2}(\Phi_2^{\dagger} \Phi_2)^2  + \lambda_3 (\Phi_1^{\dagger} \Phi_1) (\Phi_2^{\dagger} \Phi_2) + \lambda_4 (\Phi_1^{\dagger} \Phi_2) (\Phi_2^{\dagger} \Phi_1) \nonumber\\
    &\quad + \left[ \frac{\lambda_5}{2} (\Phi_1^{\dagger} \Phi_2)^2 + h.c. \right] \\
    V_{S} &= m_S^2 S^{\dagger} S + \left[ \frac{m_S'^2}{2} S^2 + h.c. \right]  \nonumber\\ 
    & \quad + \left[ \frac{\lambda_1''}{24} S^4 + h.c. \right] + \left[ \frac{\lambda_2''}{6} (S^2 S^{\dagger} S) + h.c. \right] + \frac{\lambda_3''}{4}(S^{\dagger} S)^2 \nonumber\\
    & \quad + S^{\dagger} S [\lambda_1' \Phi_1^{\dagger} \Phi_1 + \lambda_2' \Phi_2^{\dagger} \Phi_2] + [S^2(\lambda_4' \Phi_1^{\dagger} \Phi_1 + \lambda_5' \Phi_2^{\dagger} \Phi_2) + h.c.] ,
\end{align}
\end{subequations}
where $\Phi_1$ and $\Phi_2$ are the two $SU(2)$ doublets and $S$ is the $SU(2)$ singlet field. The parameters $m_{11}^2$, $m_{22}^2$, $m_{12}^2$, $m_{S}^2$ and $m_{S'}^2$ have dimension mass squared and the parameters $\lambda_{1}$, $\lambda_{2}$, $\lambda_{3}$, $\lambda_{4}$, $\lambda_{5}$, $\lambda_{1}'$, $\lambda_{2}'$, $\lambda_{4}'$, $\lambda_{5}'$, $\lambda_{1}''$, $\lambda_{2}''$ and $\lambda_{3}''$ are dimensionless. All parameters are real.

A type II structure of the model means that right-handed down-type quarks and charged leptons, couple only to $\Phi_1$ and right-handed up-type quarks couple only to $\Phi_2$.

\textbf{Symmetries and physical effects:}
To be precise, only the doublet part of the potential $V_\text{2HDM}$ is symmetric under $U(1)$ transformations. This, in combination with all parameters being real, results in the doublet part of the potential being charge and parity (CP) conserving. 
Furthermore this part of the potential is symmetric under $Z_2$ transformations of the form $\Phi_1 \overset{Z_2}{\rightarrow} - \Phi_1$, $\Phi_2 \overset{Z_2}{\rightarrow} \Phi_2$. This, in combination with the choice of a type II model, avoids flavor changing neutral currents (FCNC)\footnote{In a type II model the right handed fermion singlets transform under $Z_2$ as: $d_{jR} \rightarrow - d_{jR} $, $e_{jR} \rightarrow - e_{jR}$, $u_{jR} \rightarrow u_{jR}$, with $j=1,2,3$, where $d_{jR}$ denote the three down-type quark generations, $u_{jR}$ the three up-type quark generations and $e_{jR}$ the three charged lepton generations and the subscript $R$ denotes right handed particles. As one can see, by requiring $Z_2$ symmetry, right-handed down-type quarks and charged leptons, can couple only to $\Phi_1$ (all are odd under $Z_2$) and right-handed up-type quarks can couple only to $\Phi_2$ (both are even under $Z_2$). This way FCNC are avoided.}. 
This symmetry is spontaneously broken by the vacuum expectation values (vevs) $v_1$ and $v_2$. In order to avoid the formation of domain walls this symmetry is additionally broken softly by the term $m_{12}^2 \Phi_1^{\dagger} \Phi_2$.
The singlet part of the potential $V_S$ is symmetric under a $Z'_2$ symmetry of the form $S \overset{Z'_2}{\rightarrow} -S$, in order to stabilize the DM candidate. This symmetry is spontaneously broken by the vev $v_S$. However it is not broken softly as the $Z_2$ symmetry. The symmetries and their physical effects are summarized in Table~\ref{tab:2HDMS_symmetries}.
\begin{table}[h]
\scriptsize
    \centering
    \begin{tabular}{|p{4.1cm}|p{2cm}|p{2cm}|p{2cm}|p{2.6cm}|}
        \hline
        Symmetry & Transformation of fields & Broken spontaneously & Broken softly & Physical effect \\
        \hline
        $V_\text{2HDM}$ symmetric under $U(1)$, & $\Phi_j \overset{U(1)}{\rightarrow} e^{i\theta}\Phi_j,$ & No & No & Conservation of CP \\
        (+ all parameters real) & $\Phi_j^\dagger \overset{U(1)}{\rightarrow} e^{-i\theta}\Phi_j^\dagger$ & & &  \\
        \hline 
        $V_\text{2HDM}$ symmetric under $Z_2$ & $\Phi_1 \overset{Z_2}{\rightarrow}-\Phi_1,$ & Yes via $v_1$, $v_2$ & Yes via $m_{12}^2 $ 
        & Avoiding of FCNC \\
        (+ type II model)& $\Phi_2 \overset{Z_2}{\rightarrow}\Phi_2$ &  &  & \\
        \hline 
        $V$ symmetric under $Z^{\prime}_2$ & $S\overset{Z^{\prime}_2}{\rightarrow}-S,$ & Yes via $v_S$ & No & Stabilization of DM\\
        & $\Phi_j \overset{Z^{\prime}_2}{\rightarrow}\Phi_j$ &  & & \\
        \hline
    \end{tabular}
    \caption{Symmetries of the scalar potential and their physical effects.}
    \label{tab:2HDMS_symmetries}
\end{table} 

\textbf{Spontaneous symmetry breaking and eigenfields:}
The doublets and singlet, after electroweak symmetry breaking, obtain vevs $v_1$, $v_2$ and $v_S$ and can be {expanded} around them as:
\begin{subequations}
\begin{align}
    \Phi_i 
    &= \begin{pmatrix} \phi_i^+ \\ \frac{1}{\sqrt{2}}(v_i + \rho_i + i \eta_i) \end{pmatrix} 
    & \langle \Phi_i \rangle 
    &= \begin{pmatrix} 0 \\ \frac{v_i}{\sqrt{2}} \end{pmatrix} , \quad i=1,2 \\
    S 
    &= \begin{pmatrix} \frac{1}{\sqrt{2}}(v_S + \rho_S + i A_S) \end{pmatrix} 
    & \langle S \rangle 
    &= \begin{pmatrix} \frac{v_S}{\sqrt{2}} \end{pmatrix} ,
\end{align}
\end{subequations}
where $\phi_i^+$ (and $\phi_i^-$ in case of the transposed fields) are the complex charged components, $\rho_i$ are the real parts of the neutral components and $\eta_i$ are the imaginary parts of the neutral components of the doublets.
For the singlet there is only a neutral component, where $\rho_S$ is the real part and $A_S$ is the complex part, which will be the DM candidate in this work. 
(By charged and neutral we mean electrically charged and neutral in this case.)

The charged components of the doublets give rise to charged scalar particles $H^\pm$, the real parts of the neutral components of the doublets and the singlets give rise to neutral scalar particles $h_1$, $h_2$ and $h_3$ and the imaginary parts of the neutral components of the doublets and the singlets give rise to neutral pseudo-scalar particles $A$ and $A_S$. (There is also a charged Goldstone $G^\pm$ and a neutral pseudo-scalar Goldstone $G^0$, which are absorbed by the $W$ and $Z$ bosons after electroweak symmetry breaking.) These are the eigenfields in the mass basis{, which correspond to} physical particles. They can be obtained via rotations of the fields in the interaction basis and their eigenvalues are the squares of their masses. They are summarized in Table~\ref{tab:2HDMS_eigenfields}.
\begin{table}[h]
\scriptsize
    \centering
    \begin{tabular}{|p{2.4cm}|p{2cm}|p{2.2cm}|p{2cm}|p{2.7cm}|}
        \hline
        &\multicolumn{2}{|c|}{Mass basis} & \multicolumn{1}{|c|}{Interaction basis} & Rotation \\
        &Eigenfields & Eigenvalues & Eigenfields & \\
        \hline
        Charged & $H^\pm$, $G^\pm$ & $m_{H^\pm}^2$, $m_{G^\pm}^2=0$ & $\phi_1^\pm$, $\phi_2^\pm$ & $\begin{pmatrix} H^\pm \\ G^\pm \end{pmatrix} = R^\pm \begin{pmatrix} \phi_1^\pm \\ \phi_2^\pm \end{pmatrix}$ \\
        \hline
        Scalar & $h_1$, $h_2$, $h_3$ & $m_{h_1}^2$, $m_{h_2}^2$, $m_{h_3}^2$ & $\rho_1$, $\rho_2$, $\rho_S$ & $\begin{pmatrix} h_1 \\ h_2 \\ h_3 \end{pmatrix} = R \begin{pmatrix} \rho_1 \\ \rho_2 \\ \rho_S \end{pmatrix}$\\
        \hline
        Pseudo-scalar & $A$, $G^0$ & $m_A^2$, $m_{G^0}^2=0$ & $\eta_1$, $\eta_2$ & $\begin{pmatrix} A \\ G^0 \end{pmatrix} = R^A \begin{pmatrix} \eta_1 \\ \eta_2 \\ \end{pmatrix}$\\
        \hline
        DM & $A_S$ & $m_{A_S}^2$ & $A_S$ & $\begin{pmatrix} A_S \end{pmatrix}=\begin{pmatrix} A_S \end{pmatrix}$\\
        \hline
    \end{tabular}
    \caption{The eigenfields in the mass basis and interaction basis and the rotations to convert between them for each block (charged, scalar and pseudo-scalar). Since the vevs are real, the scalar mass basis eigenfields $h_1$, $h_2$ and $h_3$ are mixtures of the interaction basis fields $\rho_1$, $\rho_2$, $\rho_S$. However, there is no imaginary vev, hence the DM component does not mix with the other pseudo-scalar components and it remains protected under the $Z_2'$ symmetry. Note, that one of the scalars $h_1$, $h_2$ and $h_3$ has to be SM-like with a mass of about $125 \, \text{GeV}$ and the vevs $v_1$ and $v_2$ have to sum up to the electroweak vev $v=\sqrt{v_1^2 + v_2^2}\approx 246 \, \text{GeV}$.}
    \label{tab:2HDMS_eigenfields}
\end{table}

\textbf{Free parameters:}
The 19 free parameters from Eq.~\ref{eq:2HDMS_potential} are:
\begin{equation}
  \lambda_1, \lambda_2,  \lambda_3,  \lambda_4,  \lambda_5, m^2_{12}, m^2_{11}, m^2_{22}, \tan \beta, v_S, m_S^2, m_S'^2, \lambda_1', \lambda_2', \lambda_4', \lambda_5', \lambda_1'', \lambda_2'', \lambda_3'' ,
\end{equation}
where $\tan \beta = \frac{v_2}{v_1}$ and $v=\sqrt{v_1^2 + v_2^2} \approx 246 \, \text{GeV}$ is fixed by the SM.

Since there are three vevs in this model, three parameters can be replaced by the minimization conditions, which can be found in the appendix, Eq.~\ref{eq:minimization_cond}.
We choose to replace the parameters $m_{11}^2$, $m_{22}^2$ and $m_S^2$. Furthermore we set the parameter $
\lambda_2''=\lambda_1''$. {{In our tree-level calculation}}, this simplification leaves the couplings of the DM candidate to other particles unchanged as can be seen in the next section in Eq.~\ref{eq:DM_couplings}, since the parameter $\lambda_2''$ does not appear in the couplings of the DM candidate to the scalars, but only in the DM self-couplings, see Eq.~\ref{eq:2HDMS_potential}. 

This leaves us with the 15 free parameters:
\begin{align}\label{eq:inte_basis_params}
  \lambda_1, \lambda_2,  \lambda_3,  \lambda_4,  \lambda_5, m^2_{12}, \tan \beta, v_S, m_S'^2, \lambda_1', \lambda_2', \lambda_4', \lambda_5', \lambda_1'' = \lambda_2'', \lambda_3'' . 
\end{align}
These are the parameters of the interaction basis. We can change to the mass basis\footnote{To be precise we use a mixed mass basis, since still some of the parameters from the interaction basis appear. The combinations $\lambda_{14}' = \lambda^{\prime}_1-2\lambda^{\prime}_4$, $\lambda_{25}' = \lambda^{\prime}_2-2\lambda^{\prime}_5$ and $\lambda_{13}'' = \lambda''_1-\lambda''_3$ were chosen because they appear in the couplings of the DM candidate to the scalar particles $h_1$, $h_2$ and $h_3$, Eq.~\ref{eq:DM_couplings}. This leaves us with some freedom to control these couplings in our analysis.}, as in Table~\ref{tab:2HDMS_eigenfields}. The 15 free parameters in the mass basis are:
\begin{align}\label{eq:mass_basis_params}
    &m_{h_1}, m_{h_2}, m_{h_3}, m_A, m_{H^\pm}, m_{A_S}, , \tan \beta, v_S, \Tilde{\mu}^2, \alpha_1, \alpha_2, \alpha_3 \nonumber\\
    & \lambda_{14}' = \lambda^{\prime}_1-2\lambda^{\prime}_4, \lambda_{25}' = \lambda^{\prime}_2-2\lambda^{\prime}_5,\lambda_{13}'' = \lambda''_1-\lambda''_3,
\end{align} 
where $m_{h_1}$, $m_{h_2}$, $m_{h_3}$, $m_A$, $m_{A_S}$ and $m_{H^\pm}$ are the masses of the three scalars, the pseudo-scalar, the pseudo-scalar DM and the charged Higgs particle respectively as in Table~\ref{tab:2HDMS_eigenfields}, $\Tilde{\mu}^2 = \frac{m_{12}^2}{\sin(\beta) \cos(\beta)}$ and
$\alpha_1$, $\alpha_2$ and $\alpha_3$ are the mixing angles of the scalar rotation matrix in Eq.~\ref{eq:scalar_rotation_matrix}.

The equations for how to change between the interaction basis and the mass basis can be found in the appendix, Eq.~\ref{eq:basis_change}.

\subsection{Dark matter sector}
As written above, the DM candidate in this work will be the pseudo-scalar $A_S$, which arises from the imaginary component of the complex singlet $S$ and it does not mix with the imaginary components from the doublets. Its mass can be obtained from Eq.~\ref{eq:2HDMS_potential} and expressed as:
\begin{align}
    m_{A_S}^2 
    &= \frac{\partial^2 V}{\partial A_S^\dagger \partial A_S}|_{\substack{\Phi_1 = \langle \Phi_1 \rangle\\ \Phi_2 = \langle \Phi_2 \rangle\\ S = \langle S \rangle}}
    \nonumber\\ 
    &=-(2 m_S'^2 + v_S^2(\frac{\lambda_1''}{3}+\frac{\lambda_2''}{3}) +2(\lambda_4' v_1^2 + \lambda_5' v_2^2))\nonumber\\
\end{align}
Furthermore the couplings of the DM candidate to the scalar Higgs particles are of importance, since the DM does not couple to SM particles directly, but only via exchanges of scalar Higgs particles. The trilinear and quatrilinear couplings are:
\begin{subequations}\label{eq:DM_couplings}
\begin{align}
    \label{eq:DM_couplings_tril}\frac{\lambda_{h_j A_S A_S}}{v} 
    &= -i[\lambda'_{14} c_{\beta} R_{j1} + \lambda'_{25} s_{\beta} R_{j2} - \frac{v_S}{2v}\lambda_{13}'' R_{j3}] \\
    \lambda_{h_j h_k A_S A_S} 
    &= -i[\lambda'_{14} R_{j1}R_{k1} + \lambda'_{25} R_{j2}R_{k2} - \frac{1}{2}\lambda_{13}'' R_{j3}R_{k3}],\label{eq:DM_couplings_quatr}
\end{align}
\end{subequations}
where $h_j$, $h_k$ with $j,k = 1,2,3$ denote the three scalars and $R_{jk}$ with $j,k = 1,2,3$ are the entries of the scalar rotation matrix in Table~\ref{tab:2HDMS_eigenfields}. $\lambda_{h_j A_S A_S}$ is normalised to $v=\sqrt{v_1^2 + v_2^2}$ in order to obtain a dimensionless quantity and the short-hand notation $s_\beta = \sin\beta$, $c_\beta=\cos\beta$ is used. An equivalent expression in the mass basis can be found in Equation (2.24) and (2.25) of~\cite{Dutta:2023cig}.

\section{Constraints}\label{sec:const}
The model under consideration faces theoretical and experimental constraints, which are explained in detail in our earlier work \cite{Dutta:2023cig}. We shortly summarize these constraints below.

\subsection{Theoretical constraints}
\begin{itemize}
    \item \textbf{Boundedness-from-Below (bfb) Conditions:} 
We require the potential to be bounded from below, which means that at large field values it must remain positive. The behavior at large field values is dominated by the quartic terms which puts constraints on the quartic couplings of the scalar potential. To check these constraints we apply the copositivity conditions as in~\cite{Kannike:2012pe} and the Cottle-Habetler-Lemke theorem~\cite{Cottle1970OnCO}. Each point in our scans is checked for these conditions with a \texttt{Python} code using \texttt{numpy}~\cite{numpy:Harris_2020}.
    \item \textbf{Tree-level unitarity:} 
    The requirement of the model to be unitary at tree-level constrains the eigenvalues of the scattering matrices between the scalars and the longitudinal components of the gauge bosons to be lower than $\frac{1}{2}$~\cite{Goodsell:2018tti}. This condition is checked for each point of the scans using \texttt{SPheno-v4.0.5}~\cite{Porod:2003um}.
\end{itemize}

\subsection{Experimental constraints} 
\begin{itemize}  
    \item One of the scalar particles needs to comply with the SM-like Higgs with a mass of $125.25\pm 0.17 \, \text{GeV}$~\cite{ATLAS-CONF-2020-005}. We choose the second lightest scalar $h_2$ to be this particle\footnote{For some of the benchmarks the mass hierarchy is changed with the SM-like Higgs being the lightest scalar. However we stick to the convention of the SM-like Higgs being named $h_2$.}
    \item The invisible decay width of the SM-like Higgs to the DM candidate $A_S$, is constrained by ATLAS and CMS as below:
    \begin{align*}
        BR(h_{2} \rightarrow A_S A_S) &\leq 0.07^{+0.030}_{-0.022}    \text{  (ATLAS)}\text{~\cite{ATLAS:2023tkt}}
        \\&\leq 0.15 \text{ (CMS)} \text{~\cite{CMS:2023sdw}}.
    \end{align*}
    \item Constraints from flavor physics set the following bounds: 
    \begin{align*}
        BR(b \rightarrow s \gamma) 
        &= (3.55\pm0.24\pm0.09)\times 10^{-4}~\text{\cite{Lees:2012ym}},\\
        BR(B_s \rightarrow \mu^+\mu^-)
        &=(3.2^{+1.4 +0.5}_{-1.2 -0.3})\times10^{-9}~\text{\cite{Aaij:2013aka,Chatrchyan:2013bka}}.
    \end{align*}
    Furthermore the benchmark points are within the upper limit of $\Delta(g-2)_{\mu} = 261 (63) (48) \times 10^{-11}$~\cite{ParticleDataGroup:2020ssz}.
    \item The electroweak precision tests constrain the $STU$ parameters to~\cite{ParticleDataGroup:2020ssz}:
    \begin{align*}
        S &= 0.02 \pm 0.1, \\
        T &= 0.07\pm0.12, \\
        U &= 0.00\pm0.09 
    \end{align*}
    and the model predictions for the $STU$ parameters are obtained from~\cite{Grimus:2007if,Grimus:2008nb}.
    \item The DM relic density is constrained by the upper limit from Planck~\cite{Aghanim:2018eyx}: 
    \begin{align*}
        \Omega h^2 =  0.1191 \pm 0.0010 .
    \end{align*}
    The predictions of the model for the relic density are calculated using \texttt{micrOmegas-v5.2.13}~\cite{mco2}.
    \item The spin independent cross sections for scattering of DM on protons and neutrons is constrained by direct detection experiments from LUX-ZEPLIN (LZ)~\cite{LZ:2022ufs}. The predictions of the model for the direct detection cross sections are calculated using~\texttt{micrOmegas-v5.2.13}~\cite{mco2}, where we re-scale by the fraction of predicted relic density to observed relic density~\cite{Barger:2008jx,Biekotter:2021ovi,Belanger:2022esk}.
    \item The spin independent cross sections of DM annihilation are constrained by indirect detection experiments from Fermi-LAT (Fermi)~\cite{Fermi-LAT:2011vow,Fermi-LAT:2016uux}. The predictions of the model for the indirect detection cross sections are calculated using~\texttt{micrOmegas-v5.2.13}~\cite{mco2}, where we re-scale by the squared fraction of predicted relic density to observed relic density ~\cite{Barger:2008jx,Biekotter:2021ovi,Belanger:2022esk}.
    \item The constraints from LEP~\cite{Abbiendi:2013hk}, ATLAS~\cite{higgssumatlas} and CMS~\cite{higgssumcms} on the heavy Higgs searches and the $125 \, \text{GeV}$ Higgs signal strength measurements~\cite{ATLAS-CONF-2020-027} are taken into account.
\end{itemize}
For the generation of the model files we use \texttt{SARAH-v4.14.3}~\cite{Staub:2013tta} and for the particle spectra and decays \texttt{SPheno-v4.0.5}~\cite{Porod:2003um}.\footnote{The model and benchmark files 
associated with this paper are available at Ref.~\cite{dutta_2025_15240433}.}  The DM observables are computed using~\texttt{micrOmegas-v5.2.13}~\cite{mco2} and the Higgs constraints are checked using \texttt{HiggsTools}~\cite{Bahl:2022igd,Bechtle:2013xfa,Bechtle:2013wla,Bechtle:2020pkv,Bechtle:2020uwn} (which is a unification of the former \texttt{HiggsBounds-5} and \texttt{HiggsSignals-2}). \\
The constraints are applied as binary-cut, hence only parameter points passing all constraints are considered allowed.

\section{Parameter regions of interest and benchmark selection}
\label{sec:bp}

\begin{figure}[ht]
    \centering
    \includegraphics[width=1.\linewidth]{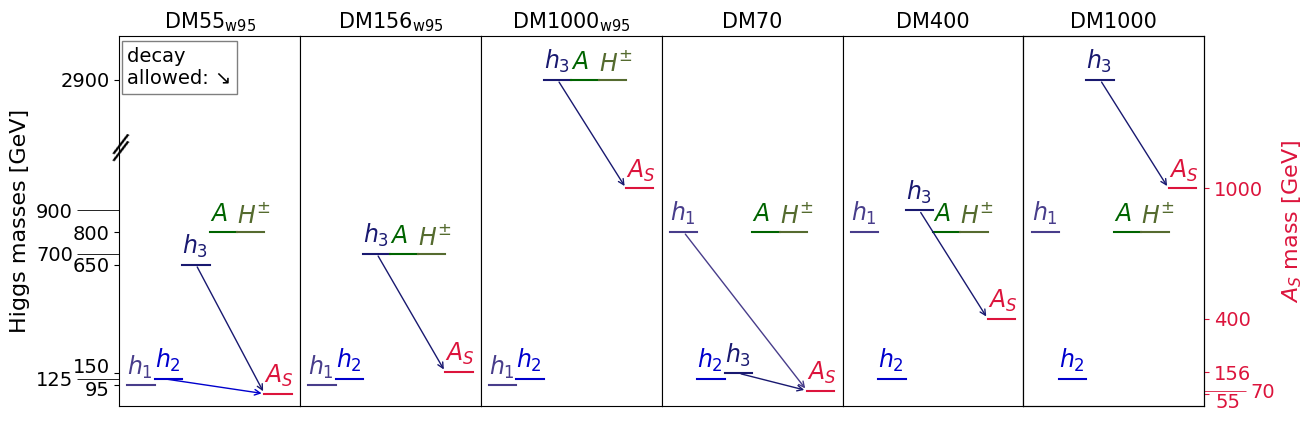}
    \caption{The masses of the scalar particles in the different benchmarks. The mass of the {Higgs bosons} (in black) are on the left y-axis while the mass of the DM candidate (in pink) are shown on right the y-axis. Please note that the point $2900 \, \text{GeV}$ is not to scale, but was lowered to make the plot more balanced.}
    \label{fig:BP_masses_overview}
\end{figure}

{In this work, we aim to perform a systematic analysis of DM phenomenology at current and future colliders. We, therefore, consider different DM mass regions: low, intermediate and heavy mass DM produced via the decay of the {Higgs bosons} and highlight the allowed parameter space consistent with all theoretical and experimental constraints. We present some representative benchmarks allowed from all constraints as shown in Fig.~\ref{fig:BP_masses_overview} in the context of our current scenario, 2HDMS with different mass hierarchies for DM and heavy Higgs masses. 
Further, we discuss the interplay of DM relic density and the invisible branching ratio of the {Higgs bosons} and discuss the dominant signal processes at different colliders and the signal topologies 
for each of the different DM mass regions. We perform an exhaustive collider analyses and highlight the efficacy of lepton colliders, particularly electron-positron colliders like ILC/CLIC/CEPC/FCC-ee for low and intermediate mass DM and muon colliders for heavy DM over HL-LHC.}

{We would like to emphasize that while our current analysis is performed in the context of 2HDMS, it is representative of generic DM models with a scalar portal. We propose signatures for detecting low, intermediate and heavy mass DM signals produced from scalar mediators at colliders and thus serves as a general guide to probe  DM at future colliders. We also compare the complimentarity  of lepton colliders to the high-luminosity LHC to further motivate the need for lepton colliders as a robust discovery probe for new physics.}

{Let us now discuss the representative benchmark scenarios chosen in our current model in more detail.}
After imposing various constraints on our model, we select a number of benchmark regions {(as shown schematically in Fig.~\ref{fig:BP_masses_overview})}, which satisfy all relevant constraints and at the same time are interesting from the point of view of discovery in the future colliders. Since we explore multiple future collider scenarios with different energy reach, we have looked into different mass-ranges of DM as well as the mediator (the scalar that acts as a major portal between DM and SM sector). Furthermore, we also take into account, in some benchmarks the recently observed excess at 95 GeV in the $b \bar{b}$ and $\gamma\gamma$ final state in our study, in order to see the impact of such a scalar on the DM phenomenology. In the rest of the benchmarks, we have not considered such an excess. Furthermore, we have explored both possibilities, namely when the DM candidate in our model accounts for the observed total relic density and also when it accounts for only a part of the total DM relic. We will see shortly that these two scenarios also have some correlation to the discovery potential at the collider experiments. In the light of the discussion above, we first discuss the benchmarks with the 95 GeV excess and will take up next the benchmarks without it.

\medskip
\noindent
{\bf Benchmarks with a 95 GeV scalar}

\medskip

\noindent
For the benchmarks, with a scalar at 95 GeV as a part of the mass spectrum, we have ensured that its signal strengths satisfy the following.

The lightest scalar $h_1$ has a mass of 95.4 GeV and plays the role of a scalar particle responsible for the observed signal strengths, which are for LEP in the $b\bar{b}$ mode ($\sim 2 \sigma$)~\cite{LEPWorkingGroupforHiggsbosonsearches:2003ing} and LHC in the $\gamma\gamma$ mode ($\sim 3 \sigma$)~\cite{CMS:2023yay,ATLAS:2023jzc}:
\begin{align}\label{eq:95_excess_signal_strength_obs}
    \mu^{b\bar{b}}_\mathrm{LEP} = 0.117^{+0.057}_{-0.057}, \qquad
    \mu^{\gamma\gamma}_\mathrm{LHC-combined} =  0.24^{+0.09}_{-0.08}.
\end{align}
We calculate the combined $\chi^2 = \chi_{b\bar{b}}^2 + \chi_{\gamma\gamma}^2$ values according to Ref.~\cite{Heinemeyer:2021msz} and \cite{Biek_tter_2024} and provide them in the Appendix, Table~\ref{tab:bpdm55_w95_mass_basis} - \ref{tab:bpdm1000_w95_mass_basis}.

\medskip
\noindent
$\bullet$ $\textbf{DM55}_\textbf{w95}$: We are interested in the low mass DM, especially, when $m_{A_S} \lesssim 62.5$ GeV. Benchmark $\textbf{DM55}_\textbf{w95}$ is in this category and fully satisfies the observed relic density. In such a scenario, the 125 GeV SM-like Higgs acts as the major portal to the dark sector. By virtue of having an extended scalar sector, it is possible to achieve such coupling between the SM-like Higgs and DM, so that, it satisfies the stringent bound from the direct detection experiments, while the observed relic density can also be accounted for. The dominant annihilation channel for the DM pair is into $b\bar b$ state. Furthermore, the upper bound from the invisible decay of the 125 GeV Higgs is also respected, since the SM Higgs only has a small branching ratio to the DM pair in this case. Although the aforementioned invisible branching ratio is small, it is highly likely, that signals from such a scenario will be visible at a future collider. Therefore, it is quite important to explore this benchmark at the colliders and make predictions.

\medskip
\noindent
$\bullet$ $\textbf{DM156}_\textbf{w95}$: We further move to a region where a heavier scalar from the extended scalar sector acts as a portal to the dark sector. In this case, we have taken the mass of the DM to be 156 GeV and the heavier scalar mediator $h_3$ of 700 GeV mass. This benchmark also accommodates the 95 GeV scalar with required signal strengths. One should note that, since the 95 GeV scalar is mostly singlet-like (in order to satisfy all the experimental bounds so far), the heavier scalar $h_3$ is dominantly doublet-like. We scan the DM portal couplings as well as the mixing angles between the two Higgs doublets, and found that in such mass range, it is typically difficult to account for the entire relic density while ensuring a significant $BR(h_3 \rightarrow A_SA_S$). We illustrate this point with results from our scanned regions in the next section. Since our focus in this work is the prospect of DM discovery at future colliders, we choose such parameters for this benchmark that lead to a considerable $BR(h_3 \rightarrow A_SA_S$) and consequently larger production rate. The dominant annihilation channel for a DM pair in this scenario is into a pair of $h_1$.

\medskip
\noindent
$\bullet$ $\textbf{DM1000}_\textbf{w95}$: We consider next a heavy DM scenario which may be detectable at high-energy future colliders. For this purpose, we choose a DM mass of 1 TeV, and the mediator ($h_3$) mass of 2.9 TeV. Our primary motivation for choosing this benchmark is to study a physics case for a future muon collider which will possibly operate at $\sqrt{s}=3$ TeV and 10 TeV and therefore can produce on-shell $h_3$ which can decay into a heavy DM pair. Since here too, a singlet-like 95 GeV scalar is part of the mass spectrum, the 2.9 TeV scalar ($h_3$) is doublet-like. However, in the heavy mass range for $h_3$ as well as DM ($A_S$), we could achieve the observed relic density with branching ratio $\lesssim 4\%$. The dominant annihilation channel for a DM pair in this scenario is into a pair of gauge bosons.

In this way we cover three mass ranges for DM as well as mediator(s), namely light, intermediate and heavy scenarios in the presence of a 95 GeV excess. 

\bigskip

\noindent
{\bf Benchmarks without a 95 GeV scalar}

\noindent
We discuss next the interesting regions of parameter space without the 95 GeV excess.

\medskip
\noindent
$\bullet$  $\textbf{DM70}$: We again start with a low mass DM. But in this case, we focus on a light mediator ($h_3$) which is singlet-dominated. In this case, we naturally have large (close to 100\%) invisible branching ratio of the singlet-like mediator to a DM pair, as well as correct relic density, especially close to the resonance region. The major annihilation channel is {into} a pair of gluons in this case. {{The gluons will further hadronize mostly into pions and and then finally lead to photons or charged leptons in the final state.}} Production of a singlet-dominated state is challenging at future lepton colliders, because of its feeble coupling to the gauge bosons. {{However, the lightness of the $h_3$-mass (150 GeV) is kinematically advantageous in this benchmark, leading to its sufficient production cross-section.}} Such scenarios may have good prospect at the LHC where GGF production dominates.

\medskip
\noindent
$\bullet$  $\textbf{DM400}$: Our next benchmark falls under the intermediate mass range category. The mediator ($h_3$) in this case is of 900 GeV and we choose it to be singlet-dominated. Here too, a substantial invisible branching ratio is achievable along with the correct relic density. DM annihilation is dominated by the $t\bar t$ final state. On the other hand, the production at colliders is naturally challenging due to the singlet nature of the mediating scalar.

\medskip
\noindent
$\bullet$  $\textbf{DM1000}$: Finally, we choose a benchmark with heavy DM (1 TeV) as well as heavy mediator $h_3$ (2.9 TeV). Interestingly, the other non-standard scalars, are at 800 GeV. We mention that, such a large mass-gap between the non-standard scalars is only possible because the heavy scalar is largely singlet-like in this case, and therefore, the strong constraints from $S,T,U$ parameters can be evaded. Achieving {the} observed relic density is possible in this scenario as well, but with a compromise on {the} invisible branching ratio. {{This benchmark has prospects to be discovered at high energy colliders.}} In this case, annihilation into {a} charged scalar and {a} pseudo-scalar opens up and dominates over other channels, owing to the mass hierarchy.

\begin{table}[h!]
    \centering
    \addtolength{\tabcolsep}{-4pt}
    \scriptsize
    \begin{tabular}{|c|c|c|c|c|c|c|}
        \hline
        & $m_{A_S}$ (GeV)& $\tan\beta$& $\Omega h^2$& $BR(h_1 \rightarrow A_S A_S)$& $BR(h_2 \rightarrow A_S A_S)$& $BR(h_3 \rightarrow A_S A_S)$ \\
        \hline 
         $\textbf{DM55}_\textbf{w95}$& 55& 2& 0.11& - & 0.0199& $3.81\cdot10^{-9}$\\
         $\textbf{DM156}_\textbf{w95}$& 156& 6.6& $1.61 \cdot 10^{-4}$ & - & - & 0.692 \\
         $\textbf{DM1000}_\textbf{w95}$& 1000& 5& 0.111& - & - & 0.0360 \\
         \textbf{DM70}& 70& 1.37 & 0.113 & $1.80 \cdot 10^{-4}$ & - & 0.999 \\
         \textbf{DM400}& 400 & 2.13 & 0.106 & - & - & 0.822 \\
         \textbf{DM1000}& 1000 & 1.34 & 0.117 & - & - & 0.00514 \\
         \hline
    \end{tabular}
    \caption{The benchmarks considered in this work. Shown are the DM mass $m_{A_S}$, $\tan\beta$, DM relic density $\Omega h^2$, and the invisible branching ratios $BR(h_{i} \rightarrow A_S A_S)$, with $i=1,2,3$, for the decays of the scalars $h_1$, $h_2$ and $h_3$ into two DM particles $A_S$. Decays which are kinematically not allowed are marked with a '-', compare with the mass hierarchy in Fig.~\ref{fig:BP_masses_overview}.
    The Benchmarks are named by the mass of the DM candidate. The ones including the $95\,\text{GeV}$ excess are labelled with \textbf{w95} as subscript. Full tables with the values for all parameters in the mass basis, along with the $\chi^2$-values for the benchmarks including the $95\,\text{GeV}$ excess, can be found in Table~\ref{tab:bpdm55_w95_mass_basis} - \ref{tab:bpdm1000_mass_basis} in the Appendix.}
    \label{tab:BPs_overview}
\end{table}

In order to summarize the various mass hierarchies pertaining to our chosen benchmarks, we provide a schematic representation in Fig.~\ref{fig:BP_masses_overview}. We also present the benchmarks with relevant masses, $\tan\beta$, corresponding relic density as well as invisible branching ratios in Table~\ref{tab:BPs_overview}.

\section{Implications from dark matter phenomenology}
\label{sec:dm}
{In this section we describe how certain parameters affect the DM phenomenology in the 2HDMS and how the benchmarks explained in \autoref{sec:bp} were selected with respect to these. We first identified regions of interest and then conducted detailed parameters scans within these regions. The effects on the parameter space from DM and collider constraints are presented in the following subsections.}

\subsection{Benchmark selection: Invisible branching ratio vs. relic density}

To find a benchmark point which has good prospects to be observed at a future collider a high invisible branching ratio (of one of the scalars, for example the heavy scalar $h_3$ to two DM particles $A_S$) is favorable. On the other hand, to explain the DM abundance in the universe one wants to obtain a relic density close to the observed limit from Planck. However for the chosen model a high branching ratio and high relic density counteract each other in many of the chosen benchmark scenarios. 

In the following we show this behavior and how we select benchmark points using the example of a benchmark with intermediate DM mass, and not including the 95 GeV excess.

\begin{figure}[ht]
    \centering
    \includegraphics[scale=0.65]{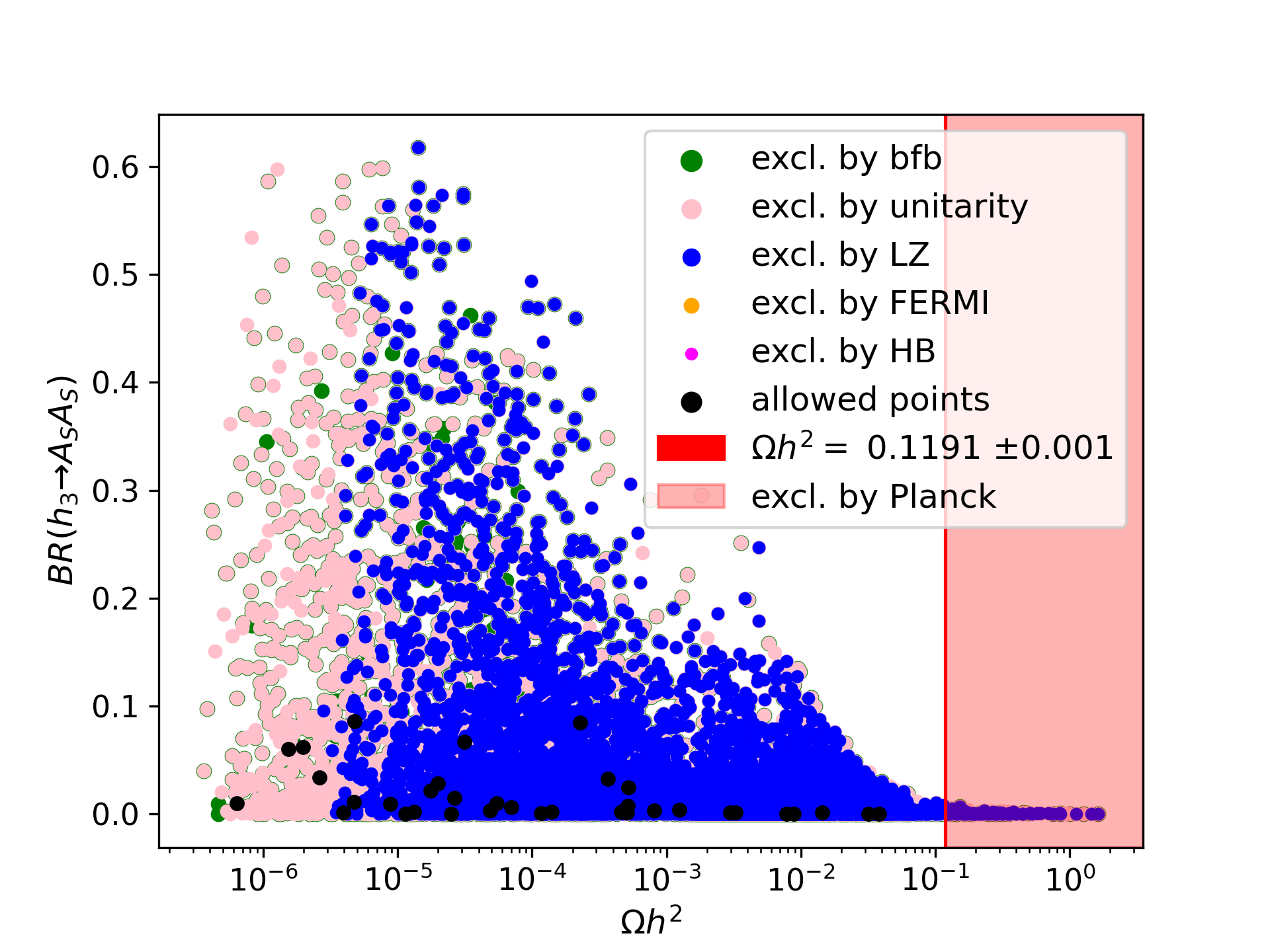}
    \caption{Invisible branching ratio $BR(h_3 \rightarrow A_S A_S)$ of the heavy scalar $h_3$ to two DM particles $A_S$ against relic density $\Omega h^2$ for randomly varying $\lambda_{14}'$, $\lambda_{25}'$, $\lambda_{13}''$, $v_S$ and $\tan \beta$ around \textbf{DM400}. The different constraints are shown by different colors. Please note that most points are excluded by more than one constraint, hence different point sizes were chosen to make overlapping constraints visible. The red line shows the relic density limit measured by Planck. Points in the red shaded area are above this limit and are excluded. The points passing all constraints are shown in black.}
    \label{fig:BR_vs_relic_random_scan_intermediate_DM_mass}
\end{figure}

To find a suitable benchmark point we start by randomly varying the parameters $\lambda_{14}' = \lambda_1' - 2 \lambda_4'$, $\lambda_{25}' = \lambda_2' - 2 \lambda_5'$, $\lambda_{13}'' = \lambda_1'' - \lambda_3''$, $v_S$ and $\tan \beta$ around the point \textbf{DM400}. The other parameters are fixed as in Table~\ref{tab:bpdm400_mass_basis} (Note that the exact values as in Table~\ref{tab:bpdm400_mass_basis} were not known before performing the scans. The point was selected from the results of these scans). In Fig.~\ref{fig:BR_vs_relic_random_scan_intermediate_DM_mass} one can see the invisible branching ratio (of the heavy scalar $h_3$ to two DM particles $A_S$) against the relic density. 
As described above a higher branching ratio results in a lower relic density. Furthermore, there are strong constraints from LZ (blue), unitarity (light pink) and bfb (green). The points passing all constraints are shown in black and all lie in an area with low branching ratio.

\begin{figure}[ht]
    \centering
    \includegraphics[scale=0.65]{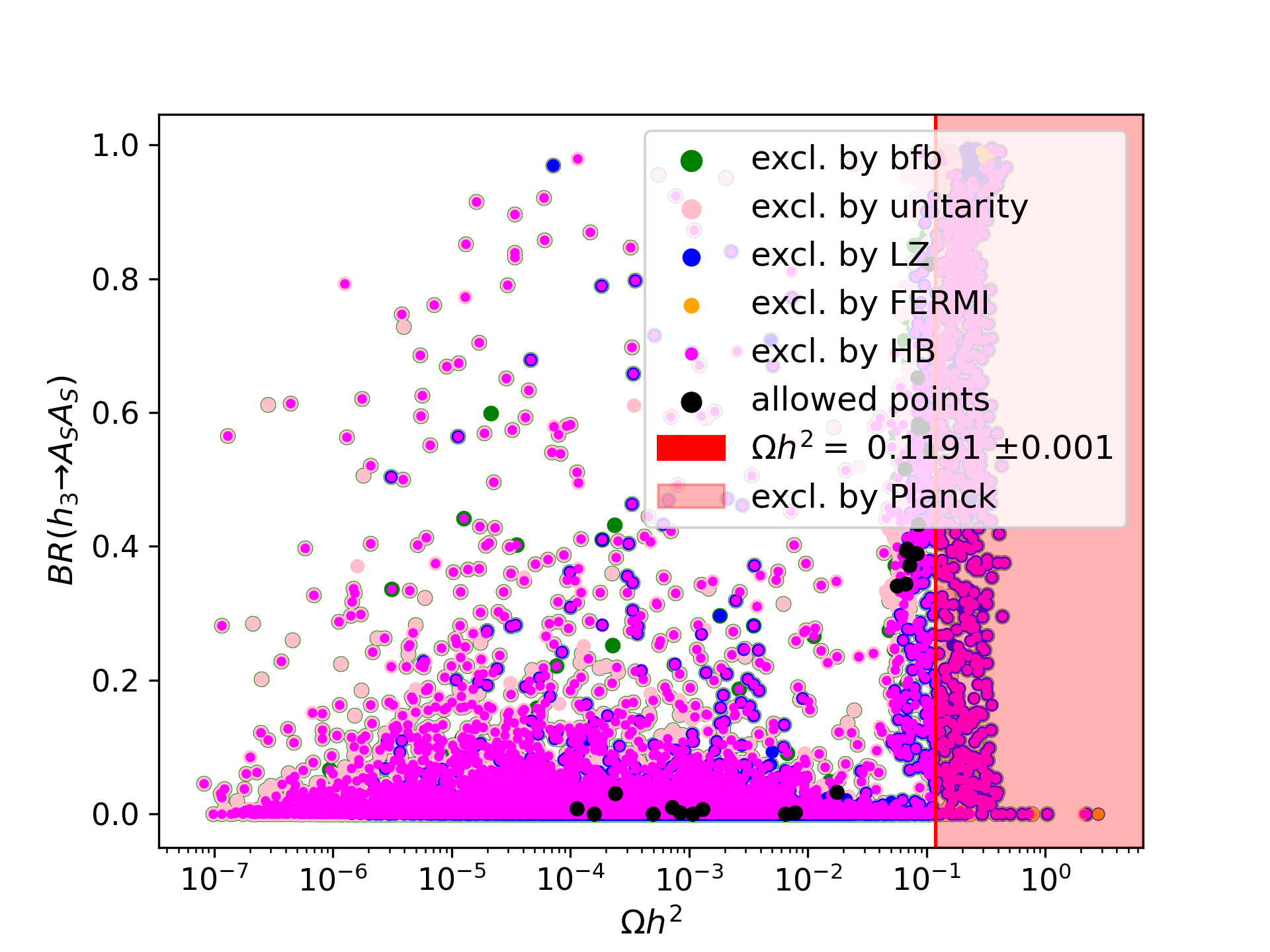}
    \caption{Invisible branching ratio $BR(h_3 \rightarrow A_S A_S)$ of the heavy scalar $h_3$ to two DM particles $A_S$ against relic density $\Omega h^2$ for randomly varying $\lambda_{14}'$, $\lambda_{25}'$, $\lambda_{13}''$, $v_S$, $\tan \beta$, $\Tilde{\mu}^2$ and the scalar mixing angles $\alpha_1$, $\alpha_2$ and $\alpha_3$ around \textbf{DM400}. The different constraints are shown by different colors. Please note that most points are excluded by more than one constraint, hence different point sizes were chosen to make overlapping constraints visible. The red line shows the relic density limit measured by Planck. Points in the red shaded area are above this limit and are excluded. The points passing all constraints are shown in black.}
    \label{fig:BR_vs_relic_random_scan_intermediate_DM_mass_2}
\end{figure}

To open up the area of high branching ratio and high relic density one needs to also vary the scalar mixing angles $\alpha_1$, $\alpha_2$, $\alpha_3$. The results of this more elaborate scan can be see in Fig.~\ref{fig:BR_vs_relic_random_scan_intermediate_DM_mass_2}.
As can be seen now there are also strong constraints from \texttt{HiggsBounds} (HB) (magenta). However the new points split into two areas, one stretched along the x-axis with low branching ratio and one stretched along the y-axis with relic density around the limit from Planck. The splitting into these two areas results from the chosen scan regions for the mixing angles. With a different choice one could also fill the sparse white area with more points. The points passing all constraints are again shown in black. From these results the point \textbf{DM400} in Table~\ref{tab:bpdm400_mass_basis} was selected.

To summarize the results one can say that, when only varying the coupling parameters $\lambda_{14}'$, $\lambda_{25}'$ and $\lambda_{13}''$, branching ratio and relic density behave inversely. To find a point with high branching ratio and high relic density one needs to also vary the scalar mixing angles $\alpha_1$, $\alpha_2$ and $\alpha_3$. We also reach the conclusion, that the allowed points (in black) that correspond to the observed relic density and also lead to a large invisible branching ratio of $h_3$, imply a combination of mixing angles, where $h_3$ is largely singlet-like. This is the scenario which is considered in {\bf DM400} in Table \ref{tab:BPs_overview} or later in detail in Table~\ref{tab:bpdm400_mass_basis}.

\subsection{Influence of dark matter mass and coupling parameters}
\begin{figure}[ht]
    \centering
    \includegraphics[scale=0.45]{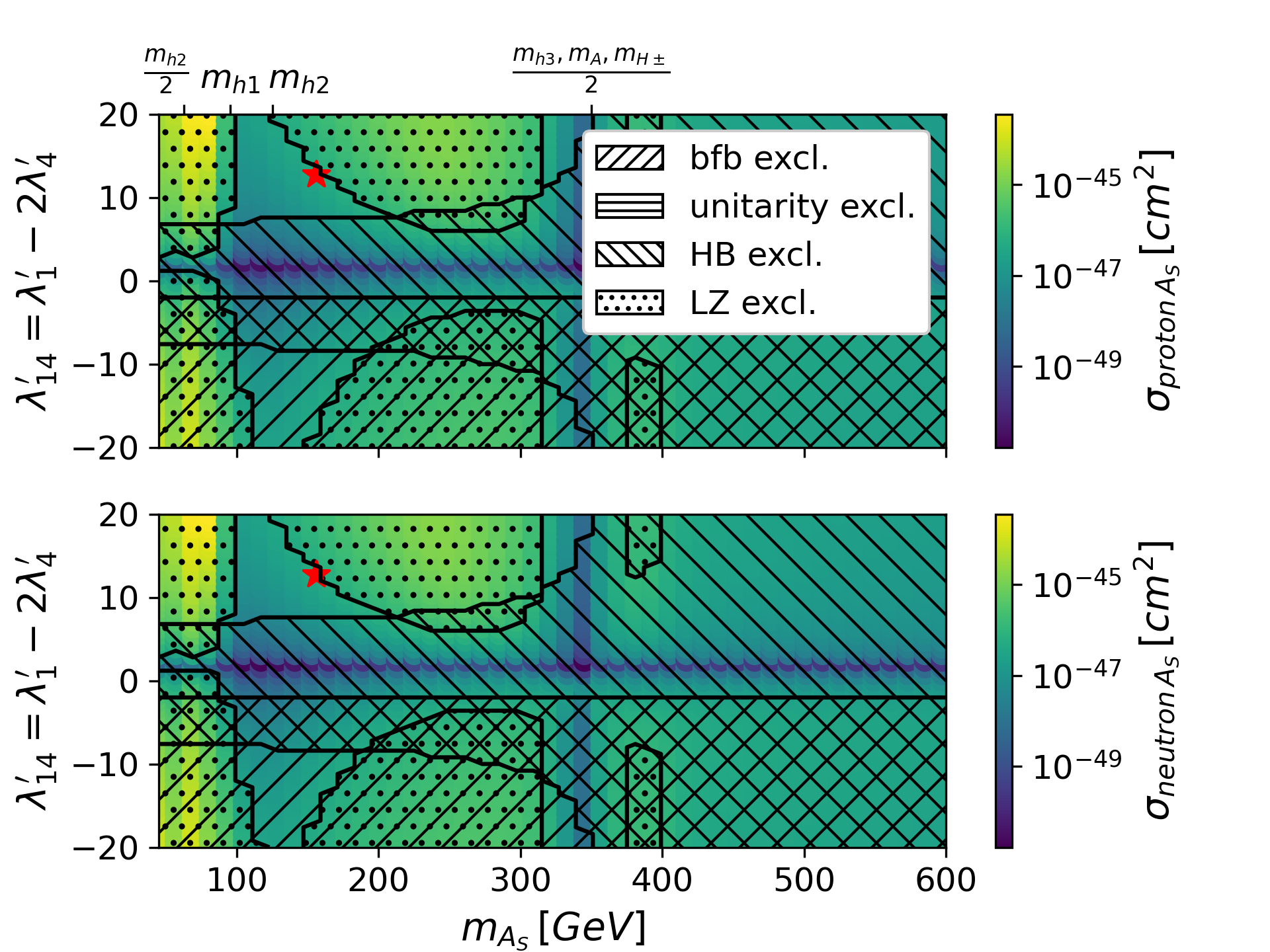}
    \includegraphics[scale=0.45]{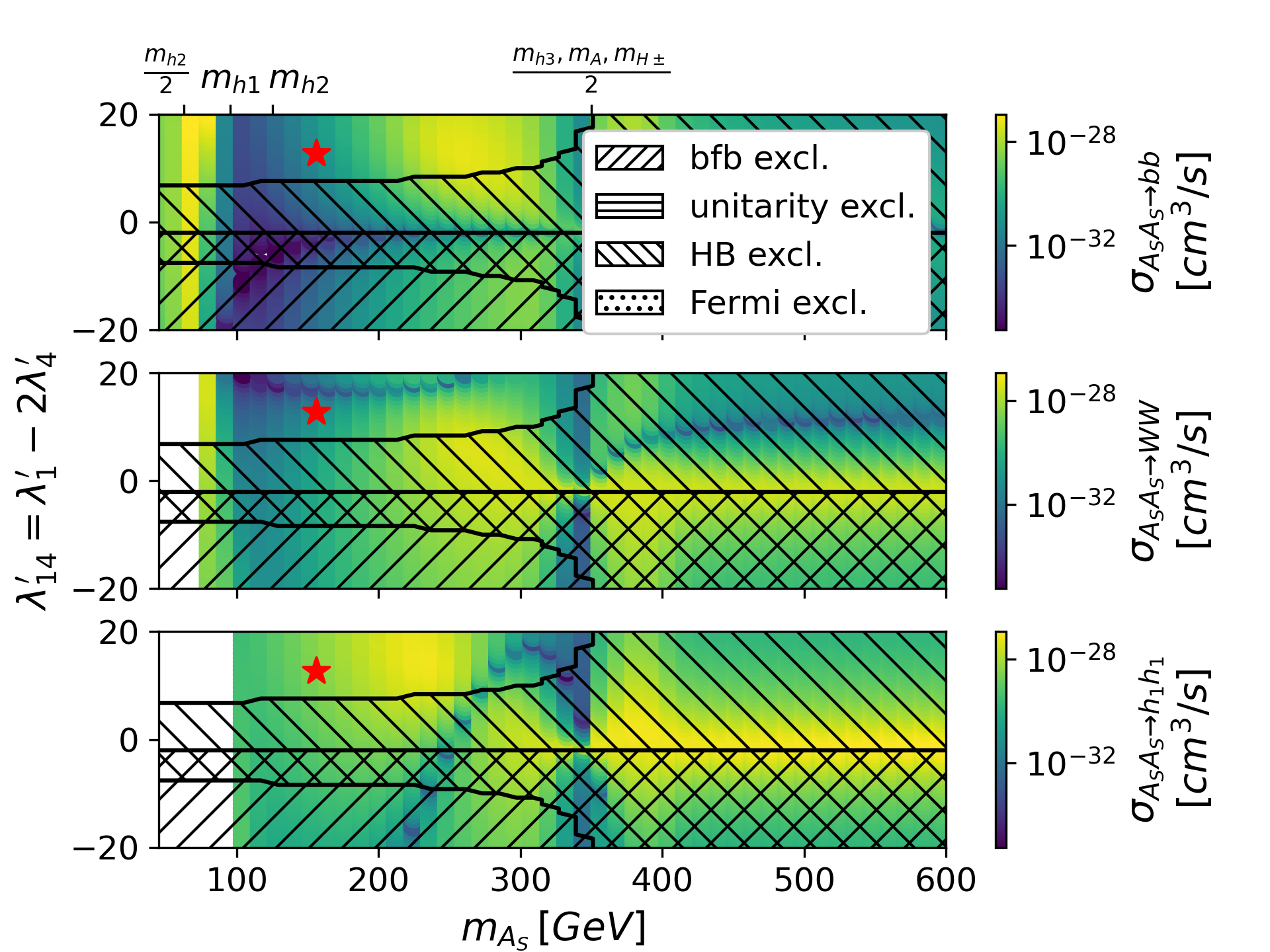}\\
    \includegraphics[scale=0.45]{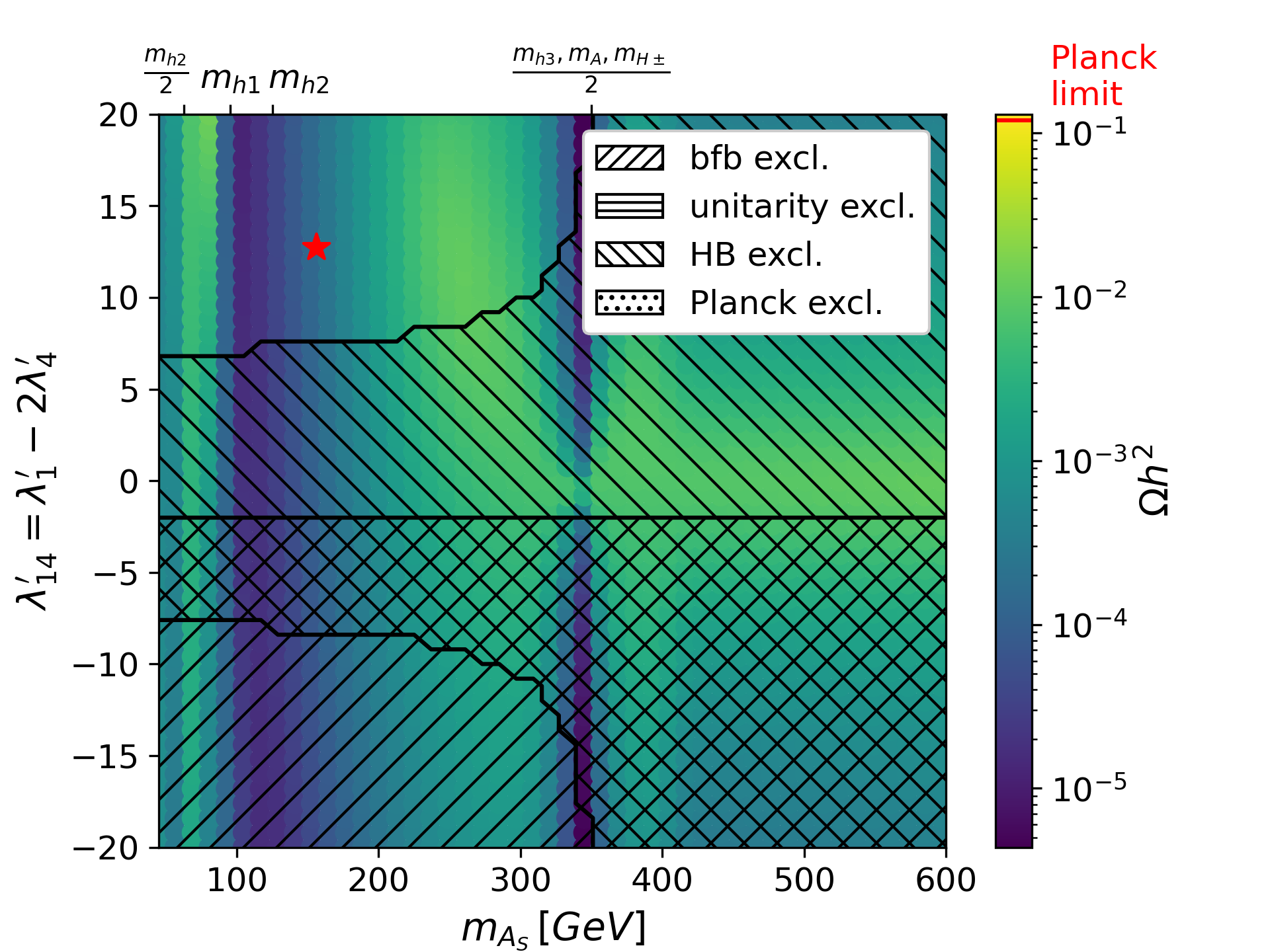}
    \includegraphics[scale=0.45]{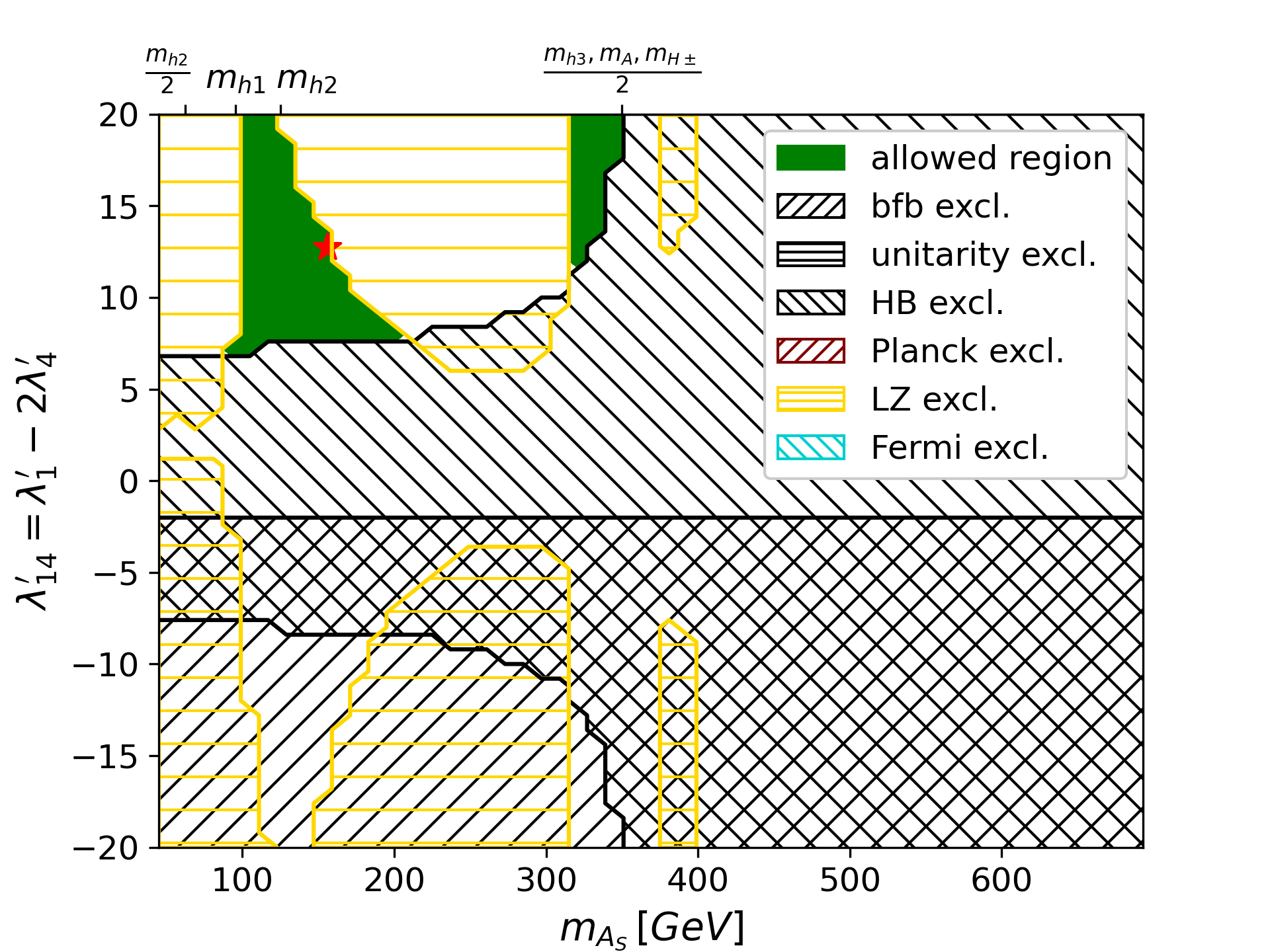}
    \caption{Influence of DM mass $m_{A_S}$ and coupling parameter $\lambda_{14}'$. The colored palette on the z-axis show the direct detection cross-section $\sigma_{\text{proton/neutron} A_S}$ for DM scattering on protons and on neutrons (top left), indirect detection cross-section $\sigma_{A_S A_S \rightarrow bb/WW/h_1h_1}$ for DM annihilating into $bb$, $WW$ and $h_1 h_1$ (top right), relic density $\Omega h^2$ (bottom left) and the allowed parameter regions under combining all constraints (bottom right). The top x-axis shows important mass thresholds. The direct and indirect detection cross-sections are re-scaled according to the relic density. $\textbf{DM156}_\textbf{w95}$ is marked with a red star.}
    \label{fig:infl_mAS_l1m24p}
\end{figure}

Apart from finding a suitable benchmark it is also instructive to see which effect each of the parameters has on the DM observables when varying continuously around the benchmark point. To keep the paper short, we show the effect of two parameters around the benchmark point $\textbf{DM156}_\textbf{w95}$ and do not go into detail on the influence of the other parameters. The effect of the DM mass $m_{A_S}$ and the coupling parameter $\lambda_{14}'$ can be seen in Fig.~\ref{fig:infl_mAS_l1m24p}.

The first plot (top left) shows the DM direct detection cross section for scattering of the DM candidate off protons (top) and off neutrons (bottom). The relevant constraints excluding some of the parameter space are shown as hatched areas. The constraints from bfb, unitarity and HB always apply and are shown in the following plots as well. For the direct detection cross section there are additional constraints from the LZ experiment shown as a dotted area in those regions where the direct detection cross section is highest.
The coupling with the largest influence on this observable is the trilinear coupling of two DM particles to the SM-like scalar $h_2$ in Eq.~\ref{eq:DM_couplings_tril}, since this mediates the coupling to other SM particles, and plays the most important role in the direct detection. As can be seen from the equation, this coupling is minimized when the parameter $\lambda_{14}'$ is zero. In the plot, this shows as a dark blue dip along this value.
Another interesting feature is a sharp dip around $m_{A_S} \approx 350 \, \text{GeV}$. On the top x-axis some important mass values are shown. At $350 \, \text{GeV}$ for example the mass of the DM candidate is half of the mass of the heavy scalar, the pseudo-scalar and the charged Higgs, which have a mass of $700 \, \text{GeV}$ in this benchmark. At this mass two DM particles can resonantly annihilate into the heavy scalar. This causes a drop in the DM relic density (see bottom left plot). Since we are plotting the direct detection cross section re-scaled with the relic density the dip shows up here as well.

In the second plot (top right) the DM indirect detection cross section is shown for the three main channels $bb$ (top), $WW$ (middle) and $h_1 h_1$ (bottom). Again the constraints from bfb, unitarity and HB are shown as hatched areas. For the indirect detection, constraints from Fermi LAT are taken into account. However in the shown scenario the indirect detection cross section is low enough to always stay below those bounds.
The indirect detection of DM is based on observing annihilation of two DM particles, hence not only the trilinear but also the quatrilinear couplings in Eq.~\ref{eq:DM_couplings_quatr} play a role. Some interesting features can be observed around the low mass region, where again resonant annihilation around $m_{A_S} \approx 63 \, \text{GeV}$ causes a peak in the top plot. The other annihilation channels $WW$ and $h_1 h_1$ instead show white areas. This is because the $W$ boson and the light scalar $h_1$ are heavier than bottom quarks $b$, hence these annihilation channels are forbidden for low DM masses.
As in the previous plot a strong dip around $m_{A_S} \approx 350 \, \text{GeV}$ can be seen. This is again caused by resonant annihilation into the heavy scalar, which lowers the relic density by which we re-scale the indirect detection cross section.

The third plot (bottom left) shows the DM relic density. Again with the constraints as in the other plots and constraints from Planck. However as the relic density is low in this scenario there are no regions excluded by Planck. 
The relic density measures the amount of DM in the universe today, hence it behaves inverse to the trilinear and quatrilinear couplings. The higher the couplings, the more DM could annihilate, hence less DM would be left today. This can be seen in the high mass region, where the area around $\lambda_{14}' = 0$ shows higher values instead of a dip, as for the direct detection. The resonant dip around $m_{A_S} \approx 350 \, \text{GeV}$, which we also observe in the direct and indirect detection, comes from resonant annihilation of DM.\footnote{The reason we also see a dip in the direct and indirect detection plots is that we can not observe the heavy particles into which the DM particles would annihilate. If these were observable particles they would cause a peak in direct and indirect detection, such as the peak around $m_{A_S} \approx 63 \, \text{GeV}$ where two DM particles resonantly annihilate into the SM-like scalar, which couples to other SM particles and can be observed.} 
For low DM masses the relic density rises slightly, because a lighter DM candidate has less annihilation channels, which means that it remains in the universe, hence the relic density is increased. When the resonant annihilation into the SM-like scalar opens up around $m_{A_S} \approx 63 \, \text{GeV}$ the relic density falls again.

The fourth plot (bottom right) shows a summary of all constraints and which allowed regions remain (green area).

As stated above we do not show the variation of all 15 free parameters in this work. We have checked that the other coupling parameters $\lambda_{25}'$ and $\lambda_{13}''$ would show similar effects as $\lambda_{14}'$. Furthermore it remains to clarify that the influence of the parameters can look different in different benchmark scenarios. Due to how the scalar mixing angles, $\alpha_1$ $\alpha_2$ and $\alpha_3$, $v_S$ or $\tan\beta$ are chosen, some parameters can have a stronger or weaker effect on the DM observables (see Eq.~\ref{eq:DM_couplings}). 

One should note that the exclusion from HB stems from the fact that, the dark-portal couplings e.g. $\lambda_{14}'$, affect the invisible branching ratio of the heavy scalars and consequently also the branching ratios to the visible final states ($\tau\tau$ final state appears to be most sensitive in our case). Therefore, we obtain a contour of exclusion in the DM mass vs portal coupling plane from HB.

\section{Collider analyses}
\label{sec:collider}

Having identified interesting regions of parameter space in the 2HDMS, we proceed to probe such regions in future colliders. We focus on low, intermediate and high DM mass, as well as mediator (the scalar that acts as a portal with DM) mass, as discussed in the previous section. We investigate, on the one hand, various possible final states such as mono-photon($\gamma$)+MET\footnote{We use the term `MET' in a generic sense in this work. At the LHC, MET is used to define missing transverse momenta. On the other hand, at the lepton collider, alongside missing transverse momenta, missing energy ($\slashed{E}$) and missing mass ($\slashed{M}$) also become important. Therefore, we use MET as a generic umbrella term to define all these quantities and present the explicit definitions of it specific to the particular collider and final state.}, mono-$Z$+MET, $b\bar b$+MET, $t\bar t$+MET. On the other hand, we consider such possibilities at various future colliders, i.e. hadron collider high-luminosity LHC (HL-LHC), $e^+e^-$ colliders such as ILC, FCC-ee and CLIC, and finally muon-collider.

The LHC is currently in its  Run 3 phase with center-of-mass energy  $\sqrt{s}=13.6$ TeV. While already impressive, further upgrades 
to the high-luminosity run of the LHC~\cite{Aberle:2749422},
planned after Run 3, is set to further enhance the center-of-mass energy up to $\sqrt{s}=14$ TeV and the integrated luminosity up to 3000 fb$^{-1}$. Along with present Run 3 results and the future HL-LHC run, several New Physics models are expected to be constrained from experimental data. Further, upcoming lepton collider machines would also play a crucial role in looking for New Physics in potentially complimentary discovery channels. Such lepton colliders would not only shed light on investigating indirect effects of New Physics via precision studies but also for high energetic leptonic colliders, via direct searches of new particles. We discuss below some of the proposed upcoming high precision and high energy future lepton colliders.

The International Linear Collider (ILC)~\cite{Behnke:2013lya,ILC:2013jhg,Adolphsen:2013jya,Adolphsen:2013kya,Evans:2017rvt,Bambade:2019fyw,LCVision-Generic}, is a linear $e^+e^-$
collider with center-of-mass energies in the range of  
$\sqrt{s}=$92, 250--550 GeV, upgradable
up to $\sqrt{s}=1000$ GeV.
The ILC provides a comprehensive and complementary physics potential~\cite{Moortgat-Pick:2015lbx} and an advantage over the HL-LHC due to its clean environment, high luminosity, tuneable energy as well as the possibility to apply precise and high  polarizations of both the electron and positron 
beams which not only leads to increased background suppression, higher rates  and better control of systematics but opens also new windows in direct as well as indirect searches for New Physics offering high precision measurements~\cite{MOORTGATPICK2008131,Behnke:2013lya}.  
Other proposed $e^+e^-$ colliders are the Compact Linear Collider (CLIC)~\cite{CLICdp:2018cto,lcfclic} with energy upgrades up to 
$\sqrt{s}=1.5,  3$ TeV, {the Linear Collider Facility at CERN (LCF)~\cite{lcfcern, LCVision-Generic}}, the Future Circular Collider (FCC-ee)~\cite{Benedikt:2651299} and Circular Electron Positron Collider (CEPC) with the latter having centre-of-mass energies up to 350 GeV and  240 GeV~\cite{CEPCStudyGroup:2018rmc}, respectively. 

A multi-TeV circular muon collider~\cite{InternationalMuonCollider:2024jyv} is also proposed to be built with center-of-mass energies $\sqrt{s}=3$ and 10 TeV with target integrated luminosity up to $\mathcal{L}=10$ ab$^{-1}$~\cite{Han:2022ubw}. While providing a cleaner environment compared to hadron machines, it also provides access to higher center-of-mass energies, thereby extending the reach for New Physics searches. Besides, the enhanced muon Yukawa coupling compared to electron Yukawa coupling significantly enhances production  cross-sections of beyond Standard Model (BSM) particles compared to electron-positron colliders, especially in context of models with extended scalars. Beam polarizations may also be useful for a muon collider~\cite{Kittel:2005ma,blöchinger2002physicsopportunitiesmumuhiggs}, but since polarized muon beams are not yet foreseen in the current design, we focus on unpolarized muon beams for the present study.  

In this work, we study the prospects of new physics searches for the 2HDMS in the high-luminosity LHC, electron-positron colliders and muon colliders in the following subsections. 

We present a detailed comparison between future colliders in terms of probing various benchmark scenarios. We also establish a complimentarity between various colliders as well as different final states. 
We generate the parton-level events at $\sqrt{s}=14$ TeV and use \texttt{MG5$\_$aMC$\_$v3.4.1}~\cite{Alwall:2014hca,Alwall:2011uj} followed by showering and hadronization using \texttt{Pythia$\_$v8.3.06}~\cite{Bierlich:2022pfr}. We have used the default parton distribution function \texttt{NNPDF2.3}\cite{Ball:2013hta}. The detector simulation for the hadron level events is performed using the fast detector simulator \texttt{Delphes-v3.5.0}~\cite{Selvaggi:2014mya}. The signal analyses at  LHC has been performed using \texttt{MadAnalysis-v5}~\cite{Conte:2012fm}.  
We generate the signal processes in \texttt{WHIZARD-v3.1.5}~\cite{Kilian:2007gr} for the $e^+e^-$ and $\mu^+\mu^-$ collider studies. 

\subsection{Benchmarks at LHC}

For the 125 GeV Higgs at the LHC, the dominant production processes are GGF and VBF. For heavy {Higgs bosons}, the associated Higgs production with $b\bar{b}$ (BBH) dominates both GGF and VBF for certain parameter regions due to the $\tan \beta$ enhancement of the bottom quark Yukawa coupling. Fig.~\ref{fig:feynlhc} - \ref{fig:feynlhc2} (in appendix~\ref{sec:appFD} shows the Feynman diagrams for GGF, VBF and BBH processes respectively. For BBH, we consider both  four-flavor and the dominant five-flavor contributions and perform the Santander matching~\cite{Harlander:2011aa} with the matched cross-section defined as a weighted average of the four- and five-flavor cross-sections.  The cross-section then reads: 
\begin{equation}
    \sigma^{w_i}=\frac{\sigma^{4f}+w_i\sigma^{5f}}{1+w_i} ,
    \label{eq:SM}
\end{equation}
where $w_i=ln(\frac{m_{h_i}}{m_b})-2$. 
The production cross-section of the {Higgs bosons}, with the Higgs decaying invisibly, for the light, intermediate and heavy benchmarks  are presented  in Table~\ref{tab:cslhc} and ~\ref{tab:cslhc2} respectively.  

For the light DM benchmarks, $\textbf{DM55}_\textbf{w95}$ the relevant Higgs decaying to the DM particles is the SM Higgs. Hence the dominant processes are GGF and VBF. While $\textbf{DM55}_\textbf{w95}$ has an invisible branching ratio of $\sim 2\%$, the cross-section is appreciable. For \textbf{DM70} the singlet dominated light Higgs with mass $m_{h_3}=$ 150 GeV almost always decays invisibly (99$\%$). Despite the singlet admixture, the cross-section for the production of $h_2$ using GGF in \textbf{DM70} is appreciable and benefits from the large invisible branching ratio which otherwise suppresses the cross-section for $\textbf{DM55}_\textbf{w95}$ with $m_{h_2}=125$ GeV.

For the intermediate benchmark $\textbf{DM156}_\textbf{w95}$ the dominant processes for the production of the heavy Higgs $h_3$ invisibly decaying into a pair of DM candidates are mainly from $b\bar{b}h_3$ which may benefit from the increase in the Yukawa coupling of the $b$ quark via its dependence on $\beta$ for large $\tan \beta$~\cite{Branco:2011iw}. While this process is suppressed for the SM Higgs production, it may be comparable or larger than traditional production channels, such as GGF or VBF and could be an alternate channel of discovery for heavy Higgs searches in Type II 2HDM.  
 
The production cross-sections for the intermediate and heavy benchmarks are shown in Table~\ref{tab:cslhc2}. For the intermediate 
DM benchmark $\textbf{DM400}$, the mass of the heavy Higgs, $m_{h_3}=900$ GeV, which decays invisibly, the dominant contribution mainly arises from GGF owing to the heavy mass of the Higgs while BBH and VBF are sub-dominant contributions. 
For the heavier DM benchmarks $\textbf{DM1000}_{w95}$ and \textbf{DM1000} with a heavy Higgs such that $m_{h_3}=$ 2900 GeV, the production cross-sections for GGF, VBF and BBH are shown in Table~\ref{tab:cslhc2}. The production cross-section greatly suffers owing to the heavy mass of the heavy Higgs, $m_{h_3}=2900$ GeV and therefore does not produce any observable number of events at $\mathcal{L}=3000$ fb$^{-1}$. Such scenarios are  beyond the  reach of the HL-LHC. However, such benchmarks have a larger production cross-section at a higher $\sqrt{s}$ such as at future hadron colliders like FCC-hh/SPPC at   $\sqrt{s}=100$ TeV as we discuss later in sec.~\ref{sec:challenge}.

\begin{table}[ht!]\begin{center}
\begin{tabular}{|c|c|c|c|}
    \hline
    Process & \multicolumn{3}{|c|}{Production cross-section (fb) at $\sqrt{s}=14$ TeV} \\
    & $\textbf{DM55}_\textbf{w95}$ & $\textbf{DM156}_\textbf{w95}$ & \textbf{DM70}  \\
    \hline
    GGF($h_2 \rightarrow A_S A_S$) & 533.9 & -  & 19.29$\times 10^3$  \\ 
    GGF($h_3 \rightarrow A_S A_S$) & - & 0.015  & - \\
    VBF($h_2 \rightarrow A_S A_S$) & 54.33 & -  & 2.72$\times 10^3$  \\
    VBF($h_3\rightarrow A_S A_S$) & - & 0.134   & 0.0022   \\ 
    BBH ($(b\bar{b}h_2 \rightarrow A_S A_S)$) & 21.6 & -   & 0.137  \\
    BBH ($(b\bar{b}h_3\rightarrow A_S A_S)$) & - & 47.24  & -   \\
    \hline
\end{tabular}
\caption{The production cross-sections  at leading order (LO) of the relevant processes at $\sqrt{s}=14$ TeV at LHC. All cross-sections below $10^{-6}$ fb are denoted by `-'.  For $b\bar{b}h_i$, with $i=2,3$, we use the Santander matched cross-section as defined in the text.}
\label{tab:cslhc}
\end{center}
\end{table}

\begin{table}[ht!]\begin{center}
\begin{tabular}{|c|c|c|c|}
    \hline
    Process & \multicolumn{3}{|c|}{Production cross-section (fb) at $\sqrt{s}=14$ TeV} \\
    & $\textbf{DM400}$ & $\textbf{DM1000}$ & $\textbf{DM1000}_\textbf{w95}$ \\
    \hline
    GGF($h_3 \rightarrow A_S A_S$)  &  0.013 & 6.35$\times10^{-7}$ &   4.5$\times10^{-6}$  \\ 
    VBF($h_3 \rightarrow A_S A_S$)  & 0.0008 & -  & -      \\
    BBH($h_3 \rightarrow A_S A_S$)   & 0.007 & -  & - \\
    \hline
\end{tabular}
\caption{The production cross-sections  at leading order (LO) of the relevant processes for the benchmarks \textbf{DM400}, \textbf{DM1000} and $\textbf{DM1000}_\textbf{w95}$ at $\sqrt{s}=14$ TeV at LHC. For $b\bar{b}h_3$ we use the Santander matched cross-section as defined in the text.}
\label{tab:cslhc2}
\end{center}
\end{table}

The possible final states at LHC for this scenario are:
\begin{itemize}
    \item Mono-jet + $\slashed{E}_T$
    \item Two forward jets + $\slashed{E}_T$
    \item Two $b$-jets + $\slashed{E}_T$
\end{itemize}

We discuss each of the final states from the production channels GGF, VBF and BBH respectively in the following subsections.  We study the prospective signals at the LHC at $\sqrt{s}=14$ TeV with integrated luminosity $\mathcal{L}=3000$ ab$^{-1}$.  

\subsubsection*{Gluon Fusion} 
We consider the final state mono-jet + MET from the GGF production channel. For the collider analyses, we use the following cuts~\cite{ATLAS:2017bfj}:
\begin{itemize}
\item \textbf{C1}: The  final state consists of  up to four jets with $p_T>30$ GeV and $|\eta|<2.8$.  
\item \textbf{C2}: We demand a large $\slashed{E}_T>250$ GeV.
\item \textbf{C3}: The hardest leading jet has $p_T>250$ GeV with $|\eta|<2.4$.
\item \textbf{C4}: We demand $\Delta \Phi ( j, \slashed{E}_T)>0.4 $ for all jets and  $\Delta \Phi ( j, \slashed{E}_T)>0.6$ for the leading jet. 
\item \textbf{C5}: A lepton-veto is imposed for electrons with $p_T>20$ GeV and $|\eta|<2.47$ and muons with $p_T>10$ GeV and $|\eta|<2.5$.
\end{itemize}
The SM background of 7.07 pb is obtained from the mono-jet + $\slashed{E}_T$ search studied in Ref.~\cite{Dey:2019lyr}.

We present the signal significance for   GGF  
in Table~\ref{tab:ggf}  
respectively using K-factors of 1.91 (NNLO+NNLL). We list only the benchmarks which are sensitive at HL-LHC with significance at least greater than 0.05$\sigma$. While both the light DM benchmarks have appreciable cross-section at LHC, their significance suffers due to the high-$p_T$ cuts as in usual mono-jet searches at LHC. 
\begin{table}[ht]
    \begin{center}
    \begin{tabular}{|c|c|}
        \hline 
        Benchmark   &  Significance  \\
        \hline 
        $\textbf{DM55}_\textbf{w95}$ & 0.30$\sigma$\\
        \textbf{DM70}  & 0.55$\sigma$\\
        \hline 
        \end{tabular}
    \caption{The signal significance for the signal benchmarks from GGF for HL-LHC at an integrated luminosity of 3000 fb$^{-1}$. }
    \label{tab:ggf}
    \end{center}
\end{table}
\subsubsection*{Vector Boson Fusion}

We consider the final state two forward-jets + MET from the VBF production channel. For the collider analyses, we use the following cuts~\cite{CMS:2018yfx}:
\begin{itemize}
\item \textbf{D1}: The final state consists of  at least two  jets with $p_T (j_1)>80$ and $p_T(j_2)>40$ GeV and $\Delta \Phi(j_i, \slashed{E}_T)>0.5$.
\item \textbf{D2}: We demand $\eta (j_1j_2)<0$ and $\Delta \Phi j_1 j_2 < 1.5$.
\item \textbf{D3}: We demand  $|\Delta \eta|_{jj}>3.0$.
\item \textbf{D4}: The invariant mass of the two forward jets is required to be large, i.e, $M_{jj}>600$ GeV.
\item \textbf{D5}: We demand  $\slashed{E}_T>200$ GeV.
\item \textbf{D6}: Furthermore, a lepton veto is imposed for   electrons with  $p_T >$ 20 GeV or muons with $p_T >$ 10
GeV.
\end{itemize}
We present the signal significance for  VBF  
in  Table~\ref{tab:vbf}   using K-factor of 1.73 (NLO QCD + NLL) for VBF.
The VBF process, as we see, are only sensitive for \textbf{DM70} with nearly 2$\sigma$ excess at HL-LHC. 

\begin{table}[ht]
    \begin{center}
    \begin{tabular}{|c|c|c|}
        \hline 
        Benchmark  & Significance   \\
        \hline 
        \textbf{DM70}  & 1.94$\sigma$ \\
        \hline 
    \end{tabular}
    \caption{The signal significance for the signal benchmarks from  VBF for HL-LHC  at an integrated luminosity of 3000 fb$^{-1}$.}
    \label{tab:vbf}
    \end{center}
\end{table}

\subsubsection*{$b\bar{b}$ Higgs associated production}

\begin{figure}
    \centering
   \includegraphics[scale=0.55]{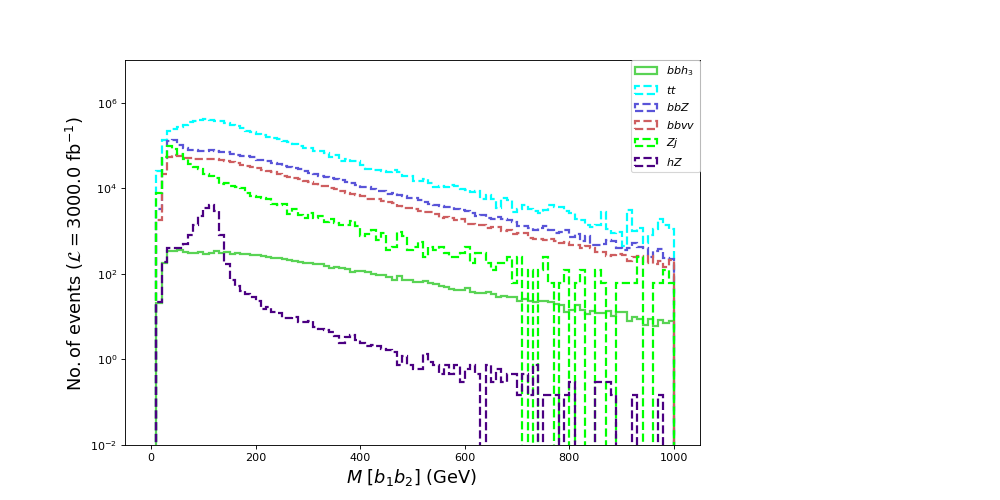}
    \caption{The distribution for the invariant mass of the $b\bar{b}$ pair for the two $b$-jet + MET $>$ 100 GeV (with photon and lepton veto) final state for the BBH signal $\textbf{DM156}_\textbf{w95}$ and dominant SM benchmarks before further analyses cuts are applied.}
    \label{fig:mbblhc}
\end{figure}

We now turn to the associated production channel of {Higgs bosons} with $b\bar{b}$, namely the BBH process for  $\textbf{DM156}_\textbf{w95}$ with the heavy Higgs decaying invisibly into a pair of DM candidates. We estimate the matched cross-section for this process using Santander matching as discussed in the text. For $\textbf{DM156}_\textbf{w95}$, the heavy Higgs mass is 700 GeV and the DM mass is 156 GeV. The heavy Higgs decays to the DM pair with a branching ratio of 69\%. The high invisible branching ratio coupled with the large production cross-section may yield a significant number of signal events at the HL-LHC. We study the prospects of the final state $b\bar{b}$+MET at $\sqrt{s}=14$ TeV. Dominant SM backgrounds to this final state arise from $t\bar{t}, Z+$jets, $b\bar{b}Z, Z(\rightarrow \nu  \bar{\nu})h$, {semi-leptonic} $Z(\rightarrow b  \bar{b})Z(\rightarrow \nu  \bar{\nu})$ and $ b\bar{b}\nu\bar{\nu}$. We generate the dominant SM backgrounds with a MET cut of 100 GeV to tame the production cross-section. We present in Fig.~\ref{fig:mbblhc} the distribution for the invariant mass of the two $b$-jets in the final state, $M(b_1b_2)$. While the SM backgrounds from Higgs and $Z$ typically have a peak at lower values of $M(b_1b_2)$ the signal has a longer tail. Keeping this in mind, we utilize the following \textbf{E1-E4} to separate the signal from the background. 
\begin{itemize}
    \item \textbf{E1}: The final state consists of two b jets and no photons or leptons. We demand $\Delta R(b_1,b_2)>0.4$, $p_T(b_1)>150$ GeV and $p_T(b_2)>100$ GeV.
    \item \textbf{E2}: We demand a large missing transverse momenta (MET) $\slashed{E}_T>200$ GeV to reduce SM background. 
    \item  {\textbf{E3}: We demand $ M(b\bar{b})>200$ GeV to reduce SM background contributions.}  
\end{itemize}
In Table~\ref{tab:bbh2}, we present the cross sections after imposing the cuts for the BBH channel for $\textbf{DM156}_\textbf{w95}$ and the dominant SM background after applying the cuts.
\begin{table}[ht]
    \begin{center}
    \begin{tabular}{|c|c|}
        \hline
        Benchmark & Cross-section after cuts (fb) \\
        \hline 
        $\textbf{DM156}_\textbf{w95}$ & 0.357  \\
        \hline
        SM Background &    \\
        \hline  
         $b\bar{b}Z$ & 18.3 \\
        $b\bar{b}\nu\bar{\nu}$ & 13.46 \\
        $t\bar{t}$ & 66.46 \\
        $Z+j$ & 2.04 \\
        $hZ$ & 0.012\\
        \hline
         Total Background  &  100.27 \\
        \hline
        \end{tabular}
             \caption{The cross-sections  for the signal and backgrounds after applying the cuts \textbf{E1-E4} as discussed in the text for signal-background distinction for BBH for HL-LHC at an integrated luminosity of 3000 fb$^{-1}$.  }
             \label{tab:bbh2}
        \end{center}
        \end{table}

        \begin{table}[ht]
        \begin{center}
    \begin{tabular}{|c|c|}
        \hline 
        Benchmark   & Significance  \\
        \hline 
    $\textbf{DM156}_\textbf{w95}$  & 1.95$\sigma$ \\ 
     \hline 
    \end{tabular}
    \caption{The signal significance for the signal from BBH for HL-LHC at an integrated luminosity of 3000 fb$^{-1}$.}
    \label{tab:bbh}
    \end{center}
\end{table} 
We observe that the matched BBH channel  for  $\textbf{DM156}_\textbf{w95}$ at the LHC has a significance  $\sim 2.65 \sigma$ at leading order at the HL-LHC. Using higher-order K-factor of 1.94 for $t\bar{t}$ to match the predicted cross-section at $\sqrt{s}$ 14 TeV~\cite{Czakon:2011xx}, 1.76 for $b\bar{b}Z$~\cite{Alwall:2014hca} and $b\bar{b}\nu\bar{\nu}$ and an approximate K-factor of 1.2 for the processes $Z+j$ and $hZ$ we obtain the total background cross-section of $\sim 100$ fb which reduces the signal significance to 1.95 $\sigma$ for leading order signal cross-section. 
Thus, we observe that while GGF and VBF are popular channels for heavy Higgs searches, BBH also is a competitive channel compared to VBF at HL-LHC as is the case here for $\textbf{DM156}_{w95}$.  
   
Future hadron colliders such as the proposed FCC-hh/SPPC to run at $\sqrt{s}=100$ TeV may further enhance the discovery potential of this scenario especially for the heavy benchmarks \textbf{DM400}, $\textbf{DM1000}_\textbf{w95}$ and \textbf{DM1000}. We defer the discussion for these benchmarks at FCC-hh/SPPC (more \textit{challenging scenarios}) to sec~\ref{sec:challenge}. 
We present the signal cross-sections {for a  $\sqrt{s}=100$ TeV collider like FCC-hh/SPPC }  in Table~\ref{tab:fcc3}. One can clearly see the prospects of improvement of the GGF, VBF and BBH signals for the benchmarks at  $\textbf{DM55}_\textbf{w95}$, $\textbf{DM156}_\textbf{w95}$ and \textbf{DM70} at FCC-hh/SPPC at $\sqrt{s}=100$ TeV.   From Table~\ref{tab:fcc3} we observe an increase in the production cross-sections at FCC-hh/SPPC by a factor of 100-1000 over LHC and thus favorable channels for heavy Higgs searches at FCC-hh/SPPC.   
\begin{table}[ht!]\begin{center}
\begin{tabular}{|c|c|c|c|}
    \hline
    Process & \multicolumn{3}{|c|}{Production cross-section (fb) at $\sqrt{s}=100$ TeV} \\
     & $\textbf{DM55}_\textbf{w95}$ & $\textbf{DM156}_\textbf{w95}$ & \textbf{DM70}  \\
    \hline
    GGF($h_2 \rightarrow A_S A_S$)   & 10.1$\times10^5$ & -  &  4.09$\times10^5$   \\ 
    GGF($h_3 \rightarrow A_S A_S$)   & - &  1.596 &  -  \\ 
    VBF($h_2 \rightarrow A_S A_S$) & 5.97$\times10^2$& -  &     81.87   \\
    VBF($h_3 \rightarrow A_S A_S$)   &-& 3.12  &     -   \\
    BBH($h_2 \rightarrow A_S A_S$)  & 6.43$\times10^2$& - &   17.2$\times10^3$  \\
     BBH($h_3 \rightarrow A_S A_S$)  & -& 5.00 & -    \\
    \hline
\end{tabular}
\caption{The production cross-sections  at leading order (LO) of the relevant processes for the benchmarks $\textbf{DM55}_\textbf{w95}$, $\textbf{DM156}_\textbf{w95}$ and \textbf{DM70} at $\sqrt{s}=100$ TeV.   For BBH  we use the Santander matched cross-section as defined in the text.}
\label{tab:fcc3}
\end{center}
\end{table}

Besides LHC, our focus will be on future lepton colliders i.e. ILC and muon collider as two possible future options, as we discuss in the following subsections. However, these results can be used in the context of other lepton colliders as well, as we will demonstrate shortly.

\subsection{Benchmarks at $e^+e^-$ colliders}

\begin{figure}[htbp]
    \centering
    \begin{subfigure}{0.49\textwidth}
        \centering
        \includegraphics[width=\textwidth]{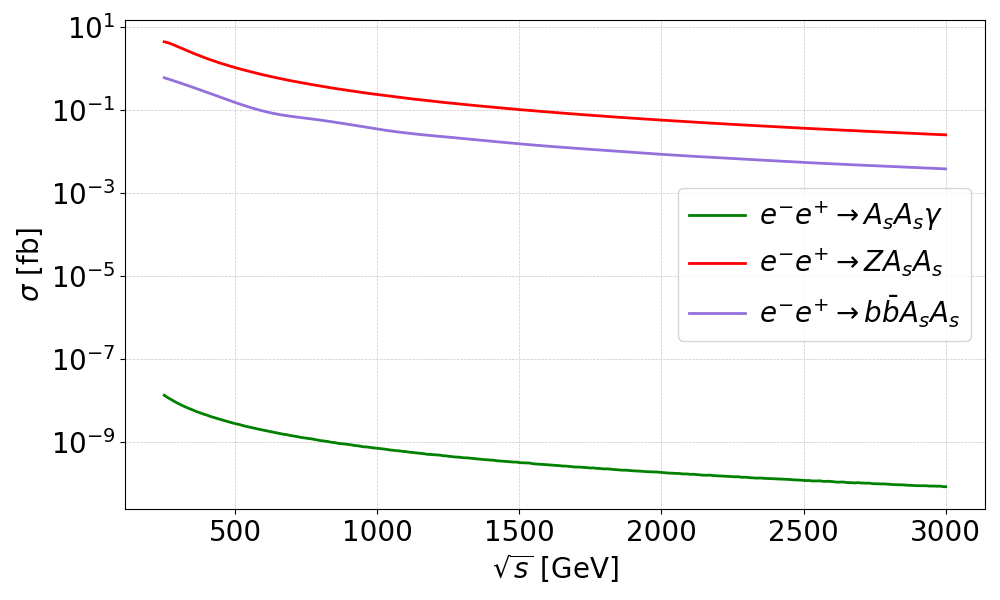}
        \caption{$\textbf{DM55}_\textbf{w95}$}
        \label{fig:prod_cs_ee_coll_DM55_w95}
    \end{subfigure}
    \hfill
    \begin{subfigure}{0.49\textwidth}
        \centering
        \includegraphics[width=\textwidth]{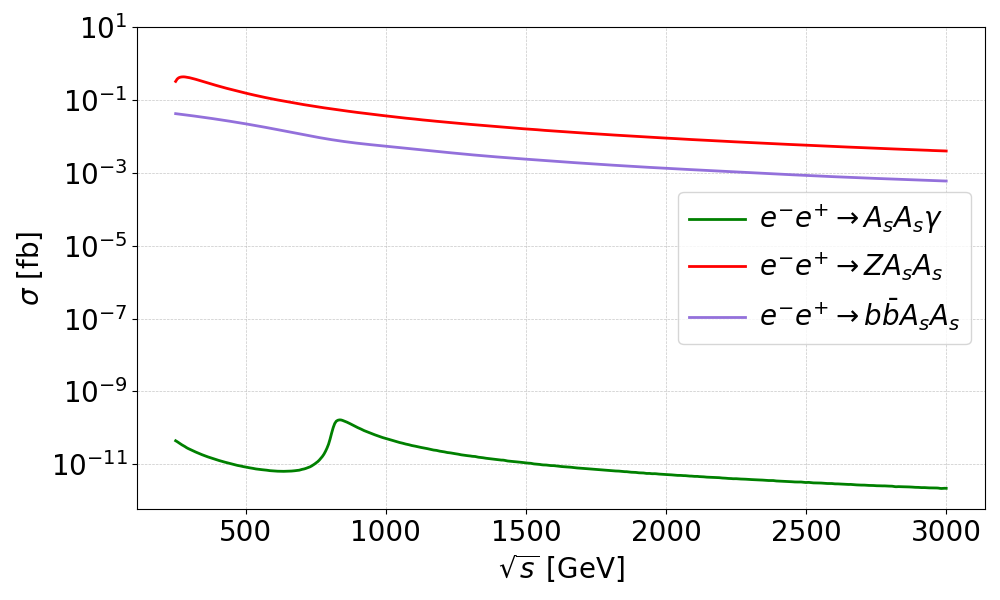}
        \caption{\textbf{DM70}}
        \label{fig:prod_cs_ee_coll_DM70}
    \end{subfigure}
     \hfill
    \vspace{2cm}   
    \begin{subfigure}{0.49\textwidth}
        \centering
        \includegraphics[width=\textwidth]{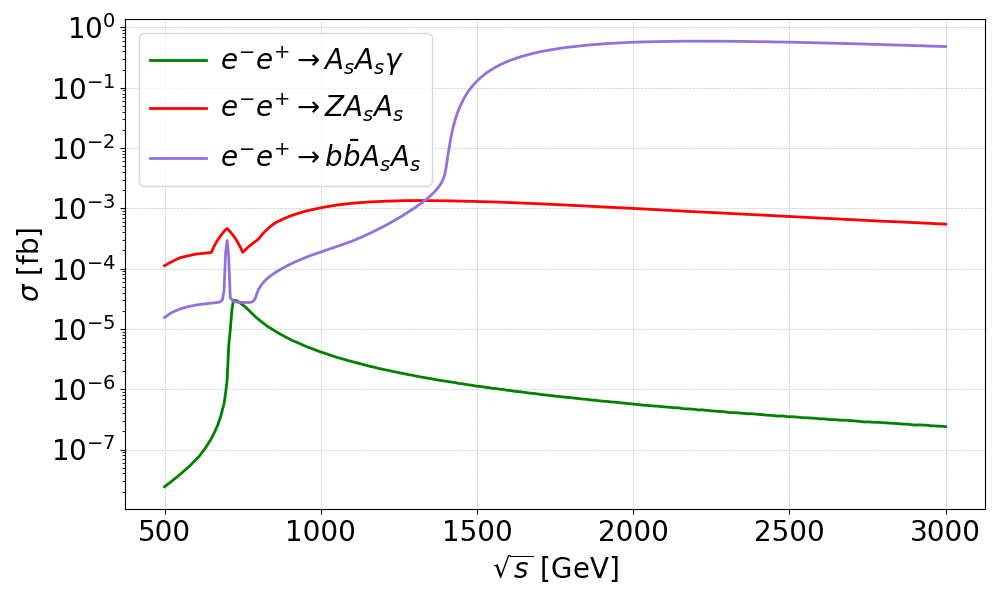}
        \caption{\textbf{$\textbf{DM156}_\textbf{w95}$}}
        \label{fig:prod_cs_ee_coll_DM156_w95}
    \end{subfigure}
    \caption{Variation of the production cross-section against $\sqrt{s}$ for the final states $A_s A_s \gamma$, $Z A_s A_s$, and $b \bar{b} A_s A_s$ at $e^+ e^-$ colliders, computed using \texttt{WHIZARD}~\cite{Kilian:2007gr} for different benchmark points.}
    \label{fig:crosssec_ilc}
\end{figure}

We first discuss the discovery prospects at $e^+e^-$ colliders, i.e. ILC, CEPC, CLIC and FCC-ee. 
Following ILC-TDR~\cite{Behnke:2013lya}, three different centre-of-mass energies, $\sqrt{s}=250$ GeV, 500 GeV and 1 TeV are considered for a comparative analysis. On the other hand, the design energy for CLIC is 3 TeV~\cite{CLICdp:2018cto}. We first show in Fig.~\ref{fig:crosssec_ilc} the production cross-sections for several benchmarks in different final states. It is clear from Fig.~\ref{fig:crosssec_ilc}, that for low mass benchmarks namely $\textbf{DM55}_\textbf{w95}$ and \textbf{DM70}, the largest cross-sections pertain to the mono-$Z$ final state at $e^+e^-$ colliders. Furthermore, the cross-section peaks at $\sqrt{s}=250$ GeV for both aforementioned benchmarks.

\begin{table}[ht]
    \begin{center}
    \begin{tabular}{|c|c|c|c|}
        \hline 
        Benchmark & \multicolumn{3}{|c|}{Production cross-section (fb)}\\
        \hline
        & at $\sqrt{s}=$ 250 GeV  & at $\sqrt{s}=$ 500 GeV & at $\sqrt{s}=$ 1 TeV \\
        \hline
        $\textbf{DM55}_\textbf{w95}$ & 4.42   & 1.1   & 0.24  \\
        \hline
        \textbf{DM70} & 0.33    & 0.15  & 0.035  \\
        \hline
        $\nu\bar\nu Z$ background & 503 &  491  & 950    \\
        \hline 
        \end{tabular}
    \caption{The Production cross-section for signal (for $\textbf{DM55}_\textbf{w95}$ and \textbf{DM70}) and background ($\nu\bar\nu Z$) for $Z$+MET final state at $\sqrt{s}=$ 250 GeV, 500 GeV and 1 TeV $e^+ e^-$ collider.}
    \label{tab:crosssec_monoz_ilc}
    \end{center}
\end{table}

In case of $\textbf{DM55}_\textbf{w95}$, the major production of DM at an $e^+e^-$ machine is via Higgsstrahlung and further decay of the 125 GeV Higgs into a DM pair. Therefore, we look for $Z$+MET 
final states to probe this scenario. The invisible branching ratio of the 125 GeV Higgs $h_2$ is 1.99\% (see Table~\ref{tab:dm_observ_all_bp}).
We further examine the benchmark \textbf{DM70}, where the DM pair results from a light singlet-like non-SM-like 
scalar (150 GeV), following Higgsstrahlung production of the mentioned scalar. The corresponding invisible branching ratio is almost 100\%.

We present the production cross-section of the $Z$+MET 
final state in Table~\ref{tab:crosssec_monoz_ilc} at different $\sqrt{s}$ for both benchmarks. The major background corresponding to the $Z$+MET 
state comes from the $\nu\bar\nu Z$ final state. 

For the signal background analysis we examine the most sensitive kinematical observable, the missing mass ($\slashed{M}$) which is defined as follows:
  
\begin{equation}
   \slashed{M}^2 = (p_{in}-p_{out})^2 ,
   \label{eq:missingmass}
\end{equation}
   

where $p_{in}$ corresponds to incoming four-momenta and $p_{out}$ corresponds to outgoing four-momenta of visible final states. 
  
\begin{figure}[!hptb]
	\centering
	\includegraphics[width=7.3cm,height=5.5cm]{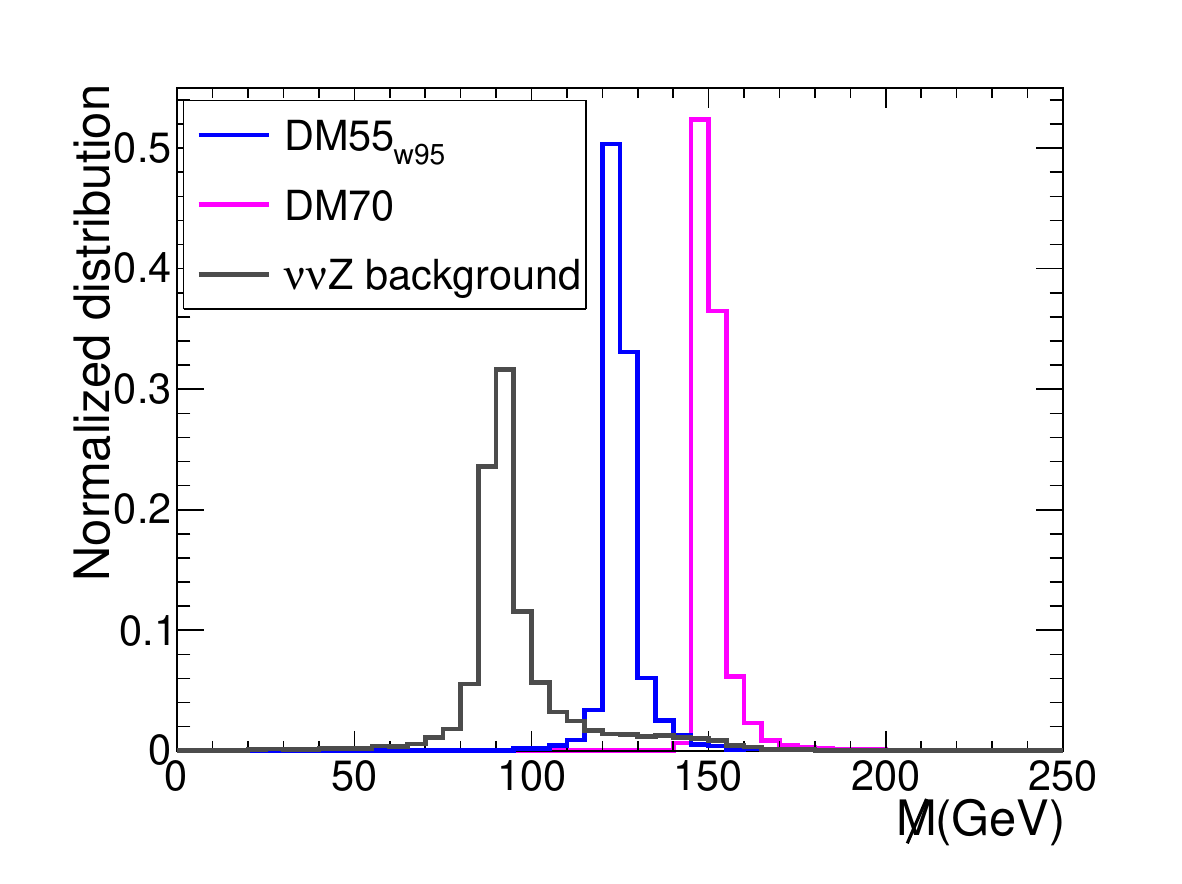}
	\includegraphics[width=7.3cm,height=5.5cm]{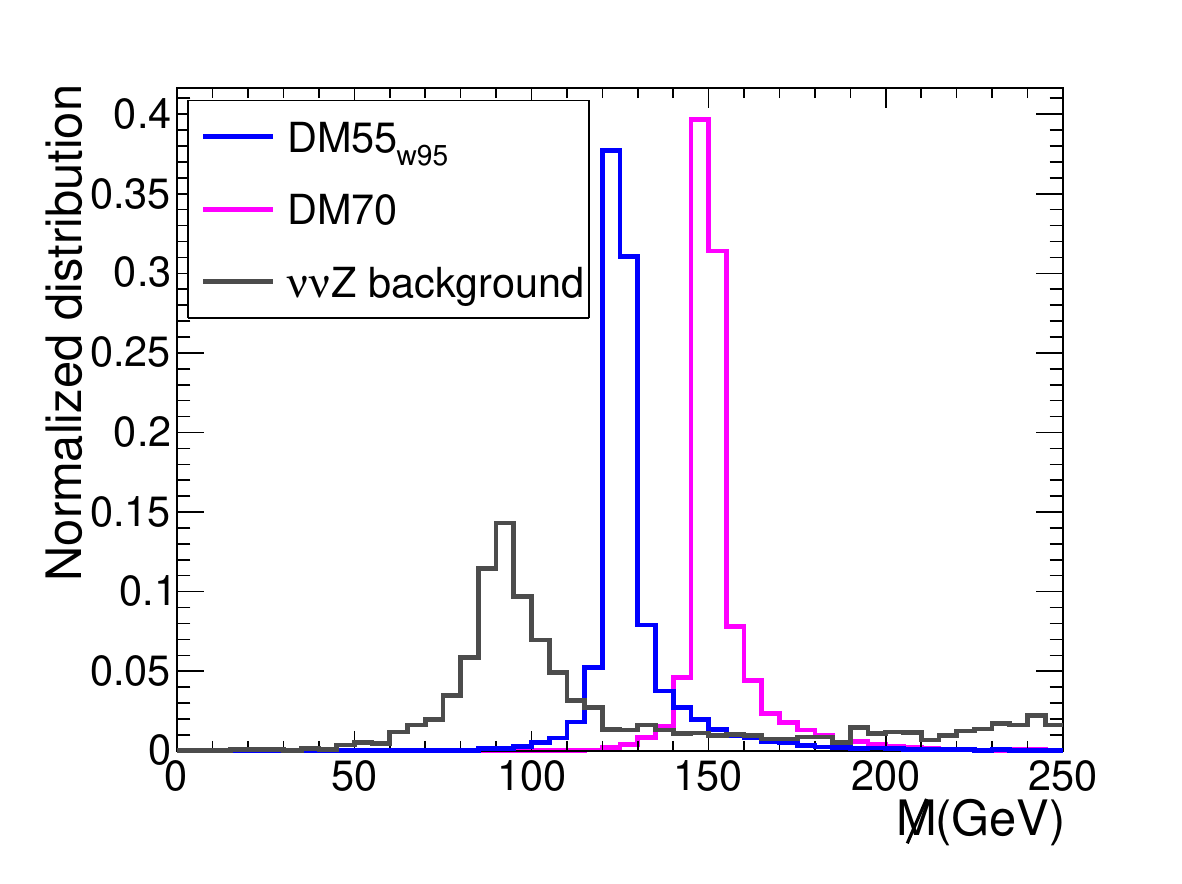}\\
	\includegraphics[width=7.3cm,height=5.5cm]{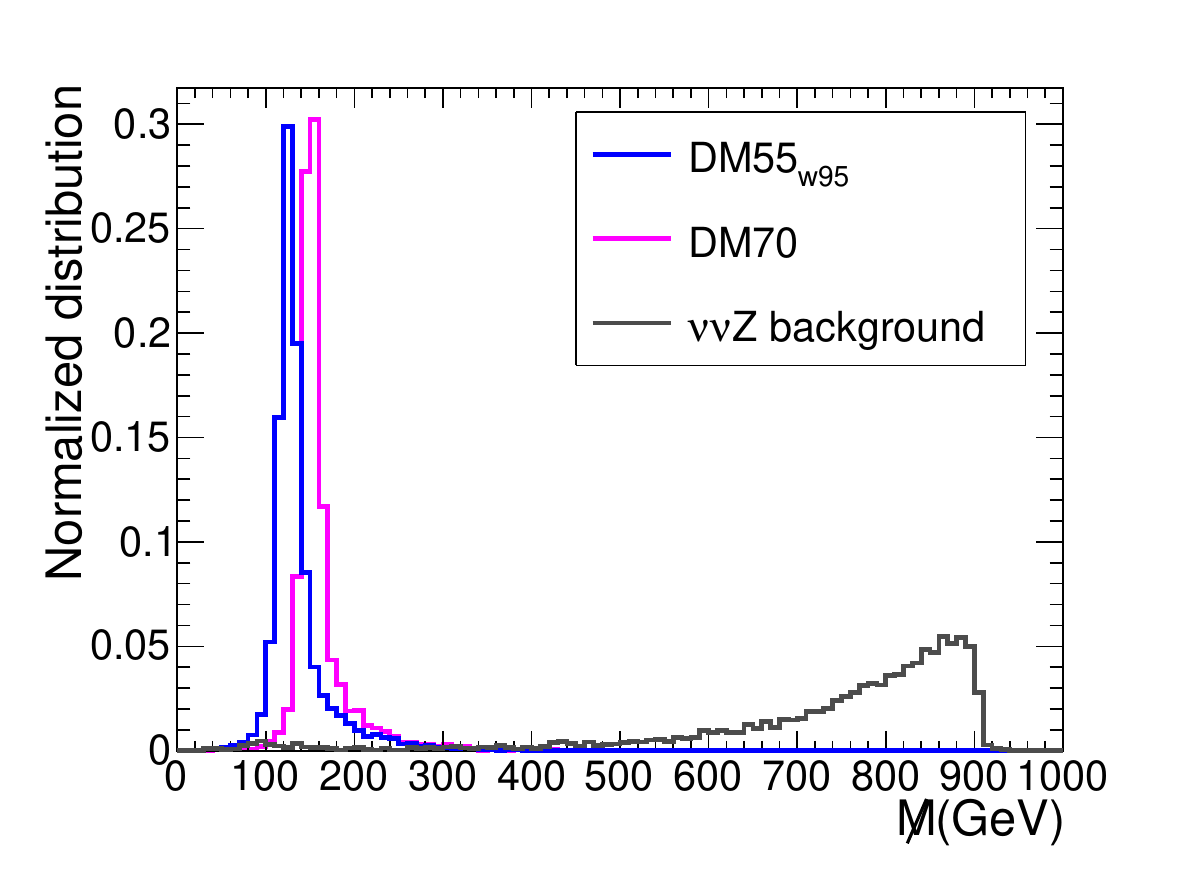}
	\caption{Distribution of missing mass $\slashed{M}$ for signal (for $\textbf{DM55}_\textbf{w95}$ and \textbf{DM70}) and background ($\nu\bar\nu Z$) for $Z$+MET final state at $\sqrt{s}$=250 GeV (top left), $\sqrt{s}$=500 GeV (top right) and $\sqrt{s}$=1 TeV (bottom) $e^+ e^-$ collider.}
	\label{fig:250_gev}
\end{figure}

\begin{table}[ht]
    \begin{center}
    \begin{tabular}{|c|c|c|c|}
        \hline 
        Benchmark & $\sqrt{s}$ & Cut & Significance  \\
        \hline
        $\textbf{DM55}_\textbf{w95}$  & 250 GeV & $\slashed{M} > 100$ GeV    &   11$\sigma$ (1 ab$^{-1}$) \\
        \hline
        \textbf{DM70}  & 250 GeV & $\slashed{M} > 130$ GeV    &   3$\sigma$ (3 ab$^{-1}$) \\
        \hline
        $\textbf{DM55}_\textbf{w95}$  & 500 GeV & $\slashed{M} > 100$ GeV and $\slashed{M} < 150$ GeV     &   3.6$\sigma$ (1 ab$^{-1}$) \\
        \hline
        \textbf{DM70}  & 500 GeV & $\slashed{M} > 140$ GeV and $\slashed{M} < 190$ GeV   &   1.5$\sigma$ (3 ab$^{-1}$) \\
        \hline
        $\textbf{DM55}_\textbf{w95}$  & 1 TeV & $\slashed{M} > 120$ GeV and $\slashed{M} < 250$ GeV     &   2.4$\sigma$ (3 ab$^{-1}$) \\
        \hline
        \textbf{DM70}  & 1 TeV &  $\slashed{M} > 120$ GeV and $\slashed{M} < 250$ GeV   &   0.36$\sigma$ (3 ab$^{-1}$) \\
        \hline
    \end{tabular}
    \caption{The Signal significance (for $\textbf{DM55}_\textbf{w95}$ and \textbf{DM70}) for $Z$+MET final state at $\sqrt{s}$=250 GeV, $\sqrt{s}$=500 GeV and $\sqrt{s}$=1 TeV $e^+ e^-$ collider.}
    \label{tab:significance_monoZ}
    \end{center}
\end{table}

\noindent
We find that $\slashed{M}$ provides best discrimination between signal and backgrounds. Therefore, we show the signal and background distribution of this variable for signal benchmarks and backgrounds. For signal benchmarks, 
$\slashed{M}$ peaks at the mass of the scalar, which decays into a DM pair. On the other hand, at low $\sqrt{s}$, the major background contribution comes from resonant $ZZ(\nu\bar\nu)$ final state, and therefore $\slashed{M}$ 
peaks at the $Z$ mass (see Fig.~\ref{fig:250_gev} top left and top right). At higher $\sqrt{s}$, the dominant contribution to the background comes from non-resonant $\nu\bar\nu Z$ 
final state, and as a result, $\slashed{M}$ peaks at a larger value (see Fig.~\ref{fig:250_gev} bottom centre).

One can infer from Fig.~\ref{fig:250_gev} and Table~\ref{tab:crosssec_monoz_ilc} that, at low $\sqrt{s}$, the event rate is larger, but the signal-background separation is best at higher $\sqrt{s}$. We present the signal significance for the two benchmarks $\textbf{DM55}_\textbf{w95}$ and \textbf{DM70} in Table~\ref{tab:significance_monoZ}.
We would like to emphasize that, the low mass DM is best probed with mono-$Z$ final state owing to a substantial event rate via Higgsstrahlung. 
Furthermore, low $\sqrt{s}$ is favored in this context. Therefore, such low mass DM scenarios are best probed at $e^+e^-$ colliders, namely ILC, FCC-ee or CEPC at $\sqrt{s}=250$ GeV.  

\subsection{Benchmarks at muon collider}

\begin{figure}[htbp]
    \centering
    \begin{subfigure}{0.49\textwidth}
        \centering
        \includegraphics[width=\textwidth]{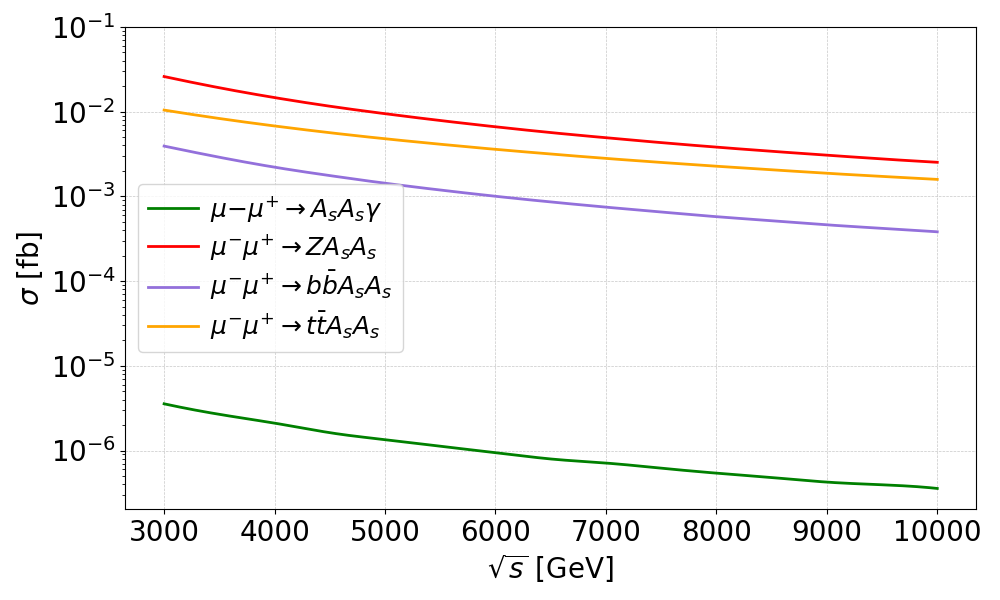}
        \caption{$\textbf{DM55}_\textbf{w95}$}
        \label{fig:prod_cs_mumu_coll_DM55_w95}
    \end{subfigure}
    \hfill
    \begin{subfigure}{0.49\textwidth}
        \centering
        \includegraphics[width=\textwidth]{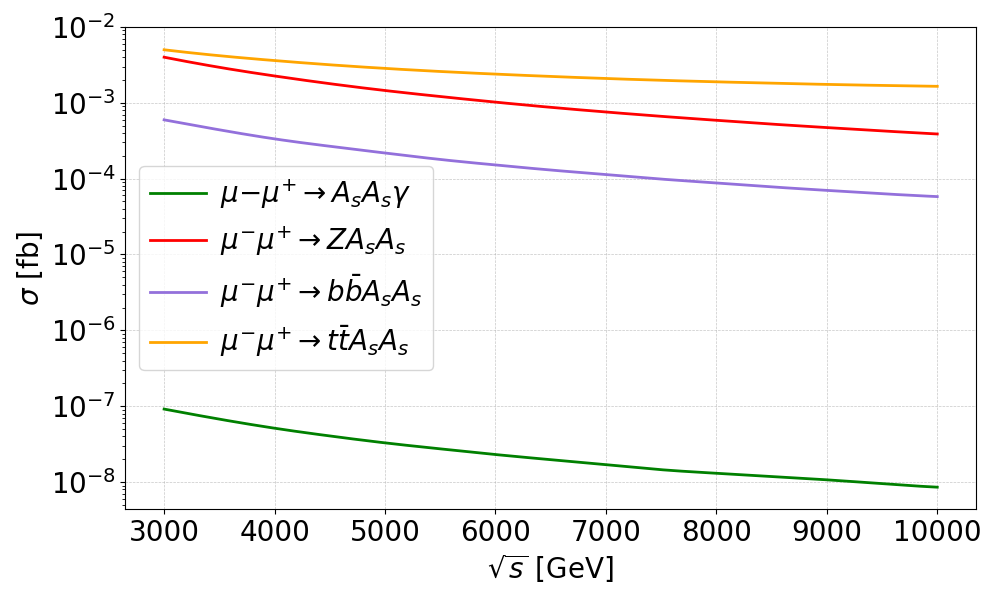}
        \caption{\textbf{DM70}}
        \label{fig:prod_cs_mumu_coll_DM70}
    \end{subfigure}
     \hfill
    \vspace{2cm}   
    \begin{subfigure}{0.49\textwidth}
        \centering
        \includegraphics[width=\textwidth]{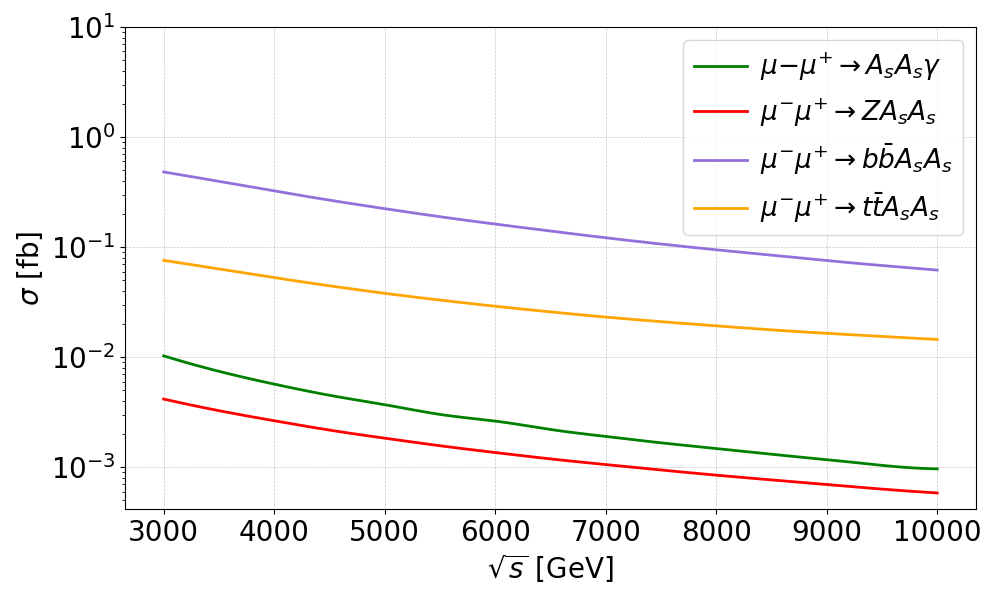}
        \caption{\textbf{$\textbf{DM156}_\textbf{w95}$}}
        \label{fig:prod_cs_mumu_coll_DM156_w95}
    \end{subfigure}
    \hfill
    \begin{subfigure}{0.49\textwidth}
        \centering
        \includegraphics[width=\textwidth]{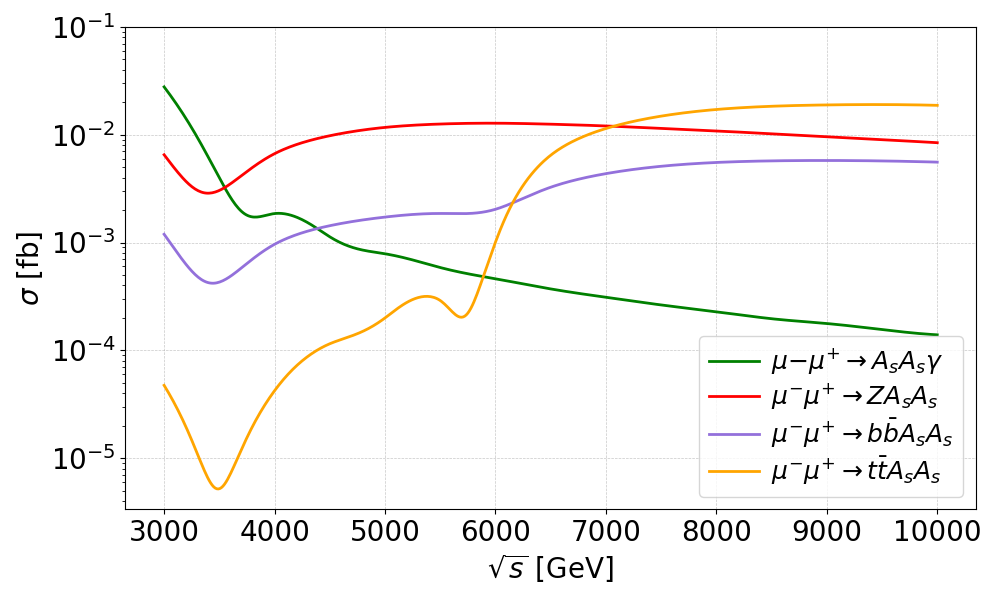}
        \caption{\textbf{$\textbf{DM1000}_\textbf{w95}$}}
        \label{fig:prod_cs_mumu_coll_DM1000_w95}
    \end{subfigure}
    \caption{Variation of the production cross-section against $\sqrt{s}$ for the final states $A_s A_s \gamma$, $Z A_s A_s$, $b \bar{b} A_s A_s$, and $t \bar{t} A_s A_s$ at a muon collider, computed using \texttt{WHIZARD}~\cite{Kilian:2007gr} for different benchmark points.}
    \label{fig:crosssec_muon}
\end{figure}

In this section, we focus on the dominant channels for study of the DM candidate at the muon collider for our chosen benchmarks, as shown in Fig.~\ref{fig:crosssec_muon}. We would like to mention that {the} conventionally favored VBF final state at the high-energy muon collider is sub-leading in our benchmarks. Furthermore, charged Higgs final states, where DM comes from the cascade decay ($H^{\pm} \rightarrow h_i W^{\pm}, h_i \rightarrow A_S A_S$) is also suppressed for our case.

\begin{figure}[!hptb]
	\centering
	\includegraphics[width=9.5cm,height=8.0cm]{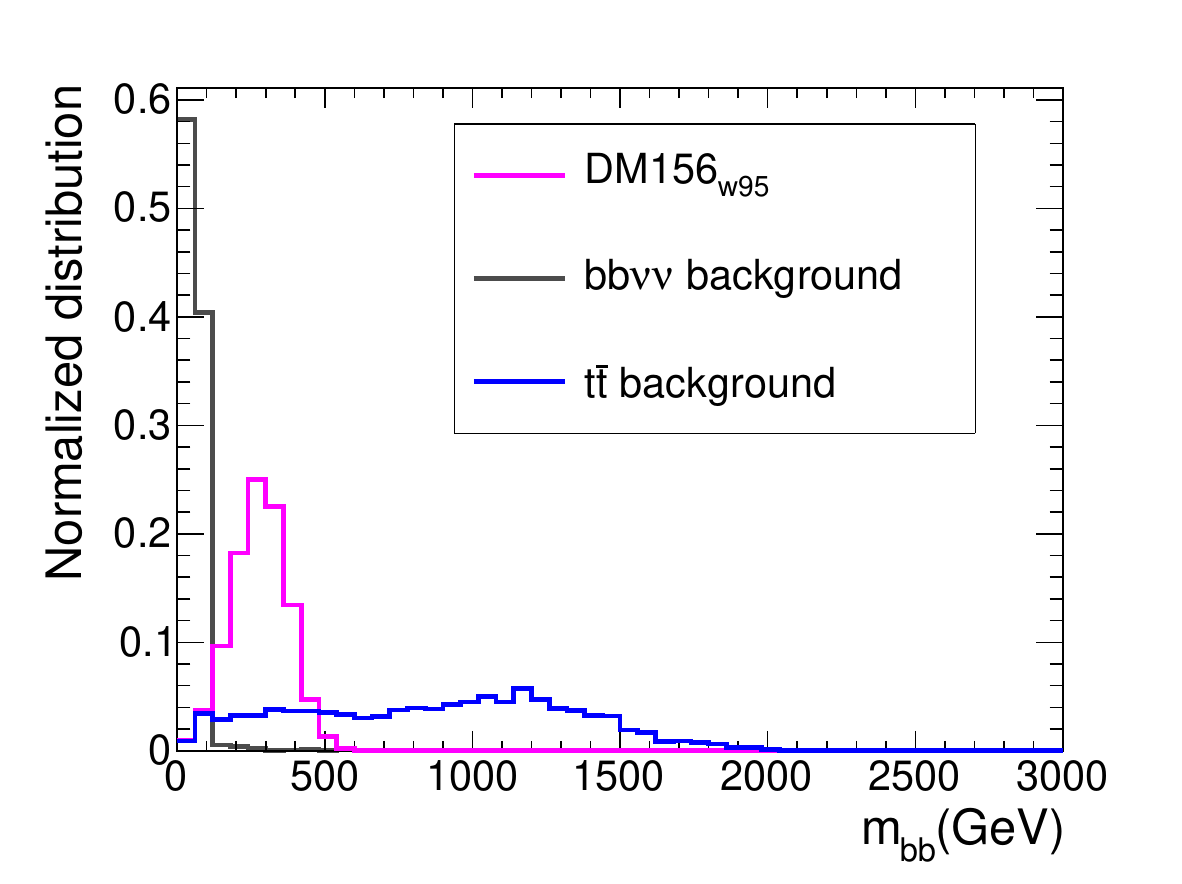}
	\caption{Distribution of invariant mass $m_{bb}$ of a $b$-jet pair for signal (for $\textbf{DM156}_\textbf{w95}$) and background ($b\bar b \nu \bar\nu$ and $t \bar t$) for $b\bar b$+MET final state at $\sqrt{s}$=3 TeV muon collider.}
  	\label{fig:3tev_BP156_w95_bbbar}
\end{figure}

We explore the possibility of probing $\textbf{DM156}_\textbf{w95}$ at the muon collider first. It is clear from Fig.~\ref{fig:crosssec_muon}, that this scenario can be probed best at a muon collider, in the final state where DM is pair-produced in association with $b\bar b$. The cross-section for the signal is largest ($\approx$ 0.5 fb) at $\sqrt{s}=3$ TeV. 
We consider the SM background contribution in the $b\bar b \nu \bar\nu$ as well as $t\bar t$ final state. The total background cross-section amounts to 800 fb. We demand 2 $b$-tagged jets in the final state for our analysis selection and look into the invariant mass distribution of the $b$-jet pair. We demonstrate that this variable offers significant signal background separation (see Fig.~\ref{fig:3tev_BP156_w95_bbbar}). After applying a suitable cut on this variable, we can achieve a signal significance of 6.3$\sigma$ at an integrated luminosity of 3 ab$^{-1}$.

\begin{table}[ht]
    \begin{center}
    \begin{tabular}{|c|c|c|}
        \hline 
        Benchmark & \multicolumn{2}{|c|}{Production cross-section (fb)}\\
        \hline
        & at $\sqrt{s}=$ 3 TeV  & at $\sqrt{s}=$ 10 TeV \\
        \hline
        $\textbf{DM156}_\textbf{w95}$ &  0.48   & 0.063   \\
        \hline
        $b\bar b \nu\nu$ background & 758   &   1.3   \\
        \hline
        $t\bar t$ background & 20  & 1.7   \\
        \hline 
    \end{tabular}
    \caption{The Production cross-section for signal (for $\textbf{DM156}_\textbf{w95}$) and background ($b\bar b \nu\nu$ and $t\bar t$) for $b\bar b$+MET final state at $\sqrt{s}=$ 3 TeV and 10 TeV muon collider.}
    \label{tab:crosssec_muon_collider}
    \end{center}
\end{table}

\begin{table}[ht]
    \begin{center}
    \begin{tabular}{|c|c|c|}
        \hline 
        Benchmark & Cut & Significance  \\
        \hline
        $\textbf{DM156}_\textbf{w95}$  &  100 GeV $< m_{bb} < 500$ GeV &   6.3$\sigma$ (3 ab$^{-1}$) \\
        \hline
        \end{tabular}
    \caption{The Signal significance and corresponding cuts (for $\textbf{DM156}_\textbf{w95}$) for $b\bar b$+MET final state at $\sqrt{s}$ = 3 TeV muon collider.}
    \label{tab:significance_bbbar_BP156_w95}
    \end{center}
\end{table}

Next we would like to draw the attention of the reader to an interesting point. We perform the analysis for $\textbf{DM156}_\textbf{w95}$ at the muon collider at $\sqrt{s}=1$ TeV. Although, the design energy for muon collider starts at 3 TeV~\cite{InternationalMuonCollider:2024jyv}, we investigate the physics case at $\sqrt{s}=1$ TeV. There is a possibility that a 1 TeV muon collider will have a stronger physics case compared to ILC or CLIC operating at 1 TeV. Here we present such an example. The reason behind this is larger muon-Yukawa couplings compared to electron-Yukawa couplings. Owing to this significant enhancement, one can achieve substantial production cross-sections (at least four orders of magnitude larger than at the $e^+e^-$-collider~\cite{Dutta:2023cig}) in the mono-photon+MET final state, mediated via $s$-channel $h_3$ production at the muon collider. 

\begin{figure}[!hptb]
	\centering
	\includegraphics[width=9.5cm,height=8.0cm]{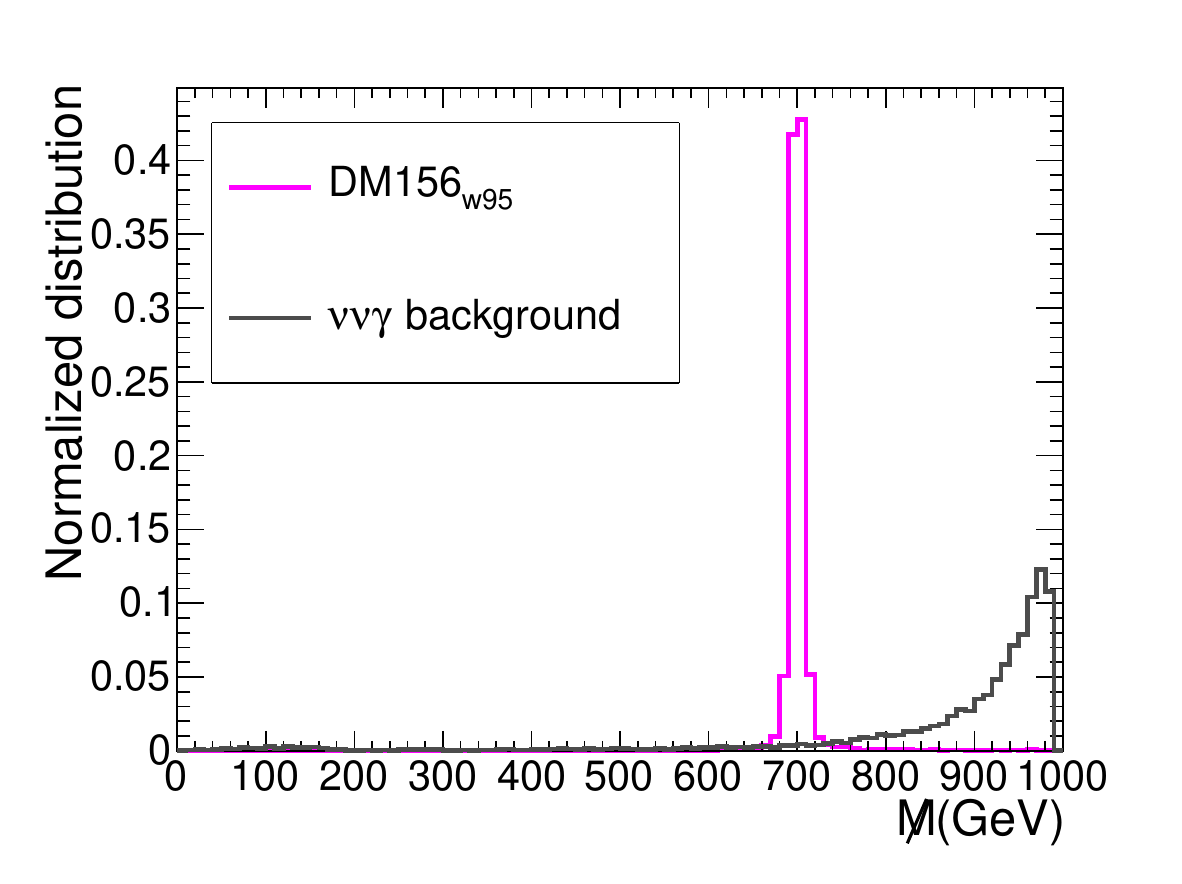}
	\caption{Distribution of missing mass $\slashed{M}$ for signal ($\textbf{DM156}_\textbf{w95}$) and background ($\nu\nu\gamma$) for $\gamma$+MET final state at $\sqrt{s}$=1 TeV muon collider.}
  	\label{fig:1tev_BP156_w95_monophoton}
\end{figure}

Keeping this fact in mind we present the detection prospects of $\textbf{DM156}_\textbf{w95}$ at a muon-collider at 1 TeV in the mono-photon final state.
We present the signal and background production cross-sections in Table~\ref{tab:crosssec_muon_collider_monophoton_1tev}. We find that the missing-mass $\slashed{M}$ 
offers best separation between signal and background. We present the distribution in Fig.~\ref{fig:1tev_BP156_w95_monophoton}. Choosing suitable cuts on this variable, we achieve appreciable detection prospects for this benchmark, which can be seen in Table~\ref{tab:significance_monophoton_BP156_w95}.

\begin{table}[ht]
    \begin{center}
    \begin{tabular}{|c|c|}
        \hline 
        Benchmark & Production cross-section (fb) at $\sqrt{s}=1$ TeV\\
        \hline
        $\textbf{DM156}_\textbf{w95}$ &  0.23   \\
        \hline
        $\nu\nu\gamma$ background & 2.45  \\
        \hline 
        \end{tabular}
    \caption{The Production cross-section for signal (for $\textbf{DM156}_\textbf{w95}$) and background ($\nu\nu\gamma$) for $\gamma$+MET final state at $\sqrt{s}=$ 1 TeV muon collider.}
    \label{tab:crosssec_muon_collider_monophoton_1tev}
    \end{center}
\end{table}

\begin{table}[ht]
    \begin{center}
    \begin{tabular}{|c|c|c|}
        \hline 
        Benchmark & Cut & Significance  \\
        \hline
        $\textbf{DM156}_\textbf{w95}$  &  690 GeV $< \slashed{M} < 710$ GeV &   3$\sigma$ (3 ab$^{-1}$), 5.3$\sigma$ (10 ab$^{-1}$) \\
        \hline
        \end{tabular}
    \caption{The Signal significance and corresponding cuts (for $\textbf{DM156}_\textbf{w95}$) for $\gamma$+MET final state at $\sqrt{s}$ = 1 TeV muon collider.}
    \label{tab:significance_monophoton_BP156_w95}
    \end{center}
\end{table}

Our motivation for this particular study may not be practical from the perspective of the current experimental designs of future colliders. However, we reiterate that a muon collider at $\sqrt{s}=1$ TeV can be beneficial for specific regions of parameter space compared to $e^+e^-$ colliders of the same $\sqrt{s}$. Furthermore, in the context of a muon collider, the mono-photon final state at $\sqrt{s}=1$ TeV can provide a good hint for the DM signal.

\begin{table}[ht]
    \begin{center}
    \begin{tabular}{|c|c|}
        \hline 
        Benchmark & Production cross-section (fb) at $\sqrt{s}=10$ TeV\\
        \hline
        $\textbf{DM1000}_\textbf{w95}$ &  0.027    \\
        \hline
        SM $t\bar t$ background & 1.66   \\
        \hline 
        \end{tabular}
    \caption{The Production cross-section for signal ($\textbf{DM1000}_\textbf{w95}$) and background ($t\bar t$) for $t\bar t$+MET final state at $\sqrt{s}=$ 10 TeV muon collider.}
    \label{tab:crosssec_muon_collider_ttbar_10tev}
    \end{center}
\end{table}

\begin{figure}[!hptb]
	\centering
	\includegraphics[width=9.5cm,height=8.0cm]{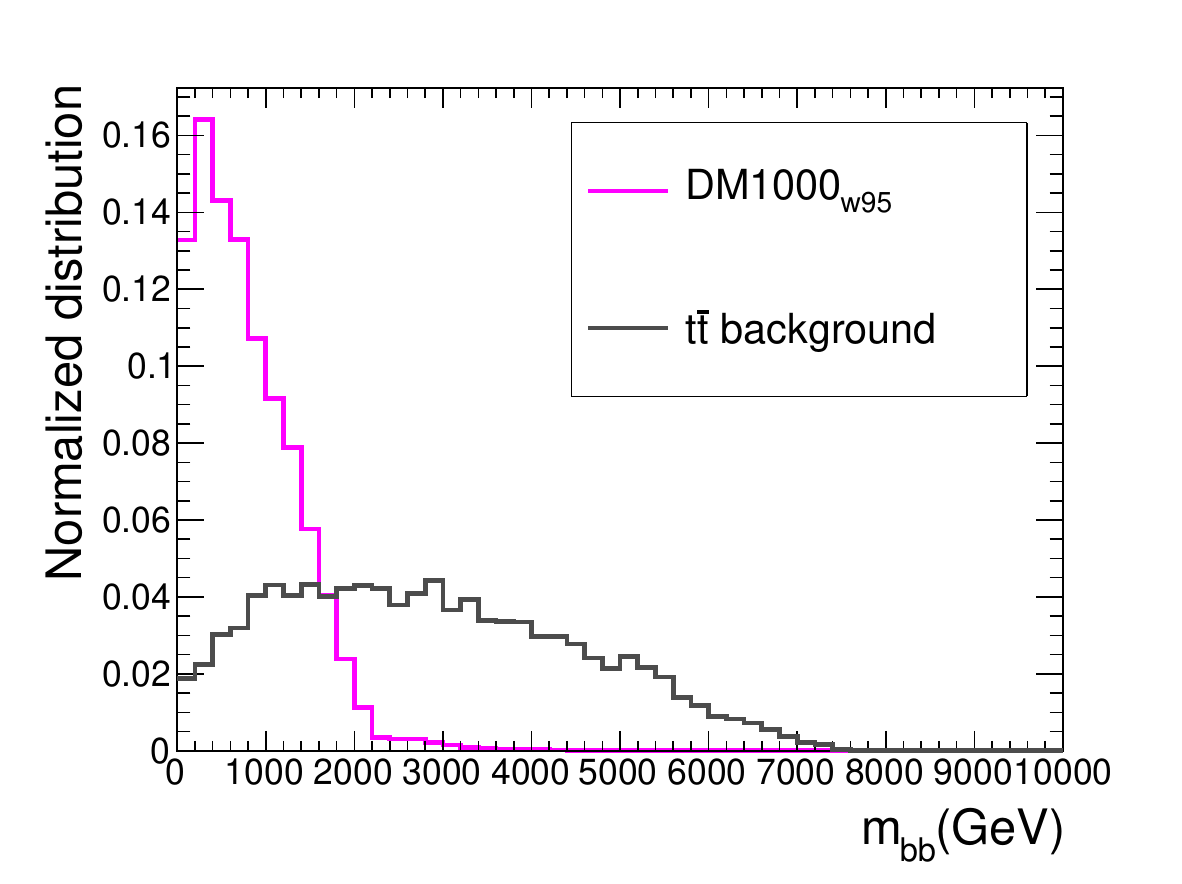}
	\caption{Distribution of invariant mass $m_{bb}$ of a $b$-jet pair for signal ($\textbf{DM1000}_\textbf{w95}$) and background (SM $t\bar t$) for $t\bar t$+MET final state at $\sqrt{s}$=10 TeV muon collider.}
  	\label{fig:10tev_BP1000_w95_ttbar}
\end{figure}

\begin{table}[ht]
    \begin{center}
    \begin{tabular}{|c|c|c|}
        \hline 
        Benchmark & Cut & Significance  \\
        \hline
        $\textbf{DM1000}_\textbf{w95}$  &  $m_{bb} < 2$ TeV &   2.9$\sigma$ (10 ab$^{-1}$) \\
        \hline
    \end{tabular}
    \caption{The Signal significance and corresponding cuts (for $\textbf{DM1000}_\textbf{w95}$) for $t\bar t$+MET final state at $\sqrt{s}$ = 10 TeV muon collider.}
    \label{tab:significance_ttbar_BP1000_w95}
    \end{center}
\end{table}

Finally we consider the benchmark in the heavy mass region $\textbf{DM1000}_\textbf{w95}$. Here the DM is of 1 TeV mass, and the non-SM-like scalars $h_3,A,H^{\pm}$ are at 2.9 TeV. We present the production cross-section of such a benchmark across various $\sqrt{s}$ in Fig.~\ref{fig:crosssec_muon}. One can see that the best prospects for this benchmark are at the high energy muon collider i.e. at $\sqrt{s}=10$ TeV. And the most suitable channel in this case turns out to be pair produced DM in association with $t\bar t$. The major reason behind this is the considerable branching ratio of $h_3$ to $t\bar t$. However, the large branching ratio of $t\bar t$ naturally leads to a somewhat reduced invisible branching ratio of $h_3$. In our scan we find the largest invisible branching ratio of $\lesssim 21$\%. However, the large branching ratio of $h_3$ to $t\bar t$ fairly compensates for this when we look at the $t\bar t$+MET final state. The major SM background in this case is the SM $t\bar t$ process. We present the signal and background cross-sections in Table~\ref{tab:crosssec_muon_collider_ttbar_10tev}.
One may note the generic feature in Fig.~\ref{fig:crosssec_ilc} and \ref{fig:crosssec_muon}, there is an $s$-channel resonant enhancement when $\sqrt{s}$ is in the vicinity of the mediator mass, this results in a peak. At higher $\sqrt{s}$ the $t$-channel processes {(see in particular the $b\bar b$ and $t\bar t$ associated production)} start to dominate and the production cross-section rises again {while for the $A_SA_S\gamma$ final state with only s-channel contributions, no such rise at higher $\sqrt{s}$ appears.}   


Next we show the invariant mass of a $b\bar b$ pair for signal and background in Fig.~\ref{fig:10tev_BP1000_w95_ttbar}. Since in case of signal the $b\bar b$ pair comes from a $t\bar t$ pair from $h_3$ decay, they are relatively close, whereas in case of $t\bar t$ background, the two $b$-quarks come from a $t\bar t$ pair, which are back-to-back. 
This leads to a larger invariant mass of the $b\bar b$ pair in case of the background compared to the signal. After choosing a suitable cut on the invariant mass, we achieve a signal significance of $\sim 3\sigma$ (see Table~\ref{tab:significance_ttbar_BP1000_w95}).

\subsection{Challenging scenarios}

\label{sec:challenge}
Having discussed benchmark regions that can possibly be probed at HL-LHC and future lepton colliders, we turn our attention to the scenarios that are extremely difficult to probe at future colliders.

\begin{table}[ht]
    \begin{center}
    \begin{tabular}{|c|c|c|}
        \hline 
        Final state & \multicolumn{2}{|c|}{Production cross-section  (fb) at muon collider}\\
        \hline
        & at $\sqrt{s}=$ 3 TeV  & at $\sqrt{s}=$ 10 TeV \\
        \hline
        $\gamma$+MET & 5.3 $\times 10^{-7}$    & 4.9 $\times 10^{-8}$   \\
        \hline 
        $Z$+MET & 1.1 $\times 10^{-5}$    & 1.5 $\times 10^{-6}$  \\
        \hline
        $b\bar b$+MET & 2.7 $\times 10^{-3}$    & 4.5 $\times 10^{-3}$    \\
        \hline
        $t\bar t$+MET & 3.7 $\times 10^{-3}$    & 8.9$\times 10^{-3}$   \\
        \hline
    \end{tabular}
    \caption{The Production cross-sections (for {\bf DM400}) for different final states at $\sqrt{s}=$ 3 TeV and 10 TeV muon collider.}
    \label{tab:crosssec_bp400}
    \end{center}
\end{table}

\begin{table}[ht]
    \begin{center}
    \begin{tabular}{|c|c|c|}
        \hline 
        Final state & \multicolumn{2}{|c|}{Production cross-section (fb) at muon collider}\\
        \hline
        & at $\sqrt{s}=$ 3 TeV  & at $\sqrt{s}=$ 10 TeV \\
        \hline
        $\gamma$+MET & 3.5 $\times 10^{-9}$    & 1.3 $\times 10^{-10}$   \\
        \hline 
        $Z$+MET & 4.4 $\times 10^{-8}$   & 2.2 $\times 10^{-6}$   \\
        \hline
        $b\bar b$+MET & 3.7 $\times 10^{-8}$    & 2.0 $\times 10^{-5}$   \\
        \hline
        $t\bar t$+MET & 7.8 $\times 10^{-9}$   & 3.7$\times 10^{-5}$   \\
        \hline
    \end{tabular}
    \caption{The Production cross-sections (for {\bf DM1000}) for different final states at $\sqrt{s}=$ 3 TeV and 10 TeV muon collider.}
    \label{tab:crosssec_bp1000}
    \end{center}
\end{table}

We find that in case of {\bf DM400} and {\bf DM1000}, where the mediator $h_3$ is singlet-dominated, the production cross-section decreases significantly, due to its suppressed couplings with SM particles. Due to the same reason the branching ratio of $h_3$ to DM states can become quite significant, which is the case for {\bf DM400} (82\%). However, since the the production cross-section is small, the resulting event rate in all the final states considered decreases significantly. We can see this in Tables~\ref{tab:crosssec_bp400} and ~\ref{tab:crosssec_bp1000} when compared with Fig.~\ref{fig:crosssec_muon}. We remind the reader that, both these benchmarks can fully account for the observed relic density, and are allowed by all relevant constraints. However, such scenarios will be extremely difficult to probe at future lepton colliders. We explored the possibility of probing these benchmarks at future high energy hadron colliders, such as FCC-hh/SPPC. Although we do not present a detailed study here, we mention the production cross-sections for these benchmarks at FCC-hh/SPPC. The production cross-sections at $\sqrt{s}=14$ TeV LHC and $\sqrt{s}=100$ TeV FCC-hh/SPPC are listed in Table~\ref{tab:lhchh} and ~\ref{tab:fcchh} respectively. We observe at least an increase in the production at FCC-hh/SPPC over LHC by a factor of $\sim 30-40$ for VBF and BBH for \textbf{DM400} while for GGF the increase in the production  cross-section is by a factor of 100. For \textbf{DM1000}, the increase in the production cross-sections are by a factor of 1000. Such a large production cross-section provides a favorable case for probing heavier Higgs masses at FCC-hh/SPPC.

\begin{table}[ht!]\begin{center}
\begin{tabular}{|c|c|c|}
    \hline
    Process & \multicolumn{2}{|c|}{Production cross-section (fb) at $\sqrt{s}=14$ TeV} \\
    &   $\textbf{DM400}$ & $\textbf{DM1000}$   \\
    \hline
     GGF &  0.016 & 1.27$\times10^{-4}$  \\
     VBF & 0.001  & 4.7$\times10^{-6}$  \\ 
     BBH & 0.008 &  1.96$\times10^{-6}$ \\
    \hline
\end{tabular}
\caption{The production cross-sections at leading order (LO) of relevant processes at $\sqrt{s}=14$ TeV at LHC.}
\label{tab:fcchh}
\end{center}
\end{table}

\begin{table}[ht!]\begin{center}
\begin{tabular}{|c|c|c|}
    \hline
    Process & \multicolumn{2}{|c|}{Production cross-section (fb) at $\sqrt{s}=100$ TeV} \\
    &   $\textbf{DM400}$ & $\textbf{DM1000}$   \\
    \hline
     GGF & 1.456  & 0.117 \\
     VBF & 0.039  & 1.182 \\ 
     BBH & 0.264 & 0.029    \\
    \hline
\end{tabular}
\caption{The production cross-sections at leading order (LO) of relevant processes at $\sqrt{s}=100$ TeV at FCC-hh/SPPC.}
\label{tab:lhchh}
\end{center}
\end{table}

\subsection{Complimentarity in search for dark matter in 2HDMS at future colliders}

We studied in detail the different regions of parameter spaces in the 2HDMS scenario and the prospects of their detection in future colliders. While analyzing complimentarity between various future machines, we summarize the {salient points highlighting the complimentarity of the different present and future colliders in the context of DM searches.}

    {The dominant production channels at }HL-LHC { are GGF, VBF and associated $b\bar{b}$ Higgs production channels. }
    In the low mass range (i.e. $\textbf{DM55}_\textbf{w95}$), while GGF and VBF processes dominate over BBH, due to small invisible branching ratio, HL-LHC would not be sensitive to these channels. In the intermediate/heavy mass range, the production cross-section is already too small. 
    Interestingly, for $\textbf{DM70}$, where the DM pair comes from a singlet-like light scalar, the production cross-section is enhanced due to kinematic reasons. In this case, LHC yields a good signal  significance in the GGF final state, owing to substantial production cross-section and almost 100\% invisible branching ratio. For the intermediate mass benchmark, $\textbf{DM156}_\textbf{w95}$, the BBH channel can provide a possible hint with $\sim 2\sigma$ excess at the HL-LHC.
     {These channels may have  better detection prospects  at a  future high energy hadron colliders such as the FCC-hh/SPPC with $\sqrt{s}=100$ TeV owing to the larger enhancements in the production cross-sections which we leave for a future study.}
   
   The low mass benchmarks (i.e. $\textbf{DM55}_\textbf{w95}$ or $\textbf{DM70}$) have best prospects at ILC/CEPC/FCC-ee operating at $\sqrt{s}=250$ GeV in the mono-$Z$ final state, due to the large Higgstrahlung cross-section. 
   Among lepton colliders the best prospect for the intermediate benchmark namely, $\textbf{DM156}_\textbf{w95}$  turns out to be at a muon collider with $\sqrt{s}=3$ TeV in the associated $b\bar b$ final state. Interesting to note that it is also possible to probe such a scenario at lower $\sqrt{s}$ at a muon collider i.e. $\sqrt{s}=1$ TeV in the mono-photon final state. 
   The heavy DM as well as heavy mediator scenario $\textbf{DM1000}_\textbf{w95}$ can only be probed at a muon collider with $\sqrt{s}=10$ TeV in the associated $t\bar t$ final states. 
   There are a few challenging scenarios, especially when the DM comes from a singlet-dominated scalar, which is not light, the production cross-section becomes too small for such scenarios to be probed at any of the future lepton colliders. {Proposed future hadron colliders such as FCC-hh/SPPC}, operating at $\sqrt{s}=100$ TeV, can be helpful in probing such challenging scenarios due to significant enhancement in the signal cross-section. 

\section{Summary and conclusion}
\label{sec:summary}

In this work, we performed a detailed analysis of DM searches, within the reference model `two Higgs doublets and a complex singlet scalar (2HDMS)' at future colliders. Such scenarios give rise to a DM candidate under the imposition of a $Z_2$ symmetry. In an earlier work~\cite{dutta_2024_10569080}, we identified regions of parameter space in this model, satisfying all theoretical and experimental constraints as well as DM constraints. In this work, we chose interesting benchmarks from such parameter regions that can be probed at future colliders{, where most of the benchmarks can provide the correct DM relic density}. In this regard, in some of our chosen benchmarks we have also considered the possibility that one of the scalars in the particle spectrum is of 95 GeV mass and can explain the recently observed excess at LEP and the LHC experiments CMS and ATLAS. However, this criterion is not central to our work. {In this work, we summarize our analysis results as the following points:}

{
\begin{itemize}
    \item We found that, the parameter space which are under-abundant in terms of relic density, are easier to probe at the collider. The reason is, smaller relic density implies larger annihilation cross-section of DM pair and consequently larger branching fraction of heavier scalars to DM pair. An example of this is $\textbf{DM156}_\textbf{w95}$ in our case. 
    \item In scenarios where the DM pair comes from a singlet-like heavy scalar, it is possible to achieve the observed relic and large invisible branching ratio simultaneously. However, such case are challenged by small production cross-section of the singlet scalar. \item We found that future $e^+e^-$ colliders such as FCC-ee, ILC and CEPC, can probe the low mass region, where the mass of the DM candidate and of the mediator which acts as a portal between DM and the SM sector, are below 250 GeV. Especially the mono-$Z$+MET final state gives 
    $11\sigma$ signal significance for DM with 55~GeV mass and $3\sigma$ significance for 70~GeV DM at $\sqrt{s}=250~$GeV.
    \item The scenario with DM and mediator in the intermediate mass region, can produce an excess of $\sim 2\sigma$ at the HL-LHC, when a DM pair is produced in association with a $b\bar b$-pair.
    \item The intermediate scenario can be probed at a muon collider operating at a high centre of mass energy ($\sqrt{s}=3$ TeV), with much larger significance of $\sim 6\sigma$ at 3~ab$^{-1}$ in associated production with a $b\bar b$ pair. 
    \item Finally, we looked into a benchmark with heavy DM and mediator mass. Such a scenario is beyond the reach of the considered $e^+e^-$ colliders as well as HL-LHC and can be probed at a muon collider ($t\bar{t}$ associated production) with $\sqrt{s}=10$ TeV, with a signal significance of $\sim 3\sigma$. 
\end{itemize}
}

{Overall, this work uses the 2HDMS as a reference model to perform the systematic DM phenomenology analysis for three scenarios, which are the light DM scenario ($m_\mathrm{DM}<100$~GeV), the intermediate DM scenario ($100<m_\mathrm{DM}<1000$~GeV) and the heavy DM scenario ($m_\mathrm{DM}>1000$~GeV).
We performed a detailed comparison between various future colliders including HL-LHC, future $e^+e^-$ colliders such as FCC-ee, ILC, CLIC and the proposed muon collider, exploring various final states which can possibly probe such DM scenarios at a substantial signal significance. We found that the muon collider ($\sqrt{s}=3$~TeV) can provide a better significance for the intermediate DM scenario compared to the HL-LHC, while heavy DM scenario can be only probed at muon collider ($\sqrt{s}=10$~TeV). In addition, $e^+ e^-$ colliders would be most promising in probing the light DM scenario.}
We believe that this work provides a thorough and exhaustive roadmap for future collider experiments aiming at discovering DM. 


\section*{Acknowledgements}
The authors (JL,GMP, SFT, JZ) acknowledge support by the Deutsche Forschungsgemeinschaft (DFG, German Research Foundation) under Germany's Excellence Strategy EXC 2121 "Quantum Universe" - 390833306.
JD acknowledges support from the computing facilities (the theory cluster, the National Analysis Facility NAF and the BIRD cluster) of the Deutsches Elektronen-Synchrotron DESY and DESY Sync $\&$ Share. JD also acknowledges support from the OUHEP cluster at University of Oklahoma and computing facilities at the Institute of Mathematical Sciences, India.
CL is supported by the Natural Science Foundation of China (NSFC) under grant numbers 12305115 and the Shenzhen Science and Technology Program (Grant No. 202206193000001, 20220816094256002).

\section*{Appendix}
\appendix
\section{Minimization conditions}
The 2HDMS potential in Eq.~\ref{eq:2HDMS_potential} has three minima at the three vevs $v_1$, $v_2$ and $v_S$. 

At a minimum the derivative of the potential is equal to zero. This leaves us with three equations, the minimization conditions, which we can use to eliminate three of the free parameters from the model:
\begin{subequations}\label{eq:minimization_cond}
\begin{align}
    0 = \frac{\partial V}{\partial \Phi_1}|_{\substack{\Phi_1 = \langle \Phi_1 \rangle\\ \Phi_2 = \langle \Phi_2 \rangle\\ S = \langle S \rangle}} &= \frac{1}{\sqrt{2}} [ m_{11}^2 v_1 -m_{12}^2 v_2 + \frac{\lambda_1}{2} v_1^3 + \frac{\lambda_{345}}{2} v_1 v_2^2 + (\frac{\lambda_1'}{2}v_1 + \lambda_4' v_1)v_S^2 ] \\
    0 = \frac{\partial V}{\partial \Phi_2}|_{\substack{\Phi_1 = \langle \Phi_1 \rangle\\ \Phi_2 = \langle \Phi_2 \rangle\\ S = \langle S \rangle}} &= \frac{1}{\sqrt{2}} [ m_{22}^2 v_2 -m_{12}^2 v_1 + \frac{\lambda_2}{2} v_2^3 + \frac{\lambda_{345}}{2} v_1^2 v_2 + (\frac{\lambda_2'}{2} v_2 + \lambda_5' v_2) v_S^2 ]\\
    0 = \frac{\partial V}{\partial S}|_{\substack{\Phi_1 = \langle \Phi_1 \rangle\\ \Phi_2 = \langle \Phi_2 \rangle\\ S = \langle S \rangle}} &= \frac{1}{\sqrt{2}} [ m_S^2 v_S + m_S'^2 v_S + \frac{\lambda_1''}{12} v_S^3 + \frac{\lambda_2''}{3} v_S^3 + \frac{\lambda_3''}{4} v_S^3 \nonumber\\ 
    & \quad  + \frac{v_S}{2}(\lambda_1' v_1^2 + \lambda_2' v_2^2) + v_S(\lambda_4' v_1^2 + \lambda_5' v_2^2) ] .
\end{align}
\end{subequations}

\section{Basis change}
We can choose to either work in the interaction basis, using the parameters as in Eq.~\ref{eq:inte_basis_params}, or in the mass basis, using the parameters as in Eq.~\ref{eq:mass_basis_params} as input parameters. Since the mass basis is more physical we choose to work in this basis. The corresponding parameters in the interaction basis can be obtained via the following basis change equations:
\begin{align}\label{eq:basis_change}
    m^2_{12} 
    &= \Tilde{\mu}^2 \cdot \sin\beta\cos\beta \nonumber\\
    \lambda_1 
    &=\frac{1}{v^2\cos^2\beta}(\Sigma^3_{i=1} m^2_i R^2_{i1}-\Tilde{\mu}^2\sin^2\beta), \nonumber \\
    \lambda_2 
    &= \frac{1}{v^2\sin^2\beta}(\Sigma^3_{i=1} m^2_i R^2_{i2}-\Tilde{\mu}^2\cos^2\beta),\nonumber \\
    \lambda_3 
    &= \frac{1}{v^2}(\frac{1}{\sin\beta\cos\beta}\Sigma^3_{i=1} m^2_i R_{i1}R_{i2}-\Tilde{\mu}^2+2m^2_{H^{\pm}}),\nonumber \\
    \lambda_4 
    &= \frac{1}{v^2}(m^2_A+\Tilde{\mu}^2-2m^2_{H^{\pm}}),\nonumber \\
    \lambda_5 
    &= \frac{1}{v^2}(-m^2_A+\Tilde{\mu}^2), \nonumber \\
    \lambda^{\prime}_1
    &=\frac{1}{2} (\frac{1}{vv_S\cos\beta}\Sigma^3_{i=1} m^2_iR_{i1}R_{i3} + \lambda_{14}'),\nonumber \\
    \lambda^{\prime}_2
    &=\frac{1}{2}(\frac{1}{vv_S\sin\beta}\Sigma^3_{i=1} m^2_iR_{i2}R_{i3} + \lambda_{25}'),\nonumber \\
    \lambda^{\prime}_4
    &= \frac{1}{4} (\frac{1}{vv_S\cos\beta}\Sigma^3_{i=1} m^2_iR_{i1}R_{i3} -\lambda_{14}'), \nonumber \\
    \lambda^{\prime}_5
    &= \frac{1}{4}(\frac{1}{vv_S\sin\beta}\Sigma^3_{i=1} m^2_iR_{i2}R_{i3} -\lambda_{25}'), \nonumber \\
    \lambda_1''& = \lambda_{13}''+\lambda_3''\nonumber\\
    \lambda_3^{{\prime}{\prime}}
    &= \frac{3}{4 v_S^2} 
    (\Sigma_{i=1}^{3} m_i^2R_{i3}^2 -  \lambda_{13}'' v_S^2 - 4\lambda_2'' v_S^2) ,\nonumber \\
    m_S'^2
    &=  - (\frac{1}{2}m_{A_S}^2 + \frac{1}{4}\Sigma_{i=1}^{3} m_i^2 (R_{i3}^2 + R_{i1}R_{i3}\frac{v \cos\beta}{v_S} + R_{i2}R_{i3}\frac{v \sin\beta}{v_S})  \nonumber\\
    &\quad  -\frac{v^2}{4}(\lambda_{14}'\cos^2\beta + \lambda_{25}'\sin^2\beta) + \frac{v_S^2}{8}\lambda_{13}'')
\end{align}
where $m_i$, with $i=1,2,3$, are the masses of the three scalars and $R_{ij}$, with $i,j=1,2,3$ are the entries of the scalar rotation matrix in Table~\ref{tab:2HDMS_eigenfields}. {The $\lambda_2''$ is a free parameter in these expressions.}

The scalar rotation matrix can be expressed via the mixing angles $\alpha_1$, $\alpha_2$ and $\alpha_3$:
\begin{align}\label{eq:scalar_rotation_matrix}
    R = 
    \begin{pmatrix}
    c_{\alpha_1}  c_{\alpha_2}  & s_{\alpha_1} c_{\alpha_2} & s_{\alpha_2} \\
    -s_{\alpha_1} c_{\alpha_3} - c_{\alpha_1} s_{\alpha_2} s_{\alpha_3} & c_{\alpha_1} c_{\alpha_3} - s_{\alpha_1} s_{\alpha_2} s_{\alpha_3}& c_{\alpha_2} s_{\alpha_3}\\
    s_{\alpha_1} s_{\alpha_3} - c_{\alpha_1} s_{\alpha_2} c_{\alpha_3} & - c_{\alpha_1} s_{\alpha_3 }- s_{\alpha_1} s_{\alpha_2} c_{\alpha_3} &  c_{\alpha_2} c_{\alpha_3}\\ 
    \end{pmatrix} ,
\end{align}
where the short-hand notation $s_\alpha = \sin(\alpha)$, $c_\alpha = \cos(\alpha)$ is used.

\section{Benchmark points}
In Table~\ref{tab:bpdm55_w95_mass_basis} - \ref{tab:bpdm1000_mass_basis} we show the mass basis parameters of the benchmark points considered in this work.
For the benchmarks including the excess at $95 \, \text{GeV}$ we also provide the $\chi^2$ values for the observed signal strength in the $b\bar{b}$ and the $\gamma\gamma$ channel given in Eq.~\ref{eq:95_excess_signal_strength_obs}.

Table~\ref{tab:dm_observ_all_bp} shows DM observables, namely relic density, direct detection cross-section, indirect detection cross-section and invisible branching ratios, for the considered benchmark points.

\begin{table}[h!]
    \centering
    \addtolength{\tabcolsep}{-3pt}
    \small
    \begin{tabular}{|c|c|c|c|c|c|}
        \hline
        $m_{h_1}$ & $m_{h_2}$ & $m_{h_3}$ & $m_A$ & $m_{H^\pm}$ & $\chi^2$\\ 
        \hline
        $95.4 \, \text{GeV}$ & $125.09 \, \text{GeV}$ & $650 \, \text{GeV}$ & $800 \, \text{GeV}$ & $800 \, \text{GeV}$ & 0.973 \\
        \hline
        $m_{A_S}$ & $\lambda_1'-2\lambda_4'$ & $\lambda_2'-2\lambda_5'$& $\lambda_1''-\lambda_3''$  &  $\tan \beta$ & \\
        \hline
        $55.596 \, \text{GeV}$ & $0.0020912$ & $0.00074611$ & $-0.025735$ & $2$ & \\ 
        \hline  
        $v_S$ & $\Tilde{\mu}$ & $\alpha_1$ & $\alpha_2$ & $\alpha_3$ & \\
        \hline
        $300 \, \text{GeV}$ & $650 \, \text{GeV}$ & $1.213$ & $-1.270$ & $-1.486$ & \\
        \hline 
    \end{tabular}
    \caption{The benchmark point $\textbf{DM55}_\textbf{w95}$ in the mass basis.}
    \label{tab:bpdm55_w95_mass_basis}
\end{table}

\begin{table}[h!]
    \centering
    \addtolength{\tabcolsep}{-3pt}
    \small
    \begin{tabular}{|c|c|c|c|c|c|}
        \hline
        $m_{h_1}$ & $m_{h_2}$ & $m_{h_3}$ & $m_A$ & $m_{H^\pm}$ & $\chi^2$\\ 
        \hline
        $95.4 \, \text{GeV}$ & $125.09 \, \text{GeV}$ & $700 \, \text{GeV}$ & $700 \, \text{GeV}$ & $700 \, \text{GeV}$ & 0.983\\
        \hline
        $m_{A_S}$ & $\lambda_1'-2\lambda_4'$ & $\lambda_2'-2\lambda_5'$ & $\lambda_1''-\lambda_3''$ &  $\tan \beta$ & \\
        \hline
        $156 \, \text{GeV}$ & $12.753$ & $-0.31351$ & $-2.6747$ & $6.6$ & \\ 
        \hline  
        $v_S$ & $\Tilde{\mu}$ & $\alpha_1$ & $\alpha_2$ & $\alpha_3$ & \\
        \hline
        $239.86 \, \text{GeV}$ & $700 \, \text{GeV}$ & $1.4661$ & $1.1920$ & $-1.5989$ & \\
        \hline 
    \end{tabular}
    \caption{The benchmark point $\textbf{DM156}_\textbf{w95}$ in the mass basis.}
    \label{tab:bpdm156_w95_mass_basis}
\end{table}


\begin{table}[h!]
    \centering
    \addtolength{\tabcolsep}{-3pt}
    \small
    \begin{tabular}{|c|c|c|c|c|c|}
        \hline
        $m_{h_1}$ & $m_{h_2}$ & $m_{h_3}$ & $m_A$ & $m_{H^\pm}$ & $\chi^2$\\ 
        \hline
        $95.4 \, \text{GeV}$ & $125.09 \, \text{GeV}$ & $2900 \, \text{GeV}$ & $2900 \, \text{GeV}$ & $2900 \, \text{GeV}$ & 0.997 \\
        \hline
        $m_{A_S}$ & $\lambda_1'-2\lambda_4'$ & $\lambda_2'-2\lambda_5'$ & $\lambda_1''-\lambda_3''$ &  $\tan \beta$ & \\
        \hline
        $1000 \, \text{GeV}$ & $7.616$ & $0$ & $-0.463$ & $5$ & \\ 
        \hline  
        $v_S$ & $\Tilde{\mu}$ & $\alpha_1$ & $\alpha_2$ & $\alpha_3$ & \\
        \hline
        $1000 \, \text{GeV}$ & $2900 \, \text{GeV}$ & $1.358$ & $-1.223$ & $1.554$ & \\
        \hline 
    \end{tabular}
    \caption{The benchmark point $\textbf{DM1000}_\textbf{w95}$ in the mass basis.}
    \label{tab:bpdm1000_w95_mass_basis}
\end{table}

\begin{table}[h!]
    \centering
    \addtolength{\tabcolsep}{-3pt}
    \small
    \begin{tabular}{|c|c|c|c|c|}
        \hline
        $m_{h_1}$ & $m_{h_2}$ & $m_{h_3}$ & $m_A$ & $m_{H^\pm}$ \\ 
        \hline
        $800 \, \text{GeV}$ & $125.09 \, \text{GeV}$ & $150 \, \text{GeV}$ & $800 \, \text{GeV}$ & $800 \, \text{GeV}$ \\
        \hline
        $m_{A_S}$ & $\lambda_1'-2\lambda_4'$ & $\lambda_2'-2\lambda_5'$ & $\lambda_1''-\lambda_3''$ &  $\tan \beta$\\
        \hline
        $70 \, \text{GeV}$ & $-0.10783$ & $0.063127$ & $-0.47818$ & $1.3728$ \\ 
        \hline  
        $v_S$ & $\Tilde{\mu}$ & $\alpha_1$ & $\alpha_2$ & $\alpha_3$\\
        \hline
        $219.05 \, \text{GeV}$ & $751.54 \, \text{GeV}$ & $-0.60016$ & $0.042445$ & $-0.054807$ \\
        \hline 
    \end{tabular}
    \caption{The benchmark point \textbf{DM70} in the mass basis.}
    \label{tab:bpdm70_mass_basis}
\end{table}

\begin{table}[h!]
    \centering
    \addtolength{\tabcolsep}{-3pt}
    \small
    \begin{tabular}{|c|c|c|c|c|}
        \hline
        $m_{h_1}$ & $m_{h_2}$ & $m_{h_3}$ & $m_A$ & $m_{H^\pm}$ \\ 
        \hline
        $800 \, \text{GeV}$ & $125.09 \, \text{GeV}$ & $900 \, \text{GeV}$ & $800 \, \text{GeV}$ & $800 \, \text{GeV}$ \\
        \hline
        $m_{A_S}$ & $\lambda_1'-2\lambda_4'$ & $\lambda_2'-2\lambda_5'$ & $\lambda_1''-\lambda_3''$ &  $\tan \beta$\\
        \hline
        $400 \, \text{GeV}$ & $0.077784$ & $0.036923$ & $-0.42725$ & $2.1309$ \\ 
        \hline  
        $v_S$ & $\Tilde{\mu}$ & $\alpha_1$ & $\alpha_2$ & $\alpha_3$  \\
        \hline
        $587.17 \, \text{GeV}$ & $755.39 \, \text{GeV}$ & $-0.41245$ & $-0.0086501$ & $-0.0055431$ \\
        \hline 
    \end{tabular}
    \caption{The benchmark point \textbf{DM400} in the mass basis.}
    \label{tab:bpdm400_mass_basis}
\end{table}

\begin{table}[h!]
    \centering
    \addtolength{\tabcolsep}{-3pt}
    \small
    \begin{tabular}{|c|c|c|c|c|}
        \hline
        $m_{h_1}$ & $m_{h_2}$ & $m_{h_3}$ & $m_A$ & $m_{H^\pm}$ \\ 
        \hline
        $800 \, \text{GeV}$ & $125.09 \, \text{GeV}$ & $2900 \, \text{GeV}$ & $800 \, \text{GeV}$ & $800 \, \text{GeV}$ \\
        \hline
        $m_{A_S}$ & $\lambda_1'-2\lambda_4'$ & $\lambda_2'-2\lambda_5'$ & $\lambda_1''-\lambda_3''$ &  $\tan \beta$\\
        \hline
        $1000 \, \text{GeV}$ & $0.32873$ & $0.21320$ & $-0.41541$ & $1.3414$ \\ 
        \hline  
        $v_S$ & $\Tilde{\mu}$ & $\alpha_1$ & $\alpha_2$ & $\alpha_3$  \\
        \hline
        $2271.3 \, \text{GeV}$ & $768.14 \, \text{GeV}$ & $-0.54917$ & $0.036530$ & $-0.056095$ \\
        \hline 
    \end{tabular}
    \caption{The benchmark point \textbf{DM1000} in the mass basis.}
    \label{tab:bpdm1000_mass_basis}
\end{table}
\clearpage

\begin{table}[h!]
    \centering
    \addtolength{\tabcolsep}{-3pt}
    \scriptsize
    \begin{tabular}{|c|c|c|c|c|c|c|}
        \hline
        &   $\Omega h^2$ & $\sigma_{\text{p}A_S}/\text{pb}$ & $\sigma_{\text{n}A_S}/\text{pb}$ & $\sigma_{A_S A_S \rightarrow XX}/\frac{\text{cm}^3}{\text{s}}$ & $BR(h_3 \rightarrow A_S A_S)$ & $BR(h_2 \rightarrow A_S A_S)$ \\ 
        \hline
        \hline
        $\textbf{DM55}_\textbf{w95}$ & 0.11& 4.21 $\times10^{-12}$ & 4.08 $\times10^{-12}$&1.98$\times 10^{-28}$   &3.81$\times10^{-9}$ &0.0199 \\
        \hline
        $\textbf{DM156}_\textbf{w95}$ &$1.61 \times 10^{-4}$ & $3.903 \times10^{-11}$ & $4.160 \times 10^{-11}$ & $3.875\times10^{-29}$ &0.69 &-\\
        \hline
      $\textbf{DM1000}_\textbf{w95}$  & 0.111 & 3.323 $\times 10^{-11}$& 3.369$\times 10^{-11}$   & 2.045$\times10^{-26}$&0.0359 & - \\
        \hline
         \textbf{DM70} & 0.113  & 8.938 $\times 10^{-16}$ & 2.651$\times 10^{-13}$ & 2.13$\times 10^{-28}$ & 0.99934 & - \\
        \hline
     \textbf{DM400} &  0.106 & 4.960$\times 10^{-11}$  & 5.101$\times 10^{-11}$ & 3.67$\times 10^{-26}$ & 0.82203 & - \\
            \hline
        \textbf{DM1000}  & 0.117  & $8.263\times 10^{-11}$ & $8.464\times 10^{-11}$ & $2.018\times 10^{-26}$ &0.005  &-   \\
        \hline
        
    \end{tabular}
    \caption{The DM observables: relic density $\Omega h^2$, direct detection cross-section $\sigma_{\text{p/n} A_S}$ for DM scattering on protons (p) and on neutrons (n), indirect detection cross-section $\sigma_{A_S A_S \rightarrow XX}$, invisible branching ratio $BR(h_i \rightarrow A_S A_S)$, with $i=2,3$, of the heavy scalar ($h_3$) and the SM-like scalar ($h_2$) to two DM particles $A_S$, for all considered benchmark points. The direct and indirect detection cross-sections are re-scaled according to the relic density.}
    \label{tab:dm_observ_all_bp}
\end{table}

\section{Feynman diagrams}
\label{sec:appFD}
\begin{figure}[ht]
    \centering
    ~\hfill\begin{tikzpicture}[scale=0.8, baseline={(0,1)}, transform shape]
    \begin{feynman}
        \vertex (l3);
        \vertex[above left=of l3] (l1);
        \vertex[below left=of l3] (l2);
        \vertex[left=of l1] (i1) {\(g\)};
        \vertex[left=of l2] (i2) {\(g\)};
        \vertex[right=of l3] (f1) {\(h_{1,2,3}\)};
        \diagram* {(i1) --[gluon] (l1) --[fermion] (l3) --[fermion] (l2) --[fermion, edge label'=\(f\)] (l1), (i2) --[gluon] (l2), (l3) --[scalar] (f1),};
    \end{feynman}
    \end{tikzpicture}\hfill\begin{tikzpicture}[scale=0.8, baseline={(0,1)}, transform shape]
    \begin{feynman}
        \vertex (l2);
        \vertex[above left=of l2] (l1);
        \vertex[below left=of l2] (l3);
        \vertex[left=of l1] (i1) {\(q\)};
        \vertex[left=of l3] (i2) {\(q\)};
        \vertex[right=of l2] (f2) {\(h_{1,2,3}\)};
        \vertex[above=of f2] (f1) {\(q\)};
        \vertex[below=of f2] (f3) {\(q\)};
        \diagram* {(i1) --[fermion] (l1) --[boson, edge label'=\(V\)] (l2) --[boson, edge label'=\(V\)] (l3), (i2) --[fermion] (l3), (l2) --[scalar] (f2), (l1) --[fermion] (f1), (l3) --[fermion] (f3)};
    \end{feynman}
    \end{tikzpicture}\hfill~
    \caption{Feynman diagrams for gluon gluon fusion (GGF) and vector boson fusion (VBF) processes, created using Ref.~\cite{Feyn_diag:Ellis_2017}}
    \label{fig:feynlhc}
\end{figure}
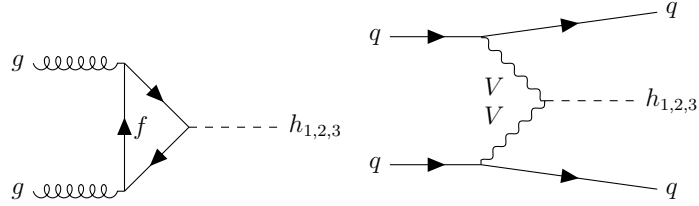

\begin{figure}[ht]
    \centering
    \hfill
    \begin{tikzpicture}[scale=0.8, baseline={(0,1)}, transform shape]
    \begin{feynman}
        \vertex (v2); 
        \vertex[above left=of v2] (v1); 
        \vertex[below left=of v2] (v3); 
        \vertex[right=of v2] (v4) {\(h_{1,2,3}\)}; 
        \vertex[left=of v1] (i1) {\(g\)};
        \vertex[left=of v3] (i2) {\(g\)};
        \vertex[above right =of v2] (f1) {\(b\)};
        \vertex[below right =of v2] (f2) {\(\Bar{b}\)};
        \diagram* {
        (i1) --[gluon] (v1),
        (i2) --[gluon] (v3),
        (f2) --[fermion] (v3) --[fermion] (v2) --[fermion] (v1) --[fermion] (f1),
        (v2) --[scalar] (v4)
        };
    \end{feynman}
    \end{tikzpicture}
    \hfill
    \begin{tikzpicture}[scale=0.8, baseline={(0,1)}, transform shape]
    \begin{feynman}
        \vertex (b) ;
        \vertex[right=of b] (c);
        \vertex[above right=0.7 cm and 0.7 cm of c] (d);
        \vertex[below right=0.7 cm and 0.7 cm of d] (e) {\(h_{1,2,3}\)};
        \vertex[above left=1.4 cm and 1.4 cm of b] (i1) {\(q\)};
        \vertex[below left=1.4 cm and 1.4 cm of b] (i2) {\(\Bar{q}\)};
        \vertex[above right=1.4 cm and 1.4 cm of c] (f1) {\(b\)};
        \vertex[below right=1.4 cm and 1.4 cm of c] (f2) {\(\Bar{b}\)};
        \vertex[below right=.2 cm and .4 cm of b] (lg) {\(g\)}; 
        \diagram* {
        (i1) --[fermion] (b) --[fermion] (i2),
        (b) --[gluon] (c),
        (f2) --[fermion] (c) --[fermion] (f1),
        (d) --[scalar] (e)
        };
    \end{feynman}
    \end{tikzpicture}
    \hfill
    \begin{tikzpicture}[scale=0.8, baseline={(0,1)}, transform shape]
    \begin{feynman}
        \vertex (b) ;
        \vertex[right=of b] (c) {\(h_{1,2,3}\)};
        \vertex[above left=of b] (i1) {\(b\)};
        \vertex[below left=of b] (i2) {\(\Bar{b}\)};
        \diagram* {
        (i1) -- [fermion] (b) -- [fermion] (i2), 
        (b) -- [scalar]  (c)
        };
    \end{feynman}
    \end{tikzpicture}\hfill
    \caption{Feynman diagrams for $bbh_i$, with $i=1,2,3$, processes at the LHC. {The left and middle diagrams correspond to 
    four-flavor (4f) scheme and the right diagram corresponds to the  five-flavor (5f) scheme}, created using Ref.~\cite{Feyn_diag:Ellis_2017}.}
    \label{fig:feynlhc2}
\end{figure}
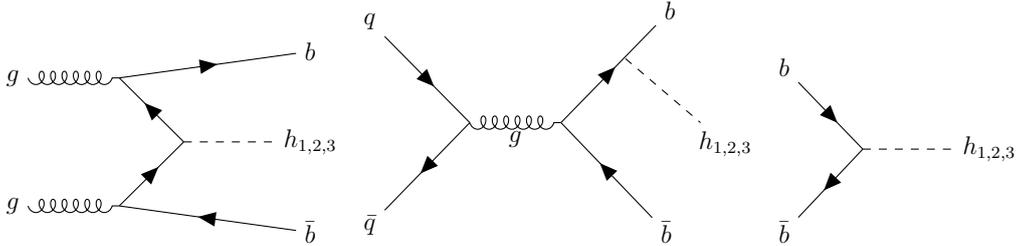

\begin{figure}[ht]
    \centering
    \begin{tikzpicture}[scale=0.8, transform shape]
    \begin{feynman}[small]
        \vertex (b) ;
        \vertex[right=of b] (c);
        \vertex[above left=of b] (i3) {\(e^+/\mu^+\)};
        \vertex[below left=of b] (i4) {\(e^-/\mu^-\)};
        \vertex[above right=of c] (f3) {\(A_S\)};
        \vertex[below right=of c] (f4) {\(A_S\)};
        \diagram* {
        (i4) -- [fermion] (b) -- [fermion] (i3), 
        (b) -- [scalar, edge label'=\(h_{1,2,3}\)]  (c), (f3) -- [scalar] (c) -- [scalar] (f4),};
    \end{feynman}
    \end{tikzpicture}
    \begin{tikzpicture}[scale=0.8, transform shape]
    \begin{feynman}[small]
        \vertex (a) ;
        \vertex[above=of a] (b);
        \vertex[right=of b] (c);
        \vertex[above left=of b] (i3) {\(e^+/\mu^+\)};
        \vertex[below left=of a] (i4) {\(e^-/\mu^-\)};
        \vertex[above right=of c] (f3) {\(A_S\)};
        \vertex[below right=of c] (f4) {\(A_S\)};
        \vertex[below=of f4] (f5) {\(\gamma\)};
        \diagram* {
        (i4) -- [fermion] (a) --[fermion] (b) -- [fermion] (i3), 
        (b) -- [scalar, edge label'=\(h_{1,2,3}\)]  (c), (f3) -- [scalar] (c) -- [scalar] (f4), (a) --[photon] (f5)};
    \end{feynman}
    \end{tikzpicture}
    \caption{Feynman diagrams for $A_SA_S$ production process and the process with additional mono-photon at lepton colliders, created using Ref.~\cite{Feyn_diag:Ellis_2017}.} 
    \label{fig:proc_xx}
\end{figure}
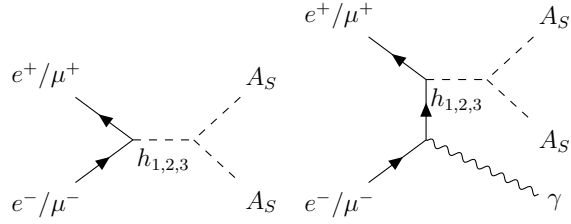

\begin{figure}
    \centering
    \begin{tikzpicture}[scale=0.8, transform shape]
    \begin{feynman}[small]
        \vertex (b) ;
        \vertex[right=of b] (c);
        \vertex[above left=of b] (i3) {\(e^+/\mu^+\)};
        \vertex[below left=of b] (i4) {\(e^-/\mu^-\)};
        \vertex[above right=of c] (d);
        \vertex[above right=of d] (f3) {\(A_S\)};
        \vertex[below right=of d] (f4) {\(A_S\)};
        \vertex[below=of f4] (f5) {\(Z\)};
        \diagram* {
        (i4) --[fermion] (b) --[fermion] (i3), 
        (b) --[boson, edge label'=\(Z\)]  (c), (c) --[scalar, edge label'=\(h_{1,2,3}\)] (d), (f3) --[scalar] (d) --[scalar] (f4), (c) --[boson] (f5)};
    \end{feynman}
    \end{tikzpicture}
    \begin{tikzpicture}[scale=0.8, transform shape]
    \begin{feynman}[small]
        \vertex (b) ;
        \vertex[right=of b] (c);
        \vertex[above left=of b] (i3) {\(e^+/\mu^+\)};
        \vertex[below left=of b] (i4) {\(e^-/\mu^-\)};
        \vertex[above right=of c] (d);
        \vertex[above right=of d] (f3) {\(A_S\)};
        \vertex[below right=of d] (f4) {\(A_S\)};
        \vertex[below=of f4] (f5) {\(Z\)};
        \diagram* {
        (i4) --[fermion] (b) --[fermion] (i3), 
        (b) --[scalar, edge label'=\(A\)]  (c), (c) --[scalar, edge label'=\(h_{1,2,3}\)] (d), (f3) --[scalar] (d) --[scalar] (f4), (c) --[boson] (f5)};
    \end{feynman}
    \end{tikzpicture}
    \begin{tikzpicture}[scale=0.8, transform shape]
    \begin{feynman}[small]
        \vertex (a) ;
        \vertex[above=of a] (b);
        \vertex[right=of b] (c);
        \vertex[above left=of b] (i3) {\(e^+/\mu^+\)};
        \vertex[below left=of a] (i4) {\(e^-/\mu^-\)};
        \vertex[above right=of c] (f3) {\(A_S\)};
        \vertex[below right=of c] (f4) {\(A_S\)};
        \vertex[below=of f4] (f5) {\(Z\)};
        \diagram* {
        (i4) -- [fermion] (a) --[fermion] (b) -- [fermion] (i3), 
        (b) -- [scalar, edge label'=\(h_{1,2,3}\)]  (c), (f3) -- [scalar] (c) -- [scalar] (f4), (a) --[photon] (f5)};
    \end{feynman}
    \end{tikzpicture}
    \caption{Feynman diagrams for $Z A_SA_S$ production processes at lepton colliders, created using Ref.~\cite{Feyn_diag:Ellis_2017}.}
    \label{fig:proc_zxx}
\end{figure}
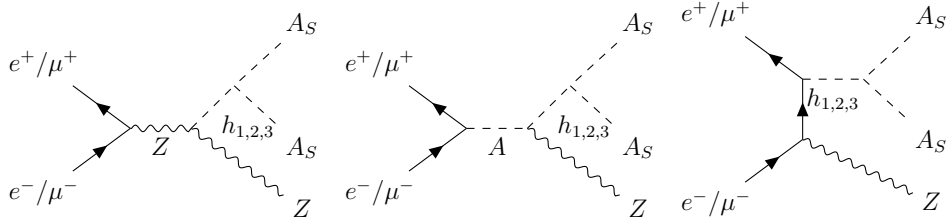

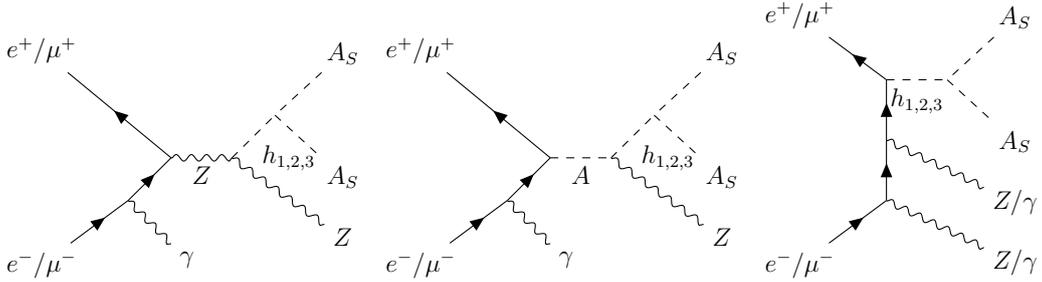
\begin{figure}
    \centering
    \begin{tikzpicture}[scale=0.8, transform shape]
    \begin{feynman}[small]
        \vertex (b) ;
        \vertex[right=of b] (c);
        \vertex[below left=of b] (a);
        \vertex[below left=of a] (i4) {\(e^-/\mu^-\)};
        \vertex[below right=of a] (f1) {\(\gamma\)};
        \vertex[above left=of b] (b2);
        \vertex[above left=of b2] (i3) {\(e^+/\mu^+\)};
        \vertex[above right=of c] (d);
        \vertex[above right=of d] (f3) {\(A_S\)};
        \vertex[below right=of d] (f4) {\(A_S\)};
        \vertex[below=of f4] (f5) {\(Z\)};
        \diagram* {
        (i4) --[fermion] (a) --[fermion] (b) --[fermion] (i3), 
        (b) --[boson, edge label'=\(Z\)]  (c), (c) --[scalar, edge label'=\(h_{1,2,3}\)] (d), (f3) --[scalar] (d) --[scalar] (f4), (c) --[boson] (f5), (a) --[photon] (f1)};
    \end{feynman}
    \end{tikzpicture}
    \begin{tikzpicture}[scale=0.8, transform shape]
    \begin{feynman}[small]
        \vertex (b) ;
        \vertex[right=of b] (c);
        \vertex[below left=of b] (a);
        \vertex[below left=of a] (i4) {\(e^-/\mu^-\)};
        \vertex[below right=of a] (f1) {\(\gamma\)};
        \vertex[above left=of b] (b2);
        \vertex[above left=of b2] (i3) {\(e^+/\mu^+\)};
        \vertex[above right=of c] (d);
        \vertex[above right=of d] (f3) {\(A_S\)};
        \vertex[below right=of d] (f4) {\(A_S\)};
        \vertex[below=of f4] (f5) {\(Z\)};
        \diagram* {
        (i4) --[fermion] (a) --[fermion] (b) --[fermion] (i3), 
        (b) --[scalar, edge label'=\(A\)]  (c), (c) --[scalar, edge label'=\(h_{1,2,3}\)] (d), (f3) --[scalar] (d) --[scalar] (f4), (c) --[boson] (f5), (a) --[photon] (f1)};
    \end{feynman}
    \end{tikzpicture}
    \begin{tikzpicture}[scale=0.8, transform shape]
    \begin{feynman}[small]
        \vertex (a) ;
        \vertex[above=of a] (b);
        \vertex[above=of b] (b2);
        \vertex[right=of b2] (c);
        \vertex[above left=of b2] (i3) {\(e^+/\mu^+\)};
        \vertex[below left=of a] (i4) {\(e^-/\mu^-\)};
        \vertex[above right=of c] (f3) {\(A_S\)};
        \vertex[below right=of c] (f4) {\(A_S\)};
        \vertex[below=of f4] (f5) {\(Z/\gamma\)};
        \vertex[below=of f5] (f6) {\(Z/\gamma\)};
        \diagram* {
        (i4) -- [fermion] (a) --[fermion] (b) --[fermion] (b2) -- [fermion] (i3), 
        (b2) -- [scalar, edge label'=\(h_{1,2,3}\)]  (c), (f3) -- [scalar] (c) -- [scalar] (f4), (b) --[boson] (f5), (a) --[boson] (f6)};
    \end{feynman}
    \end{tikzpicture}
    \caption{Feynman diagrams for $Z A_SA_S \gamma$ production processes at lepton colliders, created using Ref.~\cite{Feyn_diag:Ellis_2017}.}
    \label{fig:proc_zxxisr}
\end{figure}

\begin{figure}
    \centering
    \begin{tikzpicture}[scale=0.8, transform shape]
    \begin{feynman}
        \vertex (a);
        \vertex[below= 2.6 cm of a] (b);
        \vertex[right=of a] (c);
        \vertex[right=of b] (d);
        \vertex[above left= 0.7 cm and 1.6 cm of a] (i1) {\(\mu^+\)};
        \vertex[below left= 0.7 cm and 1.6 cm of b] (i2) {\(\mu^-\)};
        \vertex[above right= 0.7 cm and 1.2 cm of c] (f1) {\(A_S\)};
        \vertex[below right= 0.7 cm and 1.2 cm of c] (f2) {\(A_S\)};
        \vertex[above right= 0.7 cm and 1.2 cm of d] (f3) {\(\Bar{b}\)};
        \vertex[below right= 0.7 cm and 1.2 cm of d] (f4) {\(b\)};
        \diagram* {
        (i2) --[fermion] (b) --[fermion] (a) --[fermion] (i1),
        (a) --[scalar, edge label'=\(h_{1,2,3}\)]  (c), 
        (b) --[scalar, edge label'=\(A\)] (d), 
        (f1) --[scalar] (c) --[scalar] (f2), (f3) --[fermion] (d) --[fermion] (f4),
        };
    \end{feynman}
    \end{tikzpicture}    
    \begin{tikzpicture}[scale=0.8, transform shape]
    \begin{feynman}
        \vertex (b) ;
        \vertex[right=of b] (c);
        \vertex[above left= 1.6 cm and 1.6 cm of b] (i1) {\(\mu^+\)};
        \vertex[below left= 1.6 cm and 1.6 cm of b] (i2) {\(\mu^-\)};
        \vertex[above right= 1.3 cm and 1.3 cm of c] (d); 
        \vertex[above right= 0.7 cm and 1.2 cm of d] (f1) {\(A_S\)};
        \vertex[below right= 0.7 cm and 1.2 cm of d] (f2) {\(A_S\)};
        \vertex[below right= 1.3 cm and 1.3 cm of c] (e); 
        \vertex[above right= 0.7 cm and 1.2 cm of e] (f3) {\(\Bar{b}\)};
        \vertex[below right= 0.7 cm and 1.2 cm of e] (f4) {\(b\)};
        \diagram* {
        (i2) --[fermion] (b) --[fermion] (i1),
        (b) --[scalar, edge label'=\(h_{1,2,3}\)]  (c), 
        (c) --[scalar, edge label'=\(h_{1,2,3}\)] (d), 
        (f1) --[scalar] (d) --[scalar] (f2), 
        (c) --[scalar, edge label'=\(h_{1,2,3}\)] (e), 
        (f3) --[fermion] (e) --[fermion] (f4),        
        };
    \end{feynman}
    \end{tikzpicture}
    \caption{Feynman diagrams for $b \Bar{b}$+MET production processes at a muon collider, created using Ref.~\cite{Feyn_diag:Ellis_2017}. Similar diagrams will take part in producing the final state $t\bar t$ + MET.}
    \label{fig:feyn_mumu_to_bbplusmet}
\end{figure}
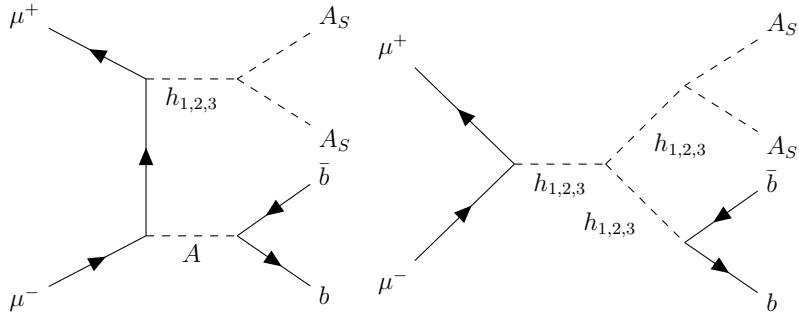
\clearpage

\bibliographystyle{JHEP}   
\bibliography{ref,ref-2}

\end{document}